\documentclass[aps,%
reprint,
superscriptaddress,
amsmath,amssymb,floatfix,
pre
]{revtex4-2}
\usepackage{lipsum}
\usepackage{hhline}
\usepackage{xtab,afterpage}
\usepackage{ upgreek }
\usepackage{mathrsfs}
\usepackage{booktabs}
\usepackage{pifont}
\usepackage{xcolor}
\newcommand{\cmark}{\textcolor{green!50!black}{\ensuremath{\pmb{\checkmark}}}}
\newcommand{\xmark}{\textcolor{red!70!black}{\ensuremath{\pmb{\times}}}}
\usepackage{graphicx}
\usepackage{comment}
\usepackage{bm}
\usepackage{xcolor}
\usepackage{makecell}
\usepackage{amsmath,amsfonts,amssymb}
\DeclareMathAlphabet{\mathbbold}{U}{bbold}{m}{n}
\definecolor{darkblue}{rgb}{0,0,0.6}
\usepackage{hyperref}
\hypersetup{colorlinks,linkcolor=darkblue,citecolor=darkblue,urlcolor=darkblue}

\usepackage[normalem]{ulem}

\newenvironment{symbollist}
  {\begin{list}{}%
    {\setlength{\labelwidth}{0.28\linewidth}%
     \setlength{\leftmargin}{0.31\linewidth}%
     \setlength{\labelsep}{0.02\linewidth}%
     \setlength{\itemsep}{0.25em}%
     \setlength{\parsep}{0pt}%
     \setlength{\topsep}{0.5em}%
     }}
  {\end{list}}
\newcommand{\sym}[2]{\item[#1] #2}

\begin{document}

\title{Kinetic and Hydrodynamic Theories of Chiral Intruder Dynamics in Nonequilibrium Baths}

\author{Raphaël Maire}
\email{maire@ub.edu}
\affiliation{Departament de Física de la Mat\`eria Condensada, Universitat de Barcelona, C. Martí Franqu\`es 1, 08028 Barcelona, Spain}
\author{Ignacio Pagonabarraga}
\email{ipagonabarraga@ub.edu}
\affiliation{Departament de Física de la Mat\`eria Condensada, Universitat de Barcelona, C. Martí Franqu\`es 1, 08028 Barcelona, Spain}
\affiliation{University of Barcelona Institute of Complex Systems (UBICS), Universitat de Barcelona, 08028 Barcelona, Spain}

\begin{abstract}
We study the chiral dynamics of an intruder immersed in a nonequilibrium bath in two complementary limits: the dilute kinetic regime and the dense hydrodynamic regime. In the dilute limit, starting from a Boltzmann-Lorentz description, we derive an effective Langevin equation whose coefficients are given explicitly by geometry-dependent boundary integrals. This formulation separates the effects of intruder-shape chirality from those of chiral intruder-bath interactions. We find that the chiral interactions generate an odd response and a torque, whereas the chirality of the intruder leads to a ratchet effect. We also show that fluctuation-dissipation-like relations exist and that certain symmetry-allowed couplings vanish in the dilute regime. In the dense regime, we argue that intruder dynamics are governed primarily by bath hydrodynamics and torque-density-driven edge currents not captured by the previous framework. These currents can generate both an antisymmetric drag and a curvature-induced torque, leading to an antisymmetric response when inertia is accounted for. Taken together, these results provide a step toward understanding the mechanisms governing the chiral dynamics of an intruder in a nonequilibrium bath across different scales.
\end{abstract}
\maketitle

\tableofcontents

\section{Introduction and overview of the results}

Chiral active matter comprises systems that are both out of equilibrium and parity-breaking~\cite{fruchart2026nonreciprocal}. Such systems are ubiquitous in biology. Sperm cells~\cite{riedel2005self,friedrich2007chemotaxis}, microorganisms~\cite{grober2023unconventional,nishiguchi2025vortex,villalobos2025active,di2011swimming, takeuchi2026variousphasesactivematter}, microtubules~\cite{afroze2021monopolar}, and cells~\cite{chen2025chirality,yashunsky2022chiral,hoffmann2020chiral} can all be chiral, giving rise to key biological mechanisms such as left-right differentiation during organ development~\cite{inaki2016cell,juan2018myosin1d}, disease regulation~\cite{fan2018cell} or the organization of microtubule bundles~\cite{novak2018mitotic}. Chirality in synthetic systems inspired by biology is also being explored for applications in cargo transport, sensing, and sorting~\cite{chan2024chiral,xu2022sorting,yang2021topologically,valecha2025active}.

This broad experimental relevance has motivated many studies of simplified models, which have revealed that chiral active systems can exhibit an antisymmetric linear response, often referred to as an ‘‘odd response.'' Such a response can manifest itself in the viscosity~\cite{fruchart2023odd,banerjee2017odd,markovich2021odd,markovich2025chiral,avron1998odd}, in the global and self-diffusivity~\cite{hargus2021odd,tan2022odd,faedi2026mobilitybasedapproachtransport,kalz2022collisions,kalz2024oscillatory,kalz2025reversal}, or in the conductivity~\cite{fruchart2022odd}; in each case, they can be derived from the microscopic dynamics~\cite{fruchart2022odd,jiao2026colloidalphoresisoddfluids,maire2026kinetic,lier2026chapman,eren2025collisional}. Odd response also arises for elastic materials, leading to unconventional elastic behavior~\cite{scheibner2020odd,lee2025odd,marconi2025spontaneous, kole2024chirality}. Other hallmarks of chirality include the emergence of edge currents~\cite{caprini2019active,caporusso2024phase,soni2019odd,uchida2026designing,shankar2022topological,langford2025phase, metzger2026equationstateedgeflow, burekovic2026bulk}, active turbulence~\cite{mecke2024chiral,beppu2022exploring,lowen2016chirality}, instabilities~\cite{caprini2025bubble,shen2023collective,digregorio2025phase,guo2025chirality,guo2026tuning}, and large-scale correlations~\cite{shee2024emergent, de2024pattern, maitra2020chiral,musacchio2026circling, zhang2020reconfigurable, gu2025emergence, PhysRevLett.127.238001}, such as hyperuniformity~\cite{kuroda2023microscopic,kuroda2025singular,kuroda2024long,lei2019nonequilibrium,maire2025hyperuniformity,huang2021circular, zhou2025large, gao2025liquidgascriticalityhyperuniformfluids} among others~\cite{levis2019activity,liebchen2022chiral,caprini2025spontaneous,pattanayak2025chirality,wang2025braided,kushwaha2025flocking,maitra2025activity}.

The behavior of an intruder immersed in a chiral medium is likewise important, both theoretically~\cite{zwanzig2001nonequilibrium, van1986brownian, van1992stochastic, widom1971velocity,marconi2008fluctuation, plyukhin2026langevin,plyukhin2006does} and for applications~\cite{hanggi2009artificial}. Intruders in achiral nonequilibrium systems have been studied extensively in both active~\cite{angelani2011active,reichhardt2017ratchet,krishnamurthy2016micrometre,anand2024transport,metzger2025exceptions,bechinger2016active,guidobaldi2014geometrical,ai2014transport,shea2024force,uchida2026designing,alston2026stochastic,benois2023enhanced,granek2022anomalous,demery2011perturbative,dean2011diffusion,granek2024colloquium, zeng2026noveldynamicsinertialpolar} and granular~\cite{costantini2007granular,eshuis2010experimental,gnoli2013brownian,plati2022collective,Semeraro_Gonnella_Lippiello_Sarracino_2023, plati2019dynamical,manacorda2014coulomb,sarracino2013ratchet,sarracino2025nonequilibrium,talbot2011kinetics,sarracino2010granular,lucente2023revealing,Mendez_Garzo_2026,q1vm-13xl, gnoli2013nonequilibrium} matter. These systems typically exhibit similar behaviors, including ratchet effects reminiscent of Brownian motors~\cite{hanggi2009artificial}. Once chirality is introduced, either through the bath or through the shape of the object, odd response can also emerge, together with spontaneous rotation of the intruder~\cite{Poggioli_Limmer_2023,kalz2022collisions,li2023chirality,Grober_Dhar_Saintillan_Palacci_2026,liao2018transport,maitra2019spontaneous,hosaka2023lorentz,puitandy2026spontaneous, goerlich2026particleresolvedrheologicalstudychirality}. Hydrodynamic theories have been developed to account for the emergence of odd response, largely based on odd viscosity~\cite{daddi2026exact,hosaka2023lorentz,lier2023lift,hosaka2024chirotactic,lier2024odd,lier2024slip,everts2024dissipative} with Ref.~\onlinecite{khain2024trading} focusing on the effects of symmetries. However, the dilute regime, ratchet effects, and the role of edge currents in the chiral dynamics of an intruder remain relatively unexplored, at least from a theoretical perspective.
Recently, Refs.~\onlinecite{Passive2025HargusPRE,hargus2025odd} used projection methods to characterize the chiral dynamics of an intruder from symmetry arguments, marking an important step toward understanding the behavior of such an intruder. Nevertheless, such an approach does not by itself disentangle the physical origins of the different effects. For example, if an odd response is observed, is it due to odd viscosity, or does it originate from chiral currents? A purely symmetry-based approach cannot answer such questions on its own. Likewise, a coefficient allowed by symmetry may be vanishingly small in a given regime, or even strictly zero. More generally, chirality can arise in three distinct ways: (i) the intruder itself can have a chiral shape; (ii) the bath can be chiral, in the sense that the particles composing it interact chirally or are themselves chiral, for instance through chiral self-propulsion; and (iii) the interaction between the intruder and the bath particles can be chiral, for example when the bath consists of noninteracting spinners that can transfer angular momentum to the intruder. Distinguishing these three mechanisms requires a direct computation rather than a symmetry argument alone.  Moreover, existing studies typically focus on either dense or dilute baths, leaving the connection between the two limits largely unexplored. This distinction matters. In dilute baths of active spinners, for instance, collisions are essentially Poissonian and uncorrelated, whereas at high density the bath behaves as a continuum and hydrodynamic effects dominate. For example, odd viscosity in the bath surrounding an intruder can induce an odd diffusivity of the intruder~\cite{lier2023lift}. This notion of viscosity is not meaningful in a very dilute bath for understanding the dynamics of an intruder smaller than the mean free path. Conversely, in the dense regime, the microscopic details of intruder-bath interactions are expected to become secondary, with the dynamics instead controlled by collective modes, hydrodynamic fields, and boundary conditions. Such concepts are foreign to a dilute kinetic approach. With this work, we aim to address these questions, at least partially, using quantitative and analytically tractable models.

This paper is accompanied by a Letter~\cite{letter} that summarizes the main results presented here and validates them through direct molecular-dynamics simulations. We focus on an intruder in two regimes. First, in the dilute regime, where the bath interacts with the intruder through uncorrelated collisions, we derive in Sec.~\ref{sec:1} a Langevin equation describing the intruder dynamics. All coefficients entering the Langevin equation, such as the damping or the noise, are obtained as explicit integrals over the geometry of the object. In this regime, the relevant sources of chirality are the chirality of the intruder shape and the chirality of the bath-intruder interaction. We then show, in Sec.~\ref{sec:2}, which coefficients vanish and which do not, based on the intruder geometry. For example, we show explicitly how the chirality of the object, together with the nonequilibrium character of the bath, is sufficient to induce a steady rotation of the intruder, how interaction chirality gives rise to odd response, and that two fluctuation-dissipation-like relations hold. These results are further illustrated through explicit computations for several intruder shapes, such as regular polygons and a chiral wheel, which are then compared with simulations in the Letter~\cite{letter}. The dilute limit is especially useful because it allows us to obtain explicit expressions for the coefficients entering the Langevin equation. It also reveals that some coefficients that are allowed by symmetry arguments are negligible in the dilute limit. For example, although symmetry arguments allow for an odd response for a chiral shape in an achiral nonequilibrium bath, such a term does not arise in our dilute model. We therefore turn to the dense regime, which requires a change of theoretical framework, from a kinetic description dominated by Poissonian collisions to a hydrodynamic one, as explained in Sec.~\ref{sec:Discussion1}. In contrast to the standard picture, where the fluid response and its back-action on the intruder are assumed to be governed primarily by odd viscosity, we argue that another key ingredient is the torque density---or the rotational viscosity in a fluid with internal spin---and the edge currents it induces. We therefore focus on this aspect, recalling in Sec.~\ref{sec:chiral} the modeling of such chiral fluids and their associated edge currents. We then show in Sec.~\ref{sec:Stokes drag} that edge currents produce a lift force, \emph{i.e.} an antisymmetric response of the intruder, but only in the presence of inertia. In Sec.~\ref{sec:torque_generation}, we further show that these edge currents can also generate a torque when they act on an intruder with varying boundary curvature. In Sec.~\ref{sec:linear_formulas}, we analyze the damping matrix using formal expressions obtained from the linear response. We conclude in Sec.~\ref{sec:discussion_conclusion} with a discussion of the remaining open questions and the substantial but exciting work that still lies ahead.

A list of symbols is provided in the Supplemental Material~\cite{supp}.

\section{Derivation of the Langevin equation of the intruder in the dilute regime}\label{sec:1}

In this section, we derive a Langevin equation for the dynamics of a heavy intruder in a dilute bath of particles. To do so, we use standard techniques that can be found, e.g., in Refs.~\onlinecite{costantini2007granular, cleuren2007granular,kanazawa2015minimal}. All assumptions and approximations performed throughout are summarized at the end, in Sec.~\ref{sec:summary}.

\subsection{Collision rule and master equation}

\begin{figure*}
    \centering
    \includegraphics[width=0.9\linewidth]{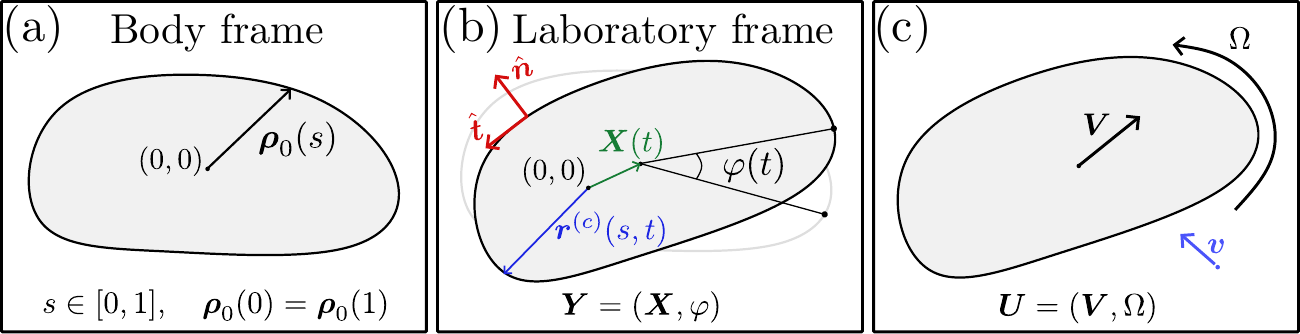}
    \caption{Cartoon of the intruder and its parametrization}
    \label{fig:cartoon}
\end{figure*}

We place an intruder in the system, with its boundary described in the body frame by the parametrization
$\bm{\rho}_0(s)$, with $s\in[0,1]$ and the origin at the center of mass (Fig.~\ref{fig:cartoon}(a)). Accordingly, we specify only the center of mass position and the body shape, without introducing an explicit mass distribution.

In the lab frame, its boundary at time $t$ is
\begin{equation}
\bm{r}^{(c)}(s,t)=\bm{X}(t)+ \bm{r}_0(s,t),\qquad \bm{r}_0(s,t)=\mathsf{R}(\varphi(t))\cdot\bm{\rho}_0(s),
\end{equation}
where $\bm{X}(t)$ and $\varphi(t)$ denote the center-of-mass position and orientation with $\mathsf R(\varphi)$ a rotation matrix of angle $\varphi$ (Fig.~\ref{fig:cartoon}(b)). The intruder has a mass $M$, moment of inertia $I$, velocity $\bm V$, and angular velocity $\Omega$. Its velocities change with time due to interactions with a surrounding bath of particles of mass $m$ and negligible radius, with which it undergoes instantaneous and binary collisions such that its post-collisional velocities $\bm V'$ and angular velocity $\Omega'$ are given by:
\begin{equation}
\quad
\bm V'=\bm V+\frac{\bm J}{M},\quad
\Omega'=\Omega+\frac{[\bm r_0\times \bm J]_z}{I},\quad\bm v'=\bm v-\frac{\bm J}{m},
\label{eq:collision_rule}
\end{equation}
where $\bm v$ is the velocity of the bath particle, $\bm J$ is the impulse change at collision, which can be decomposed into a normal $J_n$ and tangential part $J_t$ with respect to the outward unit normal $\hat{\bm{n}}$ and tangent $\hat{\bm t}=\hat{\bm z}\times \hat{\bm n}$ vector at the point of contact $\bm r^{(c)}$  (Fig.~\ref{fig:cartoon}(c)).

The intruder's contact-point velocity is:
\begin{equation}
    \bm V^{(c)}=\bm V+\Omega\hat{\bm z}\times \bm r_0.
\end{equation}
Therefore, the relative velocity between a bath particle and the intruder at contact is:
\begin{equation}
    \bm g = \bm v-\bm V^{(c)},
\end{equation}
which can be decomposed into a normal and tangential sector:
\begin{equation}
    g_n=\bm g\cdot \hat{\bm n},\quad g_t=\bm g\cdot \hat{\bm t},
\end{equation}
and related to the impulse change at collision by:
\begin{equation}
    \begin{pmatrix} g_n'\\ g_t' \end{pmatrix}=\begin{pmatrix} g_n\\ g_t \end{pmatrix}-\begin{pmatrix}\lambda_n & \kappa_n\kappa_t/I\\\kappa_n\kappa_t/I & \lambda_t \end{pmatrix}\begin{pmatrix} J_n\\ J_t \end{pmatrix},\label{eq:g_to_invert}
\end{equation}
where we defined:
\begin{equation}
    \begin{gathered}
    a=\frac{1}{m}+\frac1M, \quad\kappa_n=[\bm r_0\times \hat{\bm n}]_z,\quad \kappa_t=[\bm r_0\times \hat{\bm t}]_z,\\
    \lambda_n = a+{\kappa_n^2}/{I},\quad\lambda_t = a+{\kappa_t^2}/{I},
    \end{gathered}
\end{equation}
using Eq.~\eqref{eq:collision_rule}.

Next, we define the collision rule sector by sector. In the normal direction, we impose a simple dissipative rebound:
\begin{equation}
    g_n'=-\alpha g_n,
    \label{eq:dissipative_perp_sector}
\end{equation}
with $\alpha$ the coefficient of restitution. The dissipative rebound could equally well be imposed on $J_n$; both choices are valid and reduce to the standard dissipative collision rule for circular particles~\cite{brilliantov2010kinetic}. What matters is that, because the intruder is macroscopic, the collision may be dissipative when $\alpha<1$ and the dynamics are therefore out of equilibrium.

We now specify the collision in the tangential sector. In an achiral smooth system, we would set $J_t=0$, expressing the absence of a tangential momentum change. Note that the condition $g_t'-g_t=0$ does not imply $J_t=0$ in general, and therefore does not correctly describe a smooth interaction for an arbitrary shape, even though it does for simple cases such as disks. For this reason, we formulate the tangential sector directly at the level of the collision impulse and introduce chirality through a tangential momentum kick:
    \begin{equation}
    J_t= 2\left(a+\dfrac{\kappa_t^2}{I}\right)^{-1}\Delta,
    \label{eq:J_t}
\end{equation}
where $\Delta$ has the dimension of a velocity and applies a torque to the intruder, making it rotate in the direction set by the sign of $\Delta$. This choice of the denominator is motivated by the fact that $1/(a+\kappa_t^2/I)$ plays the role of an effective reduced mass at contact in the tangential sector. The collision rule is not uniquely defined, and we adopt a simple form that captures chirality. It may also be interpreted as an effective model for collisions between small active spinners in the bath and a large intruder, as we show in Appendix~\ref{app:spinner}.

Inverting Eq.~\eqref{eq:g_to_invert} and using Eqs.~\eqref{eq:dissipative_perp_sector} and \eqref{eq:J_t}, we find:
\begin{equation}
    J_t=\frac{2\Delta}{\lambda_t},\qquad J_n=\frac{1+\alpha}{\lambda_n}g_n-\frac{\kappa_n\kappa_t}{I\lambda_n}J_t.
    \label{eq:J}
\end{equation}
Eqs.~\eqref{eq:collision_rule} and \eqref{eq:J} fully determine the change of translational and angular velocity at collision, respectively.

When the size of the intruder is smaller than the typical mean free path of the particles, it is reasonable to assume that the velocity distribution  $f(\bm r, \bm v)$ of the bath particles is homogeneous and not affected by the intruder: $f(\bm r, \bm v)\to f(\bm v)$. The velocity distribution may be non-Gaussian if the bath is out of equilibrium. Under this approximation, the rate of change of velocity and angular velocity of the intruder can be obtained from a Boltzmann-Lorentz equation
\begin{equation}
\begin{gathered}
    \dfrac{\partial P(\bm U,\bm Y, t)}{\partial t}+\bm U\cdot \dfrac{\partial P(\bm U, \bm Y, t)}{\partial \bm Y}\!=\!\int d\bm U'\Big[W(\bm U|\bm U',\bm Y)\times\\
    P(\bm U',\bm Y,t)- W(\bm U'|\bm U,\bm Y)P(\bm U,\bm Y,t)\Big],
\end{gathered}
\label{eq:Lorentz}
\end{equation}  
where we defined the ‘‘velocity array'' $\bm U=(\bm V,\Omega)$, $\bm Y=(\bm X, \varphi)$ the ‘‘configuration array'' and $\bm  U\cdot\bm \nabla_{\bm  Y}\equiv \bm  V\cdot\bm \nabla_{\bm  X}+\Omega\partial_\varphi$. $W(\bm U'| \bm U, \bm Y)$ is the rate of change from $\bm U$ to $\bm U'$ at a given $\bm Y$ due to collisions between the bath and the intruder, and is given by~\cite{cleuren2007granular}:
\begin{equation}
\begin{gathered}
W(\bm U'|\bm U,\bm Y)=n_b\oint {ds}\int d\bm v  f(\bm v) \Theta(-g_n)(-g_n)\times\\ \delta \left(\bm V'-\bm V-\frac{\bm J}{M}\right) \delta \left(\Omega'-\Omega-\frac{[\bm r_0\times\bm J]_z}{I}\right),
\end{gathered}
\label{eq:W}
\end{equation}
with $n_b$ the density of bath particles. We introduced the integral over the shape $\oint {ds}\equiv\int_0^1 ds  |\partial_s\bm \rho_0(s)|$, and $\bm r_0(s)=\mathsf R(\varphi)\cdot\bm\rho_0(s)$ is the corresponding lever arm from the intruder center of mass where $g_n$, $\bm J$, $\hat{\bm n}$, and $\hat{\bm t}$ are evaluated.  Note that since the bath is isotropic, $W(\bm U'|\bm U,\bm Y)\equiv W(\bm U'|\bm U,\varphi)$. Rotational covariance implies the stronger identity $W\left((\bm V',\Omega')|(\bm V,\Omega),\varphi\right)=W\left((\mathsf R(-\varphi)\cdot \bm V',\Omega')|(\mathsf R(-\varphi)\cdot\bm V,\Omega),0\right).$ 

The Boltzmann-Lorentz equation \eqref{eq:Lorentz} is simply a master equation with advection. The transition rate $W$ (Eq.~\eqref{eq:W}) is the usual hard-particle collision kernel: it gives the probability per unit time for the intruder to jump from $\bm U$ to $\bm U'$ at fixed configuration $\bm Y$, and includes the incoming flux factor $\Theta(-g_n)(-g_n)$, the integration over all contact points on the boundary, and the collision rule imposed by the delta functions.

For a convex intruder in a dilute bath, it is reasonable to assume that all bath particles that collide with it are drawn from the homogeneous distribution $f(\bm v)$. However, for a non-convex object, this assumption is generally not valid: recollisions are possible, some boundary points may be shadowed and therefore not directly accessible from all incoming directions, and particles may become temporarily trapped in concavities before escaping. All these effects generate correlations between successive collisions and distort the effective velocity distribution of particles colliding with the intruder. 

The dynamics of the intruder are fully determined by Eqs.~\eqref{eq:Lorentz} and \eqref{eq:W}. We emphasize that it is not necessary to specify any collision rule for bath-bath interactions, but only for collisions between the bath particles and the intruder. Indeed, the only nontrivial information about the bath entering the Boltzmann-Lorentz equation is the bath velocity distribution $f$. In particular, any nonequilibrium features of the bath are encoded in a non-Gaussian form of $f$, even when the bath is, for example, chiral. At first glance, this may seem surprising. Even in the absence of chirality in bath-intruder collisions, chirality within the bath itself is expected to induce a torque on the intruder~\cite{puitandy2026spontaneous}. However, such effects become significant only at high densities, where bath correlations are important. This picture breaks down in dilute chiral systems where trajectories remain correlated and particles repeatedly collide with the same region of the intruder over short timescales such as in circle swimmers. We now continue our analysis of the Boltzmann-Lorentz equation by deriving from it a Langevin equation for the intruder dynamics.

\subsection{Kramers-Moyal and van Kampen expansion}

In the following, for any array $\bm Z$, we will equivalently label components by numbers $(0, 1, 2)$ or symbol $(x,y,\varphi)$: 
\begin{equation}
    Z_0 \equiv Z_x,\qquad Z_1 \equiv Z_y,\qquad Z_2 \equiv Z_{\varphi},
\end{equation}
whenever this improves clarity. For example $\bm U=(\bm V, \Omega)=(V_x, V_y, \Omega)=(U_x, U_y, U_\varphi)$. Note also that these three-component arrays do not transform as vectors in $\mathbb{R}^3$ under rotations. Instead, the pair $(Z_x,Z_y)$ rotates as a two-dimensional vector, whereas $Z_{\varphi}$ behaves as a pseudoscalar. Nevertheless, we may define the inner product
\begin{equation}
\bm Z \cdot \bm W \equiv Z_x W_x + Z_y W_y + Z_{\varphi} W_{\varphi},
\end{equation}
on $\mathbb R^2\oplus\mathbb R$, invariant under rotation and reflection. Einstein's summation convention will also be used, such that:
\begin{equation}
    \bm Z \cdot \bm W\equiv\sum_{i} Z_iW_i\equiv Z_iW_i
\end{equation}

\begin{widetext}
\subsubsection{Kramers-Moyal}
Assuming the existence of all jump moments, the Boltzmann-Lorentz equation is formally equivalent to its Kramers-Moyal expansion~\cite{van1992stochastic}. We perform the expansion in the ‘‘momentum array'' $\bm \Pi= (M V_x, M V_y, I\Omega)$:
\begin{equation}
    \dfrac{\partial P(\bm \Pi, \bm Y, t)}{\partial t}+\bm U\cdot \dfrac{\partial P(\bm \Pi, \bm Y, t)}{\partial \bm Y}=\sum_{\substack{\ell_x,\ell_y,\ell_{\varphi}\ge0\\\ell_x+\ell_y+\ell_{\varphi}\ge1}}\frac{(-1)^{\ell_x+\ell_y+\ell_{\varphi}}}{\ell_x! \ell_y! \ell_{\varphi}!}\partial_{MV_x}^{\ell_x}\partial_{MV_y}^{\ell_y}\partial_{I\Omega}^{\ell_{\varphi}}\Big[\mathcal M_{\ell_x\ell_y \ell_{\varphi}}(\bm \Pi,\varphi) P \Big],
\end{equation}
with
\begin{equation}
    \begin{split}
    \mathcal M_{\ell_x\ell_y\ell_{\varphi}}(\bm \Pi,\varphi)&=\int d\bm \Pi' [M(V_x'-V_x)]^{\ell_x}[M(V_y'-V_y)]^{\ell_y}[I(\Omega'-\Omega)]^{\ell_{\varphi}} W(\bm \Pi'|\bm \Pi,\bm Y)\\
    &=n_b\oint {ds}\int d\bm  v f(\bm  v)\Theta(-g_n)(-g_n)(\Delta \Pi_x)^{\ell_x}(\Delta \Pi_y)^{\ell_y}(\Delta\Pi_\varphi)^{\ell_{\varphi}},
    \end{split}
\end{equation}
where $\Delta\Pi_{x, y}=J_{x, y}$ is the change of translational momentum at collision and $\Delta\Pi_{\varphi}=[\bm r_0\times \bm J]_z$ is the change of angular momentum. Together with the collision rule Eq.~\eqref{eq:J}, we find:
\begin{equation}
    \Delta \bm \Pi = \bm G(s,\varphi) g_n + \bm H(s,\varphi)
    \label{eq:collision_rule_no_gt}
\end{equation}
with
\begin{equation}
\bm G=\frac{1+\alpha}{\lambda_n}\bm e^{(n)},\qquad\bm H=-\frac{\kappa_n\kappa_t}{I\lambda_n}J_t \bm e^{(n)}+J_t \bm e^{(t)},\qquad
\bm e^{(n)}=\left(\hat n_x,\hat n_y,\kappa_n\right),\qquad\bm e^{(t)}=\left(\hat t_x,\hat t_y,\kappa_t\right).
\end{equation}

\subsubsection{Partial integration}

Since $g_t$ does not enter into the momentum jumps $\Delta \bm \Pi$, we can directly integrate $f(\bm v)$ along the tangential direction:
\begin{equation}
    \phi(v_n)\equiv\int dv_t f(v_n,v_t) .
\end{equation}
After integration, $\mathcal M$ is rewritten to:
\begin{equation}
    \begin{split}
    \mathcal M_{\ell_x \ell_y \ell_{\varphi}}&=n_b\oint {ds}\int_{-\infty}^{+\infty} dv_n \phi(v_n) \Theta(-g_n)(-g_n) \prod_{i=x,y,\varphi} \left[G_{i} g_n+H_{i}\right]^{\ell_i},\\
    &=\oint {ds}\sum_{p_x=0}^{\ell_x}\sum_{p_y=0}^{\ell_y}\sum_{p_{\varphi}=0}^{\ell_{\varphi}}\binom{\ell_x}{p_x}\binom{\ell_y}{p_y}\binom{\ell_{\varphi}}{p_{\varphi}}G_x^{p_x} H_{x}^{\ell_x-p_x}G_y^{p_y} H_{y}^{\ell_y-p_y}G_{\varphi}^{p_{\varphi}}  H_{\varphi}^{\ell_{\varphi}-p_{\varphi}} \mathcal I_{p_x+p_y+p_{\varphi}}(U_n).
    \end{split}
\label{eq:moment}
\end{equation}
where we defined:
\begin{equation}
    \begin{split}
        \mathcal I_r(U_n)&\equiv n_b\int dv_n \phi(v_n) \Theta(-g_n)(-g_n) (g_n)^r=n_b\int_{-\infty}^{0} dg_n \phi(g_n+U_n) (-g_n) g_n^r,
    \end{split}
    \label{eq:I_integrals}
\end{equation}
with
\begin{equation}
    U_n\equiv\bm  V^{(c)}\cdot \hat{\bm  n}=\bm V\cdot \hat{\bm n}+\Omega(\hat{\bm z}\times \bm r_0)\cdot \hat{\bm n}=\bm V\cdot \hat{\bm n}+\Omega \kappa_n\equiv \bm U\cdot\bm e^{(n)}.
\end{equation}

\subsubsection{Van Kampen expansion}

The Kramers-Moyal expansion is exact, but truncating it at second order to obtain a Fokker-Planck equation would be an uncontrolled step, since it would simply discard all higher jump moments. For equilibrium Brownian motion, such a reduction is justified for a sufficiently heavy intruder, suggesting an expansion in inverse mass. Here, however, the dynamics are nonequilibrium, and higher-order moments need not be negligible \emph{a priori}: non-Gaussian fluctuations may persist even in the large-mass limit. We therefore proceed differently and rewrite the Kramers-Moyal series as an expansion in the small bath-to-intruder mass ratio. This yields a van Kampen expansion, which identifies the relative order of all terms, keeps possible nonequilibrium corrections under control, and provides a systematic extension beyond the Fokker-Planck level without conflicting with Pawula's theorem~\cite{van1992stochastic, PLYUKHIN2005198}.\\
In our case, an important nonequilibrium contribution (the ratchet force and torque) to the first moment will appear at the same order in small mass as the third jump moment. For simplicity, we will neglect this third-order contribution and close the equation at the Fokker-Planck level. The van Kampen expansion nevertheless makes this Gaussian approximation explicit by identifying the discarded term and clarifying its possible effect.

We introduce a bookkeeping parameter $\epsilon$ by writing the bath-particle mass as
\begin{equation}
m \to \epsilon m,
\end{equation}
while keeping $M$, $I$, $\alpha$, $\Delta$, and the tracer geometry fixed. $\epsilon$ will be set to 1 at the end of the computation. In the following, we will compute all jump moments $\mathcal M_{\ell_x\ell_y\ell_{\varphi}}$ to order $\mathcal O(\epsilon^{3/2})$. This order is chosen because it is the lowest at which a ratchet effect (\textit{i.e.}, fluctuation-driven forces and torque) arises.

To obtain all such contributions, we expand $\bm G$ and $\bm H$ as follows:
\begin{equation}
\bm G=(1+\alpha)\epsilon m \bm e^{(n)}-(1+\alpha)\epsilon^2m^2\mu_n \bm e^{(n)}+\mathcal O(\epsilon^3),\qquad
\bm H=2\epsilon m\Delta \bm e^{(t)}+\mathcal O(\epsilon^2).
\end{equation}

$\bm G$ and $\bm H$ are not expanded at the same order because $\bm G$ is multiplied by $g_n\sim \epsilon^{-1/2}$ in the jump equation. Indeed, since the bath kinetic temperature 
\begin{equation}
    T_b\equiv m\langle \bm v^2\rangle/2\equiv m/2\int d\bm v \bm v^2f(\bm v)
\end{equation}
is fixed, the typical thermal speed scales as
\begin{equation}
v_{\rm{th}} \equiv \sqrt{{T_b}/{m \epsilon}}=\epsilon^{-1/2}\sqrt{{T_b}/{m}},
\end{equation}
with which we rescale the one-dimensional normal velocity distribution:
\begin{equation}
\phi(v_n)=\frac{1}{v_{\rm th}}\psi\left(\frac{v_n}{v_{\rm th}}\right).
\label{eq:phi_scaling_general}
\end{equation}

The integrals $\mathcal I$ (Eq.~\eqref{eq:I_integrals}) admit the following expansion:
\begin{align}
    \mathcal I_0(U_n)&=n_b\left[v_{\rm th} \mathcal K_1+\mathcal K_0U_n+\frac{\psi(0)}{2v_{\rm th}}U_n^2\right]+\mathcal O(\epsilon),\\
    \mathcal I_1(U_n)&=-n_b\left[v_{\rm th}^2/2+2v_{\rm th} \mathcal K_1U_n+\mathcal K_0U_n^2\right]+\mathcal O\left(\epsilon^{1/2}\right),\\
    \mathcal I_2(U_n)&=n_b\left[v_{\rm th}^3 \mathcal K_3+3v_{\rm th}^2U_n/2\right]+\mathcal O\left(\epsilon^{-1/2}\right),\\
    \mathcal I_3(U_n)&=-n_b v_{\rm th}^4 \mathcal K_4+\mathcal O\left(\epsilon^{-3/2}\right).
\end{align}
where we defined the positive half moments
\begin{equation}
\mathcal K_r \equiv \int_0^\infty dc  c^r \psi(c),\qquad \mathcal K_0=\frac{1}{2}, \qquad \mathcal K_2 = \frac{1}{2}.
\end{equation}
The last equality follows from the definition of the temperature $T_b = m\langle \bm v^2\rangle/2$.

\subsubsection{First Kramers-Moyal coefficient}

We define the first moment as:
\begin{equation}
    A_i(\bm \Pi,\varphi)=\oint {ds}\Big[G_i \mathcal I_1(U_n)+H_i \mathcal I_0(U_n)\Big],
\end{equation}
with $\bm A=(\mathcal M_{100}, \mathcal M_{010}, \mathcal M_{001})$. It corresponds to a deterministic force for the first two components and a torque for the last one.

From our expansion and using $U_n=\bm U\cdot\bm e^{(n)}$, we find
\begin{equation}
A_i(\bm \Pi,\varphi)=R_i(\varphi)-\Gamma_{ij}(\varphi)U_j+N_{ijk}(\varphi)U_jU_k+\mathcal O(\epsilon^{3/2}),\\
\label{eq:A_vK_final}
\end{equation}
with
\begin{equation}
    R_i(\varphi)=n_b\oint {ds}\left[-v_{\rm th}^2G_i/2+v_{\rm th} \mathcal K_1 H_i\right]=2\mathcal K_1 n_b\Delta\sqrt{mT_b} \epsilon^{1/2}\oint {ds} e_i^{(t)}+\dfrac{1+\alpha}{2}n_bmT_b\epsilon\oint {ds} \mu_n e_i^{(n)}+\mathcal O(\epsilon^{3/2}),
\end{equation}
\begin{equation}
    \begin{split}
        \Gamma_{ij}(\varphi)=-n_b\oint {ds}\left[-2v_{\rm th} \mathcal K_1 G_i+\mathcal K_0 H_i\right]e_j^{(n)}=&2(1+\alpha)\mathcal K_1 n_b\sqrt{mT_b} \epsilon^{1/2}\oint {ds} e_i^{(n)}e_j^{(n)}\\&-n_bm\Delta \epsilon\oint {ds} e_i^{(t)}e_j^{(n)}+\mathcal O(\epsilon^{3/2}),\\
    \end{split}
\end{equation}
and
\begin{equation}
    N_{ijk}(\varphi)=n_b\oint {ds}\left[-\mathcal K_0 G_i+\frac{\psi(0)}{2v_{\rm th}}H_i\right]e_j^{(n)}e_k^{(n)}=-\frac{1+\alpha}{2} n_bm \epsilon\oint {ds} e_i^{(n)}e_j^{(n)}e_k^{(n)}+\mathcal O(\epsilon^{3/2}).
\end{equation}

\subsubsection{Second Kramers-Moyal coefficient}

For the second Kramers-Moyal coefficient, we find:
\begin{equation}
    B_{ij}(\bm U,\varphi)=\oint ds\Big[G_iG_j \mathcal I_2(U_n)+H_iH_j \mathcal I_0(U_n)+(G_iH_j+H_iG_j) \mathcal I_1(U_n)\Big],
\end{equation}
which, after replacing the integrals by their expansions, leads to:
\begin{equation}
B_{ij}(\bm \Pi,\varphi)=D_{ij}(\varphi)+E_{ijk}(\varphi)U_k+\mathcal O(\epsilon^{3/2}),
\end{equation}
with
\begin{equation}
    \mathsf B=
    \begin{pmatrix}
    \mathcal M_{200} & \mathcal M_{110} & \mathcal M_{101}\\
    \mathcal M_{110} & \mathcal M_{020} & \mathcal M_{011}\\
    \mathcal M_{101} & \mathcal M_{011} & \mathcal M_{002}
    \end{pmatrix}.    
\end{equation}
From our expansion, we find
\begin{equation}
\begin{split}
D_{ij}(\varphi)&=n_b\oint {ds}\left[v_{\rm th}^3\mathcal K_3 G_iG_j-v_{\rm th}^2(G_iH_j+H_iG_j)/2\right]+\mathcal O(\epsilon^{3/2}),\\
&=(1+\alpha)^2\mathcal K_3 n_bT_b\sqrt{mT_b} \epsilon^{1/2}\oint {ds} e_i^{(n)}e_j^{(n)}-(1+\alpha)n_bmT_b\Delta\epsilon\oint {ds} \left(e_i^{(n)}e_j^{(t)}+e_i^{(t)}e_j^{(n)}\right)+\mathcal O(\epsilon^{3/2}),
\end{split}
\end{equation}
and
\begin{equation}
E_{ijk}(\varphi)=\dfrac{3}{2}n_bv_{\rm th}^2\oint {ds} G_iG_je_k^{(n)}+\mathcal O(\epsilon^{3/2})=\dfrac{3}{2}(1+\alpha)^2n_bmT_b\epsilon\oint {ds} e_i^{(n)}e_j^{(n)}e_k^{(n)}+\mathcal O(\epsilon^{3/2}).
\end{equation}

\subsubsection{Third Kramers-Moyal coefficient}
At the same order, we must also keep the leading third jump tensor,
\begin{equation}
C_{ijk}(\varphi)=-n_bv_{\rm th}^4\mathcal K_4\oint {ds} G_iG_jG_k+\mathcal O(\epsilon^{3/2})=-(1+\alpha)^3\mathcal K_4 n_bmT_b^2 \epsilon\oint {ds} e_i^{(n)}e_j^{(n)}e_k^{(n)}+\mathcal O(\epsilon^{3/2}).
\end{equation}

\subsubsection{Truncated evolution equation for \texorpdfstring{$P$}{P}}

The partial differential equation resulting from this van Kampen expansion is third order in momentum derivative and is given by:
\begin{equation}
\frac{\partial P}{\partial t}+\bm U\cdot\frac{\partial P}{\partial \bm Y}=-\partial_i\left[\left(R_i-\Gamma_{ij}U_j+N_{ijk}U_jU_k\right)P\right]+\frac12\partial_i\partial_j\left[\left(D_{ij}+E_{ijk}U_k\right)P\right]-\frac16\partial_i\partial_j\partial_k\left[C_{ijk}P\right]+\mathcal O(\epsilon^{3/2}),
\label{eq:KM_order_eps}
\end{equation}
where $\partial_i\equiv \partial_{\Pi_i}$.

Although Pawula's theorem still applies, and we cannot rigorously exclude negative solutions to Eq.~\eqref{eq:KM_order_eps}, the van Kampen expansion ensures that the third-order term is subleading with respect to the drift and diffusion terms in the large-intruder limit and therefore, at least on physical grounds, mitigates this issue.
\subsection{Langevin equation}
\subsubsection{Reduced Fokker-Planck equation}

Eq.~\eqref{eq:KM_order_eps} shows that, strictly speaking, the intruder dynamics are not purely Fokker-Planck at order $\epsilon$. Indeed, the third Kramers-Moyal term $C_{ijk}$ enters at the same order as the contributions $N_{ijk}$ and $E_{ijk}$ to the drift and diffusion. Accordingly, if we rewrite the dynamics of $\bm \Pi$ as a stochastic differential equation, the resulting noise is not purely Gaussian because of $C_{ijk}$~\cite{hanggi1980langevin, akcasu1977fluctuations}, and $E_{ijk}$ further implies that it is multiplicative. Although this can be handled in principle, it makes the algebra substantially more cumbersome. We therefore proceed by neglecting $E_{ijk}$ and $C_{ijk}$ while retaining $N_{ijk}$, since the latter is essential for the ratchet effect, whereas the former are not. The resulting noise is therefore Gaussian and additive in the body frame. This manipulation is a standard approximation~\cite{costantini2007granular} rather than a controlled truncation.

We will therefore use:
\begin{equation}
\dfrac{\partial P(\bm \Pi,\bm Y,t)}{\partial t}+\bm U\cdot \dfrac{\partial P(\bm \Pi,\bm Y,t)}{\partial \bm Y}=-\dfrac{\partial}{\partial \bm \Pi}\cdot \left[\bm A(\bm\Pi,\varphi) P(\bm\Pi,\bm Y,t)\right]+\dfrac{1}{2}\dfrac{\partial}{\partial \bm\Pi }\otimes\dfrac{\partial }{\partial \bm\Pi}: \left[\mathsf D(\bm\Pi,\varphi) P(\bm\Pi,\bm Y,t)\right].
\label{eq:nonlinearForce}
\end{equation}
\end{widetext}
We recall that $\bm \Pi=(M V_x, M V_y, I\Omega)$ is the momentum array, related to the velocity array $\bm U=(V_x, V_y, \Omega)$ and $\bm Y=(\bm X, \varphi)$ is the configuration array.

Eq.~\eqref{eq:nonlinearForce} is a \textit{bona fide} Fokker-Planck equation, although the resulting force is nonlinear because the tensor $N_{ijk}$ contributes a term proportional to $U^2$ to the drift, which complicates the analysis. Once again, this term could in principle be neglected; however, it gives rise to an important ratchet effect. We therefore retain it and, to simplify the analysis, follow Ref.~\onlinecite{costantini2007granular} in linearizing $\bm A$.

\subsubsection{Linearization of \texorpdfstring{$A$}{A}}

The drift term in Eq.~\eqref{eq:A_vK_final} is quadratic in the velocity array:
\begin{equation}
A_i(\bm \Pi,\varphi)=R_i(\varphi)-\Gamma_{ij}(\varphi)U_j+N_{ijk}(\varphi)U_jU_k+\mathcal O(\epsilon^{3/2}).
\end{equation}

To close the equation at the linear level, it is convenient to replace the quadratic fluctuation by its average in the linear theory. The velocity covariance $\mathsf S$ is defined as
\begin{equation}
\mathsf S\equiv \left\langle (\bm U-\langle\bm U\rangle)\otimes(\bm U-\langle\bm U\rangle)\right\rangle.
\end{equation}
We note that by isotropy, $\langle R_x\rangle=\langle R_y\rangle=0$ and $\langle \bm V\rangle=0$. For a chiral nonequilibrium system, there might exist a persistent rotation $\langle \Omega\rangle\neq 0$.

The linear theory (neglecting $N_{ijk}$) is a multidimensional Ornstein-Uhlenbeck process whose velocity covariance  $\mathsf S^{\rm lin}$ satisfies~\cite{van1992stochastic}:
\begin{equation}
(\Upgamma\cdot\mathsf M^{-1})\cdot \left(\mathsf M\cdot \mathsf S^{\rm lin}\cdot \mathsf M\right)+\left(\mathsf M\cdot\mathsf S^{\rm lin}\cdot\mathsf M\right)\cdot\mathsf (\Upgamma\cdot\mathsf M^{-1})^{T}=\mathsf D,
\end{equation}
with 
\begin{equation}
    \mathsf M=\text{diag}(M, M, I).
\end{equation}
A simple closed-form expression follows in the regime $T_b\gg m\Delta^2\simeq 0$, where the chiral contribution of order $\mathcal O(\epsilon)$ is negligible compared with the achiral term of order $\mathcal O(\epsilon^{1/2})$. In this limit,
\begin{equation}
\mathsf S^{\rm lin} = T_I \mathsf M^{-1},
\end{equation}
with
\begin{equation}
T_I=\dfrac{1}{2} M\langle\bm V^2\rangle =I\langle \Omega^2\rangle =\dfrac{\mathcal K_3}{2\mathcal K_1}\dfrac{(1+\alpha)}{2}T_b,
\end{equation}
the kinetic temperature of the intruder. At equilibrium, $\alpha=1$ and $\mathcal K_3/2\mathcal K_1=1$ (because the bath has a Gaussian velocity distribution), and $T_I=T_b$. When $\Delta$ is not neglected, $M S^{\rm lin}_{xx}\neq I S^{\rm lin}_{\varphi\varphi}$ and even worse, $\mathsf S^{\rm lin}$ is not diagonal. In that case, no single meaningful kinetic temperature can be assigned to the intruder, as noted in Ref.~\onlinecite{Passive2025HargusPRE}. We detail this in Appendix~\ref{sec:fluctuationDissipation}.

The average contribution of the quadratic drift, neglecting quadratic average velocity $N_{ijk}\langle U_{j}U_{k}\rangle\simeq N_{ijk}S^{\rm lin}_{jk}$, is
\begin{equation}
 N_{ijk}S^{\rm lin}_{jk}\!=\!-\frac{1+\alpha}{2} n_bm \epsilon\oint {ds}  e_i^{(n)}e_j^{(n)}S^{\rm lin}_{jk}e_k^{(n)}+\mathcal O(\epsilon^{3/2}).
\end{equation}
Now,
\begin{equation}
e_j^{(n)}S^{\rm lin}_{jk}e_k^{(n)}=T_I  e_j^{(n)}\mathsf M^{-1}_{jk}e_k^{(n)}=T_I\left(\frac{1}{M}+\frac{\kappa_n^2}{I}\right)=T_I \mu_n,
\end{equation}
hence
\begin{equation}
\begin{gathered}
N_{ijk}S^{\rm lin}_{jk}=-\frac{1+\alpha}{2} n_bm \epsilon T_I\oint {ds}  \mu_n(s,\varphi) e_i^{(n)}(s,\varphi)\\+\mathcal O(\epsilon^{3/2}).
\end{gathered}
\label{eq:NS_general}
\end{equation}

Therefore, the quadratic \emph{fluctuating} term may be absorbed into an effective constant force/torque,
\begin{equation}
A_i(\bm \Pi,\varphi)\approx F_i(\varphi)-\Gamma_{ij}(\varphi)U_j,
\end{equation}
with:
\begin{equation}
    \quad F_i(\varphi)\equiv R_i(\varphi)+N_{ijk}(\varphi)S^{\rm lin}_{jk}.
\end{equation}
Combining Eq.~\eqref{eq:NS_general} with the expression of $R_i$, we obtain
\begin{align}
F_i(\varphi)&=2\mathcal K_1n_b\Delta\sqrt{mT_b}\epsilon^{1/2}\oint {ds} e_i^{(t)}\\
&+\dfrac{1+\alpha}{2}n_bm\epsilon\left(T_b-T_I\right)\oint {ds} \mu_n e_i^{(n)}+\mathcal O(\epsilon^{3/2}).\nonumber
\end{align}

With this closure, the reduced Fokker-Planck equation becomes linear in $\bm U$:
\begin{equation}
\frac{\partial P}{\partial t}+\bm U\cdot \frac{\partial P}{\partial \bm Y}=-\dfrac{\partial}{\partial \bm \Pi}\cdot\Big[\big(\bm F-\Upgamma \cdot \bm U\big)P\Big]+\frac{1}{2}\dfrac{\partial}{\partial \bm \Pi}\otimes\dfrac{\partial}{\partial \bm \Pi}\!:\!\big[\mathsf DP\big].
\label{eq:final_FP}
\end{equation}

\subsubsection{Equivalent Langevin equation}

The Fokker-Planck equation~\eqref{eq:final_FP} can be rewritten into a \textit{linear} Langevin equation:
\begin{equation}
    \begin{gathered}
    \mathsf M\cdot \dfrac{d \bm U}{dt}=\bm F-\Upgamma\cdot \bm U + \sqrt{\mathsf D}\cdot\bm \zeta,\\\dfrac{d \bm Y}{dt} = \bm U, \quad \langle \zeta_i(t)\zeta_j(t')\rangle=\delta_{ij}\delta(t-t'),
    \end{gathered}
\end{equation}
where we recall that $\bm U=(V_x,V_y,\Omega)$ collects the translational and angular velocities, while $\bm F=(F_x,F_y,F_\varphi)$ contains the forces and the torque $F_\varphi$. $\Upgamma$ is a damping matrix that also contains correlations between translational and angular velocities, such as $\Gamma_{02}\equiv \Gamma_{x\varphi}$, and $\mathsf D$ is the noise covariance. These coefficients are explicitly given by:
\begin{widetext}
\begin{align}
F_i(\varphi)&=2\mathcal K_1n_b\Delta\sqrt{mT_b}\oint {ds} e_i^{(t)}+\dfrac{1+\alpha}{2}n_bm\left(T_b-T_I\right)\oint {ds} \mu_n e_i^{(n)},\qquad T_I=\frac{2}{3}\frac{\langle |\bm  v|^3\rangle}{\langle |\bm  v|^2\rangle\langle |\bm  v|\rangle}\frac{(1+\alpha)}{2}T_b,\\
D_{ij}(\varphi)&=(1+\alpha)^2\mathcal K_3 n_bT_b\sqrt{mT_b} \oint {ds} e_i^{(n)}e_j^{(n)}-(1+\alpha)n_bmT_b\Delta\oint {ds} \left(e_i^{(n)}e_j^{(t)}+e_i^{(t)}e_j^{(n)}\right),\\
\Gamma_{ij}(\varphi)&=2(1+\alpha)\mathcal K_1 n_b\sqrt{mT_b} \oint {ds} e_i^{(n)}e_j^{(n)}-n_bm\Delta\oint {ds} e_i^{(t)}e_j^{(n)},\qquad T_b = \dfrac{m}{2}\int d\bm v \bm v^2 f(\bm v)
\end{align}
\end{widetext}
where we used:
\begin{equation}
\frac{\mathcal K_3}{2\mathcal K_1}=\frac{2}{3}\frac{\langle |\bm  v|^3\rangle}{\langle |\bm  v|^2\rangle\langle |\bm  v|\rangle},
\end{equation}
which is equal to $1$ for a Gaussian bath in $d=2$. We recall that all these expressions are obtained for a small bath particle mass $m$, and $T_I$ is additionally obtained at leading order in $m\Delta^2\ll T_b$. A larger $\Delta$ contributes to the temperature non-trivially, as we show in Appendix~\ref{sec:fluctuationDissipation}.

We note that the noise covariance is a symmetric matrix and, notably, it does not have any contribution odd in time. Typically, in chiral active systems, the noise has correlation~\cite{kuroda2023microscopic,fruchart2026nonreciprocal,han2021fluctuating, chun2018emergence}:
\begin{equation}
    \langle \bm \zeta(t)\otimes \bm\zeta(t')\rangle= \mathsf D_{\rm even}(t - t') + \mathsf D_{\rm odd}(t - t'), 
\end{equation}
with $\mathsf D_{\rm even}(t)=\mathsf D_{\rm even}^T(t)=\mathsf D_{\rm even}(-t)$ and $\mathsf D_{\rm odd}(t)=-\mathsf D_{\rm odd}^T(t)=-\mathsf D_{\rm odd}(-t)$. The Markovian character of the Boltzmann–Lorentz equation forces the noise to be white~\cite{spohn1980kinetic}, so its correlator is proportional to $\delta(t-t')$ and no odd contribution can exist~\footnote{A term proportional to $\partial_t\delta(t-t')$ is, in principle, also possible~\cite{yasuda2022time}, although it does not affect static observables. Such a term can be viewed as the distributional limit of an odd-colored-noise correlator, for instance $C(t)\propto e^{-\gamma |t|}\sin(\omega t)$,
in a scaling limit where the correlation time vanishes but the first temporal moment stays finite. Because our theory is Markovian from the outset, this odd contribution is absent from the noise.}. Odd transport may nevertheless occur via the antisymmetric part of the damping, rather than by an antisymmetric part of the noise covariance.

Finally, although the noise is multiplicative with $\mathsf D$ depending on $\varphi$, the equations are underdamped so there is no need to specify a discretization convention~\cite{van1992stochastic}.

\subsection{Approximations and validity regime}
\label{sec:summary}
Before analyzing the resulting parameters in detail, we briefly summarize the approximations used.

In the dilute regime, we adopt a Boltzmann-Lorentz description of instantaneous, binary, and uncorrelated intruder-bath collisions. The bath consists of particles drawn from a distribution $f$ that encodes its statistical properties and collide randomly with the intruder according to a prescribed collision rule. At this level, chirality enters only through the intruder shape or through the intruder-bath collision law. Any chirality internal to the bath is absorbed into a possibly non-Gaussian distribution $f$.

By rewriting the Boltzmann-Lorentz equation as a master equation, we can perform a Kramers-Moyal expansion and organize the resulting terms through a van Kampen expansion in the small bath mass. At the order retained, we obtain a nonlinear drift term $N_{ijk}$, a multiplicative-noise correction $E_{ijk}$, and a third Kramers-Moyal tensor $C_{ijk}$. Because $N_{ijk}$ contains the contributions relevant for ratchet effects, we kept it while neglecting, for simplicity, $E_{ijk}$ and $C_{ijk}$, thereby obtaining a simple Fokker-Planck equation. The drift remained nonlinear in the velocity, however, through $N_{ijk}$, so we replaced it by its average value in the linear theory and dropped the residual quadratic term. This dresses the zero-velocity force and torque by a fluctuating term that displays a ratchet effect. Finally, we assumed weak chiral driving, $m\Delta^2 \ll T_b$. This assumption is not essential, but it greatly simplifies the analysis by yielding a single effective temperature $T_I$ for both translational and angular dynamics in this regime.

We now discuss how each term entering the Langevin equation depends on the geometry of the intruder.

\section{Geometric structure of the couplings}\label{sec:2}
\begin{widetext}
It is useful to separate all couplings into the normal/dissipative $\alpha$-sector and the chiral
$\Delta$-sector. From the Langevin equation derived above, we have:
\begin{equation}
F_i=F_i^{(\Delta)}+F_i^{(\alpha)},\qquad\Gamma_{ij}=\Gamma_{ij}^{(\alpha)}+\Gamma_{ij}^{(\Delta)},\qquad D_{ij}=D_{ij}^{(\alpha)}+D_{ij}^{(\Delta)},
\label{eq:sector}
\end{equation}
with:
\begin{align}
F_i^{(\Delta)}&=2\mathcal K_1 n_b\Delta\sqrt{mT_b}\oint {ds}  e_i^{(t)},\qquad F_i^{(\alpha)}=\frac{1+\alpha}{2}n_bm(T_b-T_I)\oint {ds}  \mu_n e_i^{(n)},\qquad\mu_n=\frac1M+\frac{\kappa_n^2}{I},\\
\Gamma_{ij}^{(\alpha)}&=2(1+\alpha)\mathcal K_1 n_b\sqrt{mT_b}\oint {ds}  e_i^{(n)}e_j^{(n)},\qquad\Gamma_{ij}^{(\Delta)}=-n_bm\Delta\oint {ds}  e_i^{(t)}e_j^{(n)},\label{eq:Gamma_ij}\\
D_{ij}^{(\alpha)}&=(1+\alpha)^2\mathcal K_3 n_bT_b\sqrt{mT_b}\oint {ds}  e_i^{(n)}e_j^{(n)}=2T_I \Gamma_{ij}^{(\alpha)}, \\
D_{ij}^{(\Delta)}&=-(1+\alpha)n_bmT_b\Delta\oint {ds} \big(e_i^{(n)}e_j^{(t)}+e_i^{(t)}e_j^{(n)}\big)=(1+\alpha)T_b\big(\Gamma_{ij}^{(\Delta)}+\Gamma_{ji}^{(\Delta)}\big)\label{eq:DDelta_from_GammaDelta}.
\end{align}
Interaction chirality is encoded in the $(\Delta)$ components of the coefficients, while intruder chirality enters through the integrals. In general, the object's chirality can modify both the $(\alpha)$ and $(\Delta)$ components.
\end{widetext}
We make the geometric content of these expressions explicit in Appendix~\ref{app:geometry}, focusing on which coefficients vanish or remain nonzero for each shape. The corresponding results are collected in Table~\ref{tab:terms} and can be compared with Refs.~\onlinecite{Passive2025HargusPRE, hargus2025odd}. For instance, the coupling between a nonequilibrium bath and a chiral intruder generates a ratchet torque, but no odd translational damping in our dilute regime. Although symmetry allows such a coefficient, it emerges only at higher densities, where bath-intruder correlations become relevant~\cite{letter}. We now examine the implications of the dilute approximation and its regime of validity.

\begin{table*}[t]
  \centering
    \setlength{\tabcolsep}{.8em}
    \setlength{\heavyrulewidth}{1.5pt}
    \renewcommand{\arraystretch}{2.5}
    \normalsize
    \begin{tabular}{l c c c c c c c c}\toprule
    \makecell{\textbf{Symmetry}} & \multicolumn{2}{c}{Circular} &
    \multicolumn{2}{c}{\shortstack{Achiral +\\ $n$-fold symmetric}} &
    \multicolumn{2}{c}{\shortstack{Chiral +\\$n$-fold symmetric}} &
    \multicolumn{2}{c}{\shortstack{Polar + \\ Achiral}} \\

    \makecell{\textbf{Intruder}\\\textbf{Examples}} &
    \multicolumn{2}{c}{\raisebox{-.5\totalheight}{\includegraphics[width=0.08\textwidth]{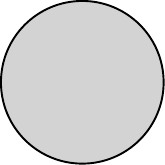}}} &

     \multicolumn{2}{c}{
     \makecell{\raisebox{-.5\totalheight}{\includegraphics[width=0.1\textwidth]{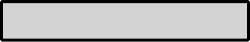}}\vspace{2mm}\\
     \raisebox{-.5\totalheight}{\includegraphics[width=0.1\textwidth]{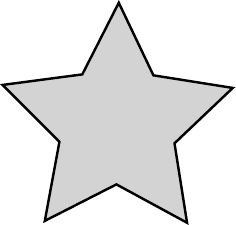}}}} &

     \multicolumn{2}{c}{
     \makecell{\raisebox{-.5\totalheight}{\includegraphics[width=0.1\textwidth]{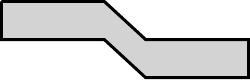}}\vspace{2mm}\\
     \raisebox{-.5\totalheight}{\includegraphics[width=0.1\textwidth]{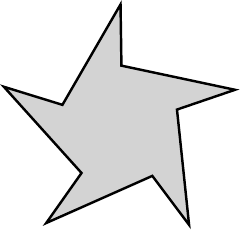}}}} &

     \multicolumn{2}{c}{
     \makecell{\raisebox{-.5\totalheight}{\includegraphics[width=0.1\textwidth]{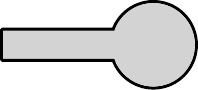}}\vspace{2mm}\\
     \raisebox{-.5\totalheight}{\includegraphics[width=0.1\textwidth]{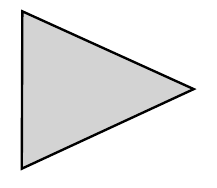}}}} \\
     \cmidrule(lr){2-3} \cmidrule(lr){4-5} \cmidrule(lr){6-7} \cmidrule(lr){8-9}
     \makecell{\textbf{Sector}}&
     \makecell{Normal\\$(\alpha)$ } & \makecell{Chiral\\$(\Delta)$} &
     \makecell{Normal\\$(\alpha)$ } & \makecell{Chiral\\$(\Delta)$} &     \makecell{Normal\\$(\alpha)$ } & \makecell{Chiral\\$(\Delta)$} &
     \makecell{Normal\\$(\alpha)$ } & \makecell{Chiral\\$(\Delta)$}   \\[0.1cm]\midrule
    \makecell{\textbf{Net torque} $F_\varphi\neq 0$} & \xmark & \cmark & \xmark & \cmark & \cmark & \cmark & \xmark & \cmark \\
    \makecell{\textbf{Net body frame}\\\textbf{force} $\tilde F_{a}\neq 0$} & \xmark & \xmark & \xmark & \xmark & \xmark & \xmark & \cmark & \cmark \\
    \makecell{\textbf{Odd translational}\\\textbf{damping}  $\Gamma_{xy}-\Gamma_{yx}\neq 0$} & \xmark & \cmark & \xmark & \cmark & \xmark & \cmark & \xmark & \cmark \\
    \makecell{\textbf{Cross couplings}\\$\Gamma_{a\varphi}\neq 0$} & \xmark & \xmark & \xmark & \xmark & \xmark & \xmark & \cmark & \cmark \\
    \makecell{\textbf{Odd cross couplings}\\$\Gamma_{a\varphi}-\Gamma_{\varphi a}\neq 0$} & \xmark & \xmark & \xmark & \xmark & \xmark & \xmark & \xmark & \cmark \\\bottomrule
  \end{tabular}
  \caption{Dependence of the coefficients entering the Langevin equation on the intruder geometry, the layout is adapted from Ref.~\onlinecite{Passive2025HargusPRE}. The ‘‘Sector'' entry corresponds to the normal contribution for $(\alpha)$ and to the chiral contribution for $(\Delta)$, as defined in Eq.~\eqref{eq:sector}. The symbols \xmark\ and \cmark\ indicate whether the term in the left column exists (\cmark) or vanishes (\xmark). For instance, a \xmark\ for $F_\varphi$ in the ‘‘Normal $(\alpha)$'' column means that $F_\varphi^{(\alpha)} = 0$ for the shape considered. We recall that $a\in\{x, y\}$ while $i, j \in \{x, y, \varphi\}$.   Unlike Ref.~\onlinecite{Passive2025HargusPRE}, we find no odd dynamics arising from the $(\alpha)$ sector (which is out of equilibrium for $\alpha<1$ or a non-Gaussian bath), even for chiral objects. In our approach, this absence is due to the uncorrelated nature of the collisions, as discussed below.}
  \label{tab:terms}
\end{table*}

We then derive explicit analytical expressions for several intruder shapes in Appendix~\ref{app:finite}. We note that \emph{a priori}, non-convex shapes are not well described by the Boltzmann-Lorentz equation since the concavity can induce recollisions and subsequent collisions may therefore not be fully uncorrelated. However, the molecular simulations reported in the Letter~\cite{letter}, where the bath is modeled explicitly, remain in good agreement even for the non-convex shape considered.

\section{Discussion and limitations of the dilute theory}
\label{sec:Discussion1}
We have thus fully characterized the intruder dynamics in the limits of large mass and a dilute bath. A natural next step is to carefully study the terms neglected within this treatment, namely the multiplicative noise and the non-Fokker-Planck contributions that were Gaussianized here but become important at finite mass. In particular, we have omitted a genuinely non-Gaussian noise with a nontrivial third moment that arose at the same order as the ratchet effects. It would be especially interesting to determine whether chirality breaking leaves a clear signature in this term. Finally, although our calculations were carried out for a specific collision rule, we do not expect the resulting coefficients to change qualitatively for other chiral collision rules, but this should be verified.

In the Letter~\cite{letter}, we showed that the theory performs very well for a dilute bath without momentum conservation to evade long-time tails~\cite{hansen2013theory}. It breaks down, however, once the bath mean free path becomes smaller than the characteristic intruder size, because bath-intruder correlations can no longer be neglected. A particularly clear illustration is the damping of a circular intruder. In Sec.~\ref{sec:circular} (or Eq.~\eqref{eq:Gamma_ij}), we obtained
\begin{equation}
    \Gamma_{xx}=(1+\alpha)2\pi \mathcal K_1 n_b\sqrt{mT_b} R,
    \label{eq:epstein}
\end{equation}
for a circular intruder of radius $R$. This is the Epstein drag~\cite{stoyanovskaya2020simulations}. By contrast, a two-dimensional Stokes-Oseen drag derived from the low-Reynolds-number Navier-Stokes equation reads
\begin{equation}
    \Gamma^{\rm{Stokes}}_{xx}=c\eta R,
    \label{eq:stokes_damping}
\end{equation}
where $\eta$ is the bath viscosity and $c$ is a dimensionless constant set by the boundary conditions and the Reynolds number in 2D~\cite{khalili2017stokes}. The difference in physical origin is very telling. Epstein drag arises from uncorrelated impacts of individual particles, whereas Stokes drag reflects viscous momentum transport by the collective flow of the surrounding fluid. The former depends on the microscopic details of the interaction via the coefficient of restitution $\alpha$, while the latter depends on the viscosity $\eta$, a collective property.

A second example of non-trivial correlations, which even happen in the dilute limit, concerns odd response. Our theory predicts that a chiral object immersed in a nonequilibrium but achiral bath should not display odd linear response. Indeed, at $\Delta=0$, we find a symmetric translational damping $\Gamma_{ab}^{(\alpha)}=\Gamma_{ba}^{(\alpha)}$. Yet the simulations reported in the Letter~\cite{letter} reveal a very small but finite odd response. Such a response is also allowed on symmetry grounds~\cite{Passive2025HargusPRE}. The resolution of this apparent discrepancy again lies in bath-intruder correlations. In the dilute limit, we found that a chiral intruder in a nonequilibrium bath can spontaneously rotate through a ratchet effect. This rotation is transferred to the nearby bath particles, which then develop a biased circulation around the intruder. As a result, collisions become statistically asymmetric, and the bath behaves as if it were effectively chiral and can therefore induce odd dynamics on the intruder exactly as the $\Delta$ sector does in our case. Such a mechanism is necessarily absent from a Boltzmann-Lorentz description, which assumes perfectly uncorrelated collisions. This distinction is nevertheless useful, because our calculation cleanly separates correlation-induced effects from genuinely kinetic ones, something a purely symmetry-based approach cannot do.

Hence, correlations may matter in two distinct regimes. The first is a still-dilute bath, where the mean free path is large but weak correlations persist, as discussed above. This regime is analytically difficult because the bath cannot be treated as a continuum: we must remain at the Boltzmann level and describe it as a dynamical field. Doing so requires the joint distribution of the intruder position and velocity together with the local one-particle bath velocity distribution, yielding a nonlinear integro-differential equation that seems analytically intractable.

The second regime is more tractable and corresponds to a short mean free path. There, one may start from the Boltzmann equation for the bath alone, derive the hydrodynamic equations for the slow modes through a Chapman-Enskog expansion~\cite{dorfman2021contemporary}, and then introduce the intruder through boundary conditions~\cite{batchelor2000introduction}. Intermediate approaches, in which the intruder is retained explicitly as an additional degree of freedom throughout the derivation, are also possible~\cite{cukier1980microscopic}. These seem the most promising route for describing ratchet effects within a hydrodynamic framework. Indeed, it is not obvious how the ratchet mechanism identified above should emerge at the hydrodynamic scale, especially since it is fluctuation-driven. In that respect, a nonequilibrium fluctuating-hydrodynamic description, perhaps akin to treatments used for Casimir forces~\cite{rodriguez2011dynamical}, may prove useful.

We now turn to this second regime and focus on a chiral bath. The hydrodynamics of a simply chiral bath, whose particles interact through a transverse impulse $\Delta_b$ (here describing bath-bath rather than bath-intruder interactions), has recently been derived from a Chapman-Enskog expansion~\cite{maire2026kinetic}. We will use that framework to study the behavior of an intruder in a bath with intrinsic chiral interactions. This limit is particularly interesting because, in our dilute description, the intruder feels bath chirality only through the non-Gaussianity of the bath velocity distribution. Since that distribution carries no imprint of chirality, a chiral bath is, for the intruder, indistinguishable from an achiral nonequilibrium bath. We stress that this concerns the chirality of bath-bath interactions, governed by $\Delta_b$, not that of bath-intruder interactions, governed by $\Delta$. The latter is accounted for in the dilute theory but neglected in the hydrodynamic approach we will use.

In a dense bath, at the hydrodynamic level, we will see that the edge currents generated by the flow lead to an odd response with an antisymmetric $\Upgamma$ and an applied torque $F_\varphi$ generated by a varying curvature. 

\section{Intruder, boundary conditions and edge currents in a chiral hydrodynamic environment}
\label{sec:chiral}
In this section, we first formulate the hydrodynamic description of the bath and then introduce the intruder through boundary conditions. In this approach, the microscopic bath-intruder parameters $\alpha$ and $\Delta$ play only a secondary role. For simplicity, we therefore set $\alpha=1$ and $\Delta=0$, all the more so since encoding them in boundary conditions is not straightforward. We return to this point later.

\subsection{Chiral Navier-Stokes equation}

To obtain the hydrodynamics equations for the bath \emph{alone}, we need to specify the equation of motion and the bath-bath particle interaction. We can use the same kind of collision as for the one we specified between a bath particle and the intruder. When two bath particles $\alpha$ and $\beta$ of diameter $\sigma$ collide, they undergo the following change of velocity:
\begin{subequations}
\begin{align}
    \bm v_\alpha'&= \bm v_\alpha - \dfrac{1+\alpha_b}{2}(\bm v_{\alpha\beta}\cdot \hat{\bm\sigma}_{\alpha\beta})\hat{\bm\sigma}_{\alpha\beta} -  \Delta_b(\bm\varepsilon\cdot\hat{\bm\sigma}_{\alpha\beta}), \\
    \bm v_\beta'&= \bm v_\beta + \dfrac{1+\alpha_b}{2}(\bm v_{\alpha\beta}\cdot \hat{\bm\sigma}_{\alpha\beta})\hat{\bm\sigma}_{\alpha\beta} + \Delta_b(\bm\varepsilon\cdot\hat{\bm\sigma}_{\alpha\beta}),
\end{align}
\label{eq: collision_rule_b}
\end{subequations}
where $\hat{\bm\sigma}_{\alpha\beta}$ is the unit vector joining the particle centers, $\alpha_b$ is a bath-bath coefficient of restitution, and $\Delta_b$ is a transverse kick (they should not be confused with $\alpha$ and $\Delta$, which rule the bath-intruder interactions rather than the bath-bath interactions). This provides a simple model of an active chiral bath. Other systems are possible, such as ones with chiral self-propulsion~\cite{kuroda2023microscopic, lei2019nonequilibrium} or spin~\cite{gao2025liquidgascriticalityhyperuniformfluids, eren2025collisional, eren2025collisional}. We expect the qualitative conclusion to remain unchanged by the nature of the underlying system, although the precise hydrodynamics equations may change. We discuss this point in greater detail later.

Once again, the particles are described by a particle distribution, although now, we allow for a spatial and temporal dependence to capture correlations $f(\bm r,\bm v,t)$. On hydrodynamic scales, these dependencies are expected to be slaved to the slow modes. Retaining only the density $n(\bm r,t)$ and velocity $\bm u(\bm r,t)$, we write
\begin{equation}
    f(\bm r,\bm v,t)\simeq f[\bm v | n(\bm r,t), \bm u(\bm r,t)].
    \label{eq:f_non_homogeneous}
\end{equation}
This spatiotemporal dependence is essential for capturing collective effects, while the explicit collision rule~\eqref{eq: collision_rule_b} accounts for chirality in the bath. Both effects were neglected in the Boltzmann-Lorentz description. Hydrodynamic equations then follow by taking velocity moments of the time derivative of Eq.~\eqref{eq:f_non_homogeneous}~\cite{maire2026kinetic}:
\begin{align}
        \partial_t n + \bm u \cdot \bm \nabla n &= -n\bm\nabla \cdot \bm{u},\\
        \partial_t \bm{u} + \bm{u}\cdot\bm \nabla \bm{u}&= \dfrac{1}{mn}\bm \nabla\cdot \left(\mathsf \Sigma^{\rm h}+\mathsf \Sigma^{\eta}\right), 
        \label{eq:chiral_NS}
\end{align}
where the homogeneous stress
\begin{equation}
    \mathsf \Sigma^{\rm h}=-p\bm 1 + \tau\bm\varepsilon=\begin{pmatrix}
    -p & \tau \\
    -\tau & -p 
    \end{pmatrix}.
\end{equation}
contains a pressure $p$ on its diagonal, but also a torque-density $\tau$ on its off-diagonal part to include the non-conservation of angular momentum induced by (effective) chiral forces. Microscopically, for example, it can come from the Virial stress:
\begin{equation}
\begin{gathered}
    \mathsf \Sigma^{\rm virial} = -\dfrac{\sum_{\alpha} m(\bm v_\alpha-\bm u) \otimes  (\bm v_\alpha-\bm u)}{L^2}\\+\dfrac{  \sigma\sum_{\alpha<\beta}^{N}\hat{\bm\sigma}_{\alpha\beta}\otimes \bm F_{\alpha\beta}}{L^2},
    \end{gathered}
    \label{eq: virial}
\end{equation}
which has a non-diagonal component if the force $\bm F_{\alpha \beta}$ between particle $\alpha$ and $\beta$ is not solely directed toward the vector joining the particles' centers $\hat{\bm\sigma}_{\alpha\beta}$, as ours~\cite{maire2026kinetic}:
\begin{equation}
\begin{split}
    \bm F_{\alpha\beta}=-m\Big(\dfrac{1+\alpha_b}{2}&(\bm v_{\alpha\beta}\cdot \hat{\bm\sigma}_{\alpha\beta})\hat{\bm\sigma}_{\alpha\beta} \\&+ \Delta_b(\bm\varepsilon\cdot\hat{\bm\sigma}_{\alpha\beta})\Big)\delta(t - t^{\rm coll}_{\alpha\beta}),
    \end{split}
    \label{eq: force}
\end{equation}
with $\delta(t - t^{\rm coll}_{\alpha\beta})$ a Dirac delta at the time of the collision. The homogeneous stress follows from averaging the virial stress over a ‘‘local equilibrium'' velocity distribution~\cite{maire2026kinetic}. The pressure $p$ and torque density $\tau$ are then expressed as functions of $\alpha_b$, $\Delta_b$ and the local density $n(\bm r)$~\cite{maire2026kinetic}. The torque density can also emerge from the coarse-graining of other fields that encode the system's chirality. Whereas pressure exerts a force normal to a wall, the torque density generates a force parallel to it. It is therefore responsible for the edge currents of mass observed in many chiral systems~\footnote{In systems where the internal spin is retained, the torque density is replaced by a term proportional to the so-called rotational viscosity. In systems where chirality arises instead from a biased self-propulsion, such as chiral active Brownian particles, the torque density arises from the so called ‘‘active stress''.}.

The viscous stress $\Sigma^\eta$ is assumed to depend on the first velocity gradient:
\begin{equation}
\Sigma^{\eta}_{ij}\equiv \eta_{ijkl}\partial_k u_l.
\end{equation}
In our isotropic system with broken time-reversal and parity symmetries, the viscosity tensor $\eta_{ijkl}$ can be decomposed into a basis of 6 independent isotropic tensors built from $\delta_{ij}$ and $\varepsilon_{ij}$~\cite{fruchart2023odd}:
\begin{equation}
    \begin{split}
        \eta_{ijkl}&= \eta(\delta_{i k}\delta_{j l}+\delta_{il}\delta_{jk}-\delta_{ij}\delta_{kl})+   \zeta\delta_{ij}\delta_{kl}\\
        &+\eta_o(\varepsilon_{ik}\delta_{jl}+\varepsilon_{j l}\delta_{ik})-\eta_B\delta_{ij}\varepsilon_{kl}\\
        &-\eta_A\varepsilon_{ij}\delta_{kl}+\eta_R\varepsilon_{ij}\varepsilon_{kl}.
    \end{split}
    \label{eq: viscosity decomposition}
\end{equation}
The shear viscosity $\eta$ and bulk viscosity $\zeta$ exist in achiral fluids, but the other viscosities arise from the chirality of our system. 

In dilute fluids dominated by angular-momentum-conserving free flight, $\eta$ and $\eta_o$ are typically the only relevant viscosities, while the others are negligible or irrelevant for incompressible flows. In the 2d incompressible limit, parity-breaking viscosities mainly shift the mechanical pressure and enter only through stress boundary conditions, while edge currents near boundaries are set by the torque density $\tau$. At intermediate density, the fluid is still compressible, and all transport coefficients are important. Nevertheless, to capture the phenomenology relevant here, it is enough to consider the incompressible limit. Although the boundary conditions considered below impose a stress and could therefore activate odd viscosity, we neglect this contribution because it is typically much smaller than the ordinary viscosity and decreases at intermediate density~\cite{maire2026kinetic}. By contrast, the torque density grows with density and drives strong chiral currents~\cite{Marconi2026hydrodynamics,maire2026kinetic}.
In any case, odd viscosity can be reinstated easily through a pressure shift~\cite{fruchart2023odd}. Although we omit the details, we indicate in each section how it would modify the results.

In the stationary regime, we obtain the stationary Navier-Stokes equation:
\begin{equation}
    \rho_b(\bm u\cdot\bm\nabla \bm u) = -\bm  \nabla \tilde{\mathtt{p}}+ \bm\varepsilon\cdot \bm  \nabla\tau+ \eta\bm\nabla^2\bm u,
     \qquad \bm\nabla\cdot\bm u=0,
   \label{eq:stokes}
\end{equation}
where $\rho_b=mn=mn_b$ is the constant mass density in the bath and $\mathtt p$ is the mechanical pressure which enforces incompressibility and is defined up to a constant, taken to be the homogeneous ‘‘thermodynamic'' pressure at infinity: $\mathtt p(\bm r\to\infty)=p(n(\bm r\to\infty))$. Since the density $n$ is constant throughout the fluid because of incompressibility, the torque density $\tau$ is also constant and therefore can only act at boundaries. The torque density can nevertheless contribute in the bulk of non-isothermal or compressible fluids.

\subsection{Boundary problem for an intruder}
We place an intruder with boundary $\partial\mathcal B$ given by $\bm r^{(c)}(s)$. The set of points inside the intruder is denoted by $\mathcal B$, and the fluid occupies the exterior domain $\mathcal{D}$. In our incompressible limit, this can be modeled by taking $n(\bm r)=n_0$ for $\bm r\in\mathcal{D}$ and $n(\bm r)=0$ for $\bm r\in\mathcal B$. Accordingly, the torque density is $\tau(\bm r)=\tau_0$ in the fluid and $\tau(\bm r)=0$ inside the intruder, where $\tau_0=\tau(n_0)$ is fixed by the equation of state at density $n_0$. Since $\tau$ is constant in each bulk region, its gradient is supported only at the interface:
\begin{equation}
\bm \varepsilon\cdot\bm\nabla\tau(\bm r)=-\tau_0\hat{\bm t}\delta_{\partial\mathcal B},
\end{equation}
where $\hat{\bm t}$ is a unit tangent vector to $\partial\mathcal B$ and $\delta_{\partial\mathcal B}$ is the Dirac delta on the boundary. 

In the sharp-interface incompressible limit of the underlying compressible problem, Eq.~\eqref{eq:stokes} reduces to:
\begin{equation}
\begin{gathered}
\rho_b(\bm u(\bm r)\cdot\bm\nabla \bm u(\bm r))=-\bm\nabla \mathtt p(\bm r)+\eta\bm\nabla^2\bm u(\bm r)-\tau_0\hat{\bm t}\delta_{\partial\mathcal B},\\\qquad\bm\nabla\cdot\bm u(\bm r)=0,\qquad \bm r\in\mathcal{D}.
\end{gathered}
\label{eq:interface_1}
\end{equation}
\textit{A priori}, the convective term need not be small due to the edge currents setting $\bm u\neq 0$ around boundaries. This bulk equation is supplemented by the impermeability condition and a perfect slip boundary condition, since our intruder is smooth and the collision rule in Eq.~\eqref{eq:collision_rule_no_gt} contains no term proportional to $g_t$:
\begin{equation}
\begin{gathered}
\big(\bm u(\bm r) -\bm V^{(c)}(\bm r)\big)\cdot\hat{\bm n}(\bm r)=0,\\ \hat{\bm t}(\bm r)\cdot\mathsf \Sigma(\bm r)\cdot\hat{\bm n}(\bm r)=0\qquad \bm r\in\partial\mathcal B,
\label{eq:slip}
\end{gathered}
\end{equation}
where $\hat{\bm n}$ is the unit normal to the interface pointing into the fluid, $\bm V^{(c)}$ is the velocity of the intruder at the point considered, and $\mathsf \Sigma=-\mathtt p\bm 1+\eta\left(\bm\nabla\otimes\bm u+(\bm\nabla\otimes\bm u)^T\right)$.  In the absence of convection, the problem reduces to a Stokes flow driven by the singular body force $-\tau_0 \hat{\bm t}\delta_{\partial \mathcal B}$, and can therefore be solved directly using the Stokes Green's function.

For our purposes, however, it is more convenient to use an equivalent formulation by rewriting the problem as a pure Navier-Stokes flow in the exterior domain $\mathcal{D}$:
\begin{equation}
\begin{gathered}
-\bm\nabla \mathtt p(\bm r)+\eta\bm\nabla^2\bm u(\bm r)=\rho_b(\bm u(\bm r)\cdot\bm\nabla \bm u(\bm r)),\\\bm\nabla\cdot\bm u(\bm r)=0,\qquad \bm r\in\mathcal{D}.
\end{gathered}
\label{eq:NS_with_conve}
\end{equation}
The singular chiral force is then encoded in the boundary condition:
\begin{equation}
\begin{gathered}
\big(\bm u(\bm r) -\bm V^{(c)}(\bm r)\big)\cdot\hat{\bm n}(\bm r)=0,\\ \hat{\bm t}(\bm r)\cdot\mathsf \Sigma(\bm r^+)\cdot\hat{\bm n}(\bm r)=\tau_0,\quad \bm r\in\partial\mathcal B,
\end{gathered}
\label{eq:boundary_condition_stokes}
\end{equation}
where the stress is evaluated infinitesimally close to the interface $\bm r^+=\lim_{\tilde\epsilon\to 0}\bm r+\tilde\epsilon\hat{\bm n}$, rather than exactly on the interface $\bm r$ since $\tau_0$ is a bulk term, generated by a boundary layer, that is pushed at the interface in the incompressible limit. At the interface itself $\bm r$, the perfect-slip boundary condition~\eqref{eq:slip} must still hold. The distinction between $\bm r^+$ and $\bm r$ is unimportant for deriving the velocity field, but matters for the force and torque applied to the intruder, as we will later see.

We emphasize that, although $\tau_0$ enters only through the intruder boundary, it is a \emph{bulk property} of the fluid, generated by angular-momentum-non-conserving interactions \emph{within} the fluid. It should therefore not be confused with the term $\Delta$ introduced in the previous section, which denotes the transverse kick imparted by bath particles to the intruder and which, for simplicity, we set here to $0$, although it could \emph{a priori} be incorporated into the boundary condition. In this dense regime, the precise collision rule between the intruder and the bath plays only a secondary role and is partially encoded in the boundary condition. Indeed, in deriving Stokes drag, it is sufficient to prescribe boundary conditions at the intruder-fluid interface, and the detailed microscopic form of the intruder-fluid interaction is largely irrelevant. We return to this point in the final discussion.

We begin by examining the chosen boundary conditions and their consequences for the intruder motion.

\subsection{Force and torque applied by the flow on an intruder}

The fluid force $F_a$ and torque $F_\varphi$ on the intruder follow from the traction
$\bm f_{\partial\mathcal B}=\mathsf\Sigma(\bm r)\cdot\hat{\bm n}$ evaluated on its boundary~\cite{batchelor2000introduction}:
\begin{equation}
    F_a = \oint ds f_{\partial\mathcal B,a}, \quad F_\varphi = \oint ds \big[\bm r_0 \times \bm f_{\partial\mathcal B}\big]_z.
    \label{eq:traction_force}
\end{equation}
Importantly, this traction must be taken at $\bm r\in\partial\mathcal B$, not at the
$\bm r^+=\lim_{\tilde\epsilon\to0}\bm r+\tilde\epsilon \hat{\bm n}$.
Indeed, the bulk torque density acts in the fluid just outside the surface, so that
\begin{equation}
    \mathsf\Sigma(\bm r^+) \cdot \hat{\bm n}=\mathsf\Sigma(\bm r) \cdot \hat{\bm n}+\tau_0\hat{\bm t},
\end{equation}
with $\hat{\bm t}\cdot\mathsf\Sigma(\bm r) \cdot \hat{\bm n}=0$ from the perfect slip boundary condition. Because of this perfect slip, the tangential contribution $\tau_0$ acting only at $\bm r^+(s)$ is not transmitted to the intruder at $\bm r(s)$.

The distinction between $\bm r^+$ and $\bm r$ is irrelevant for the velocity field, as we may impose the boundary condition at $\bm r^+$ and recover the same velocity at $\bm r$ by continuity. It is, however, essential for the force and torque applied to the intruder, which depend on the stress. Using $\mathsf\Sigma(\bm r^+)\cdot\hat{\bm n}$ as the traction on the intruder would incorrectly predict that even a circular intruder rotates under the bath torque density, in contradiction with the perfect-slip assumption, which forbids tangential stress transfer to the intruder.

Under this perfect-slip condition, the traction on the intruder is purely normal, so the stress can be projected onto $\hat{\bm n}$ and written directly in terms of $\bm r^+$:
\begin{equation}
    \bm f_{\partial\mathcal B}
    =\mathsf\Sigma(\bm r)\cdot\hat{\bm n}
    =\big(\hat{\bm n}\cdot \mathsf\Sigma(\bm r^+)\cdot\hat{\bm n}\big)\hat{\bm n}.
\end{equation}
This removes the distinction between $\bm r$ and $\bm r^+$ in the traction. We therefore drop it from now on, keeping in mind that forces must still be computed from the normal projection of the stress. This simplification is specific to perfect slip.

\subsection{Edge currents}\label{sec:edge}

Before turning to the dynamics of an intruder in such a bath, we briefly recall how edge currents arise in these systems~\cite{lou2022odd, mecke2025obstacle}, as they underlie several of the chiral effects acting on the intruder. In the two examples below, we retain the convective term. Although it does not modify the velocity profiles, it may still affect other fields, such as the pressure.

\subsubsection{Flat walls}

Let us consider a chiral bath confined between two walls at $y=-L/2$ and $y=L/2$ and with otherwise periodic boundary conditions along $\hat{\bm e}_x$. A solution of Eq.~\eqref{eq:NS_with_conve} is
\begin{equation}
    \bm u(y)=y\frac{\tau_0}{\eta}\hat{\bm e}_x, \qquad y\in[-L/2, L/2],
\end{equation}
which corresponds to a Couette flow. Under these boundary conditions, the velocity field grows without bound as the wall separation increases. Indeed, in the absence of a second wall or any additional source of shear or dissipation, boundary particles can be accelerated indefinitely.

\subsubsection{Circular boundaries}

Let us now consider a circular boundary at $r=R_0$. The fluid may lie either inside the boundary, corresponding to confinement in a circular box, or outside it, in which case the boundary represents a circular intruder. In both cases, Eq.~\eqref{eq:NS_with_conve} admits the solution~\cite{Marconi2026hydrodynamics}
\begin{equation}
    \bm u(r)=(Ar + B/r)\hat{\bm e}_\theta,
\end{equation}
where $\hat{\bm e}_\theta$ is the azimuthal unit vector, tangent to $\hat{\bm e}_r$ and $A$ and $B$ are constant to be determined.

For particles confined inside this circle, no steady solution can be found. Indeed, $B$ must vanish to avoid a divergence at $r=0$, leaving a flow of the form $\bm u\sim r\hat{\bm e}_\theta$. Such a flow does not dissipate energy through shear stresses, so the particles keep accelerating under the uncompensated torque injected at the boundary. This contrasts, for instance, with a square box, where the curvature variation at the corners generates a nonzero shear contribution that dissipates the energy injected by the torque.

For a circular intruder, \emph{i.e.} particles in $\mathcal{D}=\{r>R_0\}$, the velocity field remains finite~\cite{Marconi2026hydrodynamics}:
\begin{equation}
    \bm u(r)=-\dfrac{\tau_0R_0^2}{2\eta}\dfrac{1}{r}\hat{\bm e}_\theta.
    \label{eq:circular}
\end{equation}
Its amplitude grows with $R_0$, while decaying as $1/r$. As in the case of parallel walls, including the convective term does not modify the velocity profile. However, in this case, it adds a centrifugal contribution to the mechanical pressure~\cite{Marconi2026hydrodynamics}. Similarly, the inclusion of an odd viscosity would modify the mechanical pressure but not the velocity field.

\section{Stokes drag and lift force via Oseen's equations in a uniform stream}\label{sec:Stokes drag}

\subsection{General remarks on Stokes and Oseen flow}

Having established the effect of a stationary circular intruder in a chiral bath, we now consider a moving intruder, or equivalently, an intruder in a uniform background flow, $\bm u(\bm r\to\infty)=\bm U_\infty$. In an ordinary fluid, a fixed intruder in a uniform background flow $\bm U_\infty$ experiences a drag force along the imposed flow, $\bm F^{\rm drag}\propto \bm U_\infty$. Equivalently, for an intruder translating with velocity $\bm V$ in a fluid at rest, the drag force opposes the intruder's motion, $\bm F^{\rm drag}\propto -\bm V$. In a chiral fluid, we expect not only this dissipative contribution but also a transverse lift force, $\bm F^{\rm lift}\propto -\bm\varepsilon\cdot\bm V$, perpendicular to the imposed flow. That is, we expect the damping matrix $\Upgamma$ to have an antisymmetric term. Transverse forces of this kind also arise from odd viscosity and, in two dimensions, vanish in the incompressible limit under no-slip boundary conditions~\cite{hosaka2021nonreciprocal,lier2023lift,daddi2026exact,ganeshan2017odd}. Here, however, our focus is on their generation through edge currents.

In two dimensions, Stokes flow is pathological because its Green's function does not decay at infinity~\cite{lamb1993hydrodynamics}. This can be regularized either by external damping, which cuts off the divergence beyond $r\sim \ell_\gamma$~\cite{leiderman2016swimming}, or by retaining convection. As noted by Oseen and Lamb~\cite{lamb1993hydrodynamics}, the Stokes approximation assumes a small Reynolds number and therefore breaks down at sufficiently large distances, where convective effects inevitably become important. In our case, convection is even more relevant due to the finite velocities generated by the edge currents, which may even drive instabilities~\cite{Marconi2026hydrodynamics}. Second, in the absence of convection and edge currents, the Stokes equation obeys the Lorentz reciprocity theorem, which forbids an antisymmetric mobility and thus any odd response~\cite{kim2013microhydrodynamics}. These reciprocity constraints can be broken by odd viscosity~\cite{hosaka2023lorentz}, but also by convection or inertia~\cite{lier2024odd}.

We now compute the drag and lift forces on the intruder.

\subsection{General setup}

We consider a circular intruder of radius $R_0$ in the intruder frame, in the presence of a uniform flow at infinity
\begin{equation}
    \bm u(\bm r\to\infty)=\bm U_\infty=U_\infty \hat{\bm e}_x.
\end{equation}
The boundary conditions remain
\begin{equation}
    u_r(R_0,\theta)=0,\qquad\Sigma_{r\theta}(R_0,\theta)=\tau_0,
\end{equation}
with
\begin{equation}
    \Sigma_{r\theta}=\eta\left(\partial_r u_\theta+\frac1r\partial_\theta u_r-\frac{u_\theta}{r}\right).
\end{equation}

\begin{widetext}
\subsection{Decomposition of the flow}

We decompose the solution as
\begin{equation}
    \bm u=\bm U_\infty+\bm u^{(T)}+\bm u^{(\chi)},
    \qquad
    \mathtt p=p_\infty+\mathtt p^{(T)}+\mathtt p^{(\chi)},
\end{equation}
where $\bm u^{(T)}$ is the translational disturbance created by the uniform stream, and $\bm u^{(\chi)}$ is the previously found circular chiral flow induced by the torque density $\tau_0$ around the circular intruder:
\begin{equation}
    u_r^{(\chi)}=0,\qquad u_\theta^{(\chi)}=-\frac{c}{r}, \qquad  c=\frac{\tau_0 R_0^2}{2\eta}.
    \label{eq:nucleation_paper}
\end{equation}

Starting from the steady incompressible Navier-Stokes equation
\begin{equation}
    -\bm\nabla \mathtt p+\eta \bm\nabla^2\bm u=\rho_b(\bm u\cdot\bm\nabla)\bm u,\qquad \bm\nabla\cdot\bm u=0,
\end{equation}
we neglect the self-advecting quadratic term $\bm u^{(T)}\cdot\bm \nabla\bm u^{(T)}$ and therefore:
\begin{equation}
    \begin{split}
    (\bm u\cdot\bm\nabla)\bm u\simeq (\bm U_\infty\cdot\bm\nabla)\bm u^{(T)}+(\bm U_\infty\cdot\bm\nabla) \bm u^{(\chi)}
    + (\bm u^{(\chi)}\cdot\bm\nabla)\bm u^{(\chi)}+(\bm u^{(\chi)}\cdot\bm\nabla)\bm u^{(T)}
    +(\bm u^{(T)}\cdot\bm\nabla)\bm u^{(\chi)}.
    \end{split}
\end{equation}
We first note that
\begin{equation}
    (\bm u^{(\chi)}\cdot\bm\nabla)\bm u^{(\chi)}=\dfrac{1}{2}\bm\nabla\left({\bm u^{(\chi)}}^2\right) = \bm\nabla\left(\frac{c^2}{2r^2}\right),
\end{equation}
and
\begin{equation}
    (\bm U_\infty\cdot\bm\nabla) \bm u^{(\chi)}
    =\bm\nabla\left(U_\infty c\frac{\sin(\theta)}{r}\right).
    \label{eq:U_inu_chi}
\end{equation}
Since these are gradient terms, they can be incorporated into the pressure
\begin{equation}
    \mathtt p^{(\chi)}(r,\theta)=-\rho_b U_\infty c\frac{\sin(\theta)}{r}-\frac{\rho_b c^2}{2r^2}.
\end{equation}
Then,
\begin{equation}
\begin{split}
    (\bm u^{(\chi)}\cdot\bm\nabla)\bm u^{(T)}+(\bm u^{(T)}\cdot\bm\nabla)\bm u^{(\chi)}&= \bm\nabla\left(-\frac{c}{r}u_\theta^{(T)}\right)
    +\frac{c}{r}(\bm\nabla\times \bm u^{(T)})\hat{\bm e}_r.
\end{split}
    \label{eq:vorticity}
\end{equation}
Hence, while the first term on the right-hand side simply shifts the pressure due to its gradient nature:
\begin{equation}
    \bar{\mathtt p}^{(T)}= \mathtt p^{(T)}-\rho_b\frac{c}{r}u_\theta^{(T)},
\end{equation}
the second contribution in Eq.~\eqref{eq:vorticity} is nontrivial and the Oseen problem:
\begin{equation}
    \rho_b(\bm U_\infty\cdot\bm\nabla)\bm u^{(T)}=-\bm\nabla\bar{\mathtt p}^{(T)}+\eta\bm\nabla^2\bm u^{(T)}-\rho_b\frac{c}{r}\omega^{(T)}\hat{\bm e}_r,\qquad \bm\nabla\cdot\bm u^{(T)}=0,\qquad \omega^{(T)}\equiv \bm\nabla \times\bm u^{(T)},
    \label{eq:oseen_translational}
\end{equation}
with $\bm u^{(T)}(\bm r\to\infty)\to0$ must be solved perturbatively. We treat it in Appendix~\ref{app:oseen} and only provide the final result here.

\begin{figure}
    \centering
    \includegraphics[width=0.99\linewidth]{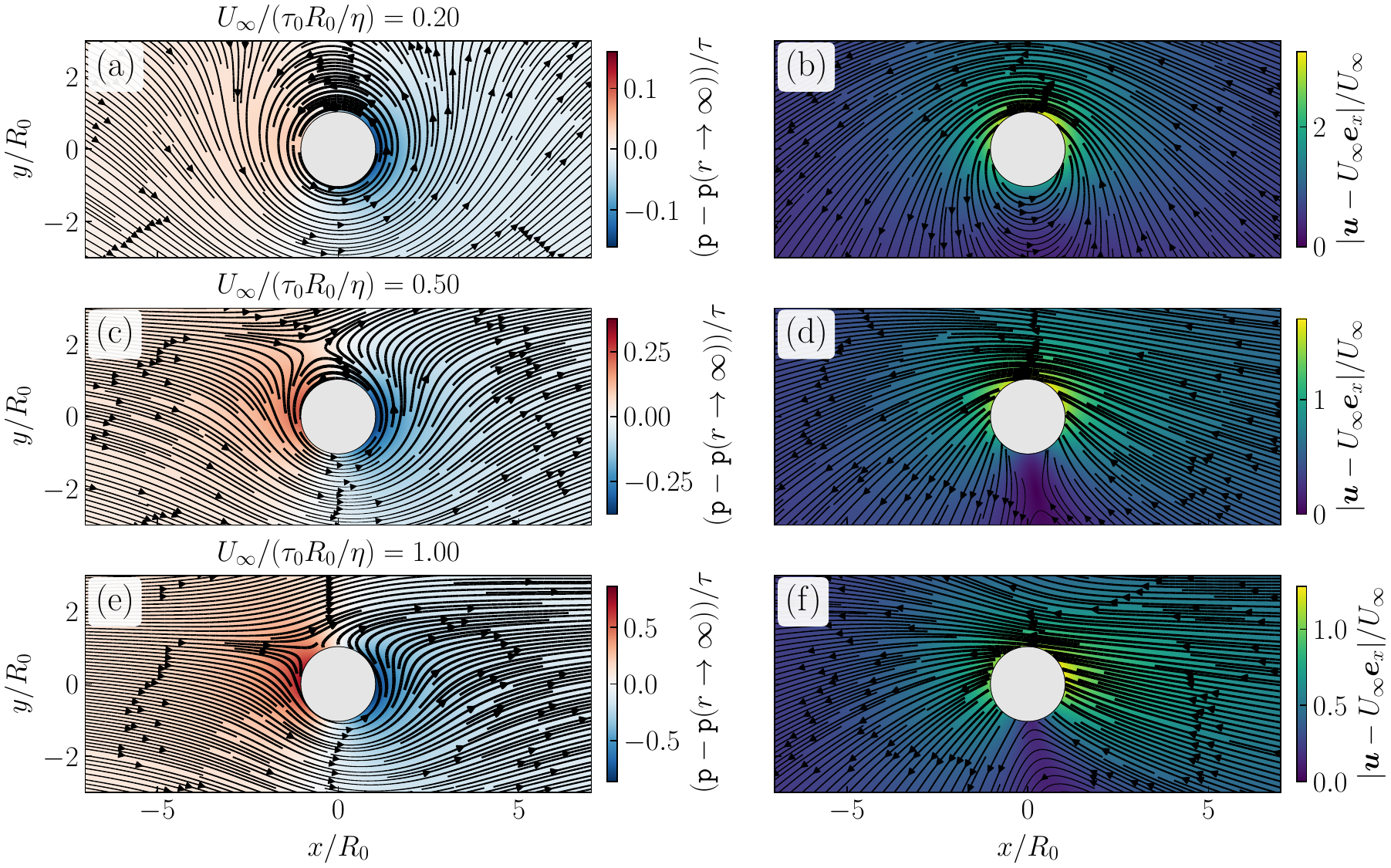}
    \caption{Pressure and velocity field (streamline plot) (a, c, e) around an intruder in a chiral bath using Eqs.~\eqref{eq:a}, \eqref{eq:b}, and ~\eqref{eq:c}. The corresponding body frame velocity field is also shown in (b, d, f). For (a) and (b) $U_\infty\eta/\tau_0R_0=0.2$, for (c) and (d) $U_\infty\eta/\tau_0R_0=0.5$ and for (e) and (f) $U_\infty\eta/\tau_0R_0=1$.}
    \label{fig:oseen}
\end{figure}

The final velocity and pressure fields are given in Fig.~\ref{fig:oseen} for various speeds of the intruder, we see that when the speed of the intruder is comparable to the velocity generated by the active torque, a clear biased flow is generated, which is also found to induce an odd-response via an off-diagonal translational drag (Eq.~\eqref{eq:appendix_important}):

\begin{equation}
    \Upgamma_{\rm trans}=
    \begin{pmatrix}
    \dfrac{4\pi\eta}{\Lambda} & -\pi\rho_bc\\[2mm]
    \pi\rho_bc & \dfrac{4\pi\eta}{\Lambda}
    \end{pmatrix}
    +\mathcal O\left(\frac{\eta}{\Lambda^2},\frac{\rho_bc}{\Lambda},\dfrac{\rho_b^2 c^2}{\eta}\right),\qquad \Lambda=1+\log\left(\frac{4\eta}{\rho_bR_0U_\infty}\right)-\gamma_E.
    \label{eq:upvarGamma_hydro}
\end{equation}
with $\gamma_{\rm E}$ the Euler–Mascheroni constant. A circular intruder translating with velocity $\bm V$ therefore feels a force:
\begin{equation}
    \bm F_{\rm trans} =-\Upgamma_{\rm trans}\cdot \bm V.
\end{equation}
The even drag retains the logarithmic dependence associated with the Stokes paradox. By contrast, the leading odd coefficient $\Gamma_{xy}\propto \tau_0$, which induces an odd force, is independent of $\Lambda$. Hence, although the translational flow remains singular, the corresponding transverse force has a finite leading value in the stated limit.

Including odd viscosity would likewise produce a nontrivial flow and an odd response~\cite{hosaka2023lorentz}. In this case, unlike in the simple settings of Sec.~\ref{sec:edge}, the edge currents would also be affected by odd viscosity. This interplay is interesting and deserves to be worked out in detail.

\section{Torque generation in a chiral hydrodynamic bath}\label{sec:torque_generation}

We now consider the torque generated on an object initially at rest, in the absence of any externally imposed flow. As we will show, a circular object experiences no torque, again because our collision rule between the intruder and the bath particles does not transmit tangential forces. Instead, the motion is driven by pressure and viscous-stress gradients near points where the intruder curvature varies, which are induced by edge currents.

To compute this effect, we first determine the hydrodynamic fields around a weakly non-axisymmetric object, initially neglecting the convective term.

\subsection{General setup}

We consider a small deformation of a circle and parametrize the boundary $\bm \rho_0$ by an angle $\theta$:
\begin{equation}
\begin{gathered}
    \bm \rho_0=\begin{pmatrix}
        R(\theta)\cos(\theta)\\
        R(\theta)\sin(\theta)
    \end{pmatrix},\quad R(\theta)=R_0[1+\varepsilon h(\theta)],\quad \varepsilon\ll1,
\end{gathered}
    \label{eq:rounded_shape}
\end{equation}
with:
\begin{equation}
    h(\theta)=\cos(m(\theta-\phi))
\end{equation}
where $R_0$ is the mean radius, $m=3$ corresponds to a rounded triangle, $m=4$ to a rounded square, and $\phi$ fixes the orientation of the shape.

We neglect the convective term $\bm u\cdot\bm\nabla \bm u$ at first, and work with the forced Stokes equation:
\begin{equation}
\begin{aligned}
-\bm\nabla \mathtt p(\bm r)+\eta\bm\nabla^2\bm u(\bm r)=0,\qquad \bm\nabla\cdot\bm u(\bm r)=0,\qquad \bm r\in\mathcal{D}.\\
\bm u(\bm r)\cdot\hat{\bm n}(\bm r)=0,\qquad \hat{\bm t}(\bm r)\cdot\mathsf \Sigma(\bm r)\cdot\hat{\bm n}(\bm r)=\tau_0,\qquad \bm r\in\partial\mathcal B,
\end{aligned}
\end{equation}
Because the flow is incompressible, we introduce a streamfunction $\psi$ such that the velocity field in polar coordinates $\bm u=u_r\hat{\bm e}_r+u_\theta \hat{\bm e}_\theta$ decomposes as:
\begin{equation}
    u_r=\frac{1}{r}\partial_\theta \psi,\qquad
    u_\theta=-\partial_r \psi.
    \label{eq:psivsV}
\end{equation}
The steady Stokes equation then reduces to the biharmonic equation
\begin{equation}
    \bm\nabla^4 \psi =0
\end{equation}
in the exterior domain. We seek a perturbative solution of the form
\begin{equation}
    \psi(r,\theta)=\psi^{(0)}(r)+\varepsilon \psi^{(1)}(r,\theta)+\mathcal O(\varepsilon^2),
\end{equation}
where the circular solution is
\begin{equation}
    \psi^{(0)}(r)=\frac{\tau_0 R_0^2}{2\eta}\log\left(\frac{r}{R_0}\right),
\end{equation}
which gives the purely azimuthal flow previously found
\begin{equation}
    u_r^{(0)}=0,\qquad
    u_\theta^{(0)}(r)=-\frac{\tau_0 R_0^2}{2\eta r}.
\end{equation}

The most general decaying $m$-fold biharmonic solution is:
\begin{equation}
    \psi^{(1)}(r,\theta)=\Big(A_m r^{2-m}+B_m r^{-m}\Big)\cos\big(m(\theta-\phi)\big),
    \label{eq:here}
\end{equation}
which is the most general decaying biharmonic mode with $m$-fold symmetry.

\subsection{Boundary conditions}

The tangent and normal at the boundary are
\begin{equation}
\begin{gathered}
    \hat{\bm t}=\hat{\bm e}_\theta+\varepsilon \frac{h'(\theta)}{R_0}\hat{\bm e}_r+\mathcal O(\varepsilon^2),\\\hat{\bm n}=\hat{\bm e}_r-\varepsilon \frac{h'(\theta)}{R_0}\hat{\bm e}_\theta+\mathcal O(\varepsilon^2).
\end{gathered}
\end{equation}
The boundary conditions $\bm u\cdot \hat{\bm n}=0$ and $\hat{\bm t}\cdot\mathsf \Sigma\cdot\hat{\bm n}=\tau_0$ are therefore given by:
\begin{equation}
\begin{gathered}
    u_r^{(0)}(R_0)+\varepsilon u_r^{(1)}(R_0)-\varepsilon \frac{h'(\theta)}{R_0}u_\theta^{(0)}(R_0)+\mathcal O(\varepsilon^2)=0\\
    \Sigma_{r\theta}^{(0)}(R_0)+\varepsilon\Big[h(\theta)\partial_r \Sigma_{r\theta}^{(0)}(R_0)\!+\!\Sigma_{r\theta}^{(1)}(R_0,\theta)\Big]+\mathcal O(\varepsilon^2)=\tau_0
\end{gathered}
\end{equation}
Together, they yield:
\begin{equation}
\begin{gathered}
    A_m=-\frac{\tau_0 R_0^m}{4\eta m}(m-1)(m+2),\\ B_m=\frac{\tau_0 R_0^{m+2}}{4\eta m}(m-2)(m+1),
    \end{gathered}
\end{equation}
and therefore:
\begin{equation}
    \psi(r,\theta)=\frac{\tau_0 R_0^2}{2\eta}\log \frac{r}{R_0}+\frac{\varepsilon\tau_0}{4\eta m}\Big[
    -(m-1)(m+2)R_0^m r^{2-m}+(m-2)(m+1)R_0^{m+2}r^{-m}\Big]\cos\big(m(\theta-\phi)\big)+\mathcal O(\varepsilon^2).
    \label{eq:psifullrounded}
\end{equation}

\subsection{Velocity and pressure fields}
From Eqs.~\eqref{eq:psifullrounded} and \eqref{eq:psivsV}, we obtain the velocity field
\begin{align}
    u_r(r,\theta)&=\frac{\varepsilon\tau_0 R_0}{4\eta}\Bigg[(m-1)(m+2)\left(\frac{R_0}{r}\right)^{m-1} -(m-2)(m+1)\left(\frac{R_0}{r}\right)^{m+1}\Bigg]\sin(m(\theta-\phi))+\mathcal O(\varepsilon^2),
    \label{eq:urapp}\\
    u_\theta(r,\theta)&=-\frac{\tau_0 R_0^2}{2\eta r}+\frac{\varepsilon\tau_0 R_0(m-2)}{4\eta}\Bigg[-\frac{(m-1)(m+2)}{m}\left(\frac{R_0}{r}\right)^{m-1}+(m+1)\left(\frac{R_0}{r}\right)^{m+1}\Bigg]\cos\big(m(\theta-\phi)\big)+\mathcal O(\varepsilon^2).
    \label{eq:uthetaapp}
\end{align}
The mechanical pressure follows from the integration of the Stokes equation using the known velocity field:
\begin{equation}
    \mathtt p(r,\theta)=p_\infty+\varepsilon\tau_0 \frac{(m-1)^2(m+2)}{m}\left(\frac{R_0}{r}\right)^m\sin\big(m(\theta-\phi)\big) +\mathcal O(\varepsilon^2),
    \label{eq:papp}
\end{equation}
where $p_\infty$ is the pressure at infinity.

The value of any field $f$ on the boundary of the intruder is: $f\big|_{\partial\mathcal B} = f^{(0)}(R_0+\varepsilon h(\theta)) +\varepsilon f^{(1)}(R_0,\theta) +\mathcal O(\varepsilon^2)$. Therefore, Eqs.~\eqref{eq:urapp}, \eqref{eq:uthetaapp} and \eqref{eq:papp} give:
\begin{align}
    u_r\big|_{\partial\mathcal B}&=\varepsilon \frac{\tau_0 R_0}{2\eta} m \sin\big(m(\theta-\phi)\big)+\mathcal O(\varepsilon^2),\\
    u_\theta\big|_{\partial\mathcal B}&=-\frac{\tau_0 R_0}{2\eta}+\varepsilon \frac{\tau_0 R_0}{\eta} \frac{m-1}{m} \cos\big(m(\theta-\phi)\big)+\mathcal O(\varepsilon^2),\\
    \mathtt p\big|_{\partial\mathcal B}&=p_\infty+\varepsilon\tau_0 \frac{(m-1)^2(m+2)}{m} \sin\big(m(\theta-\phi)\big)+\mathcal O(\varepsilon^2).\label{eq:p_boundary}
\end{align}
\end{widetext}

\begin{figure*}
    \centering
    \includegraphics[width=0.9\linewidth]{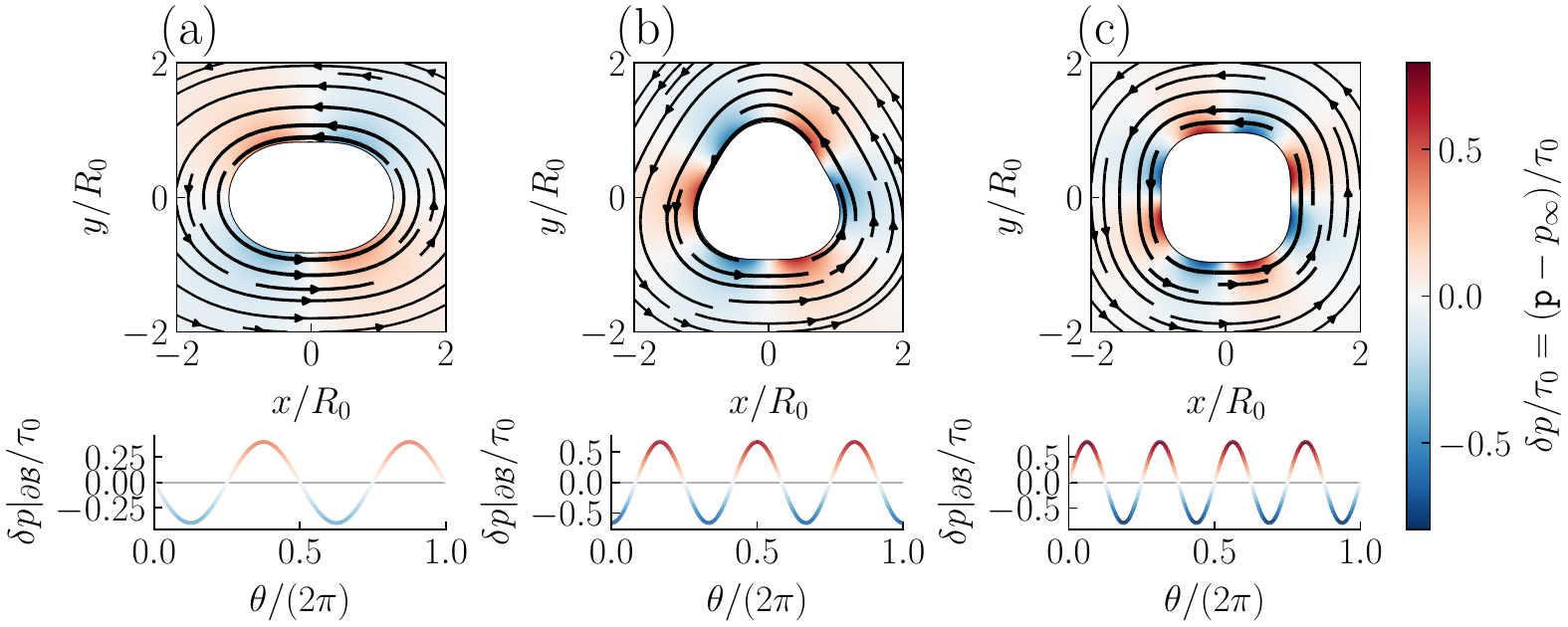}
    \caption{Pressure and velocity fields around a \emph{convex} intruder (top), together with the pressure distribution along its boundary (bottom), for an ellipse (a: $m=2$, $\epsilon = 1/5$), a triangle (b: $m=3$, $\epsilon = 1/10$), and a square (c: $m=4$, $\epsilon = 1/17$).}
    \label{fig:pressure_field_theo}
\end{figure*}

Fig.~\ref{fig:pressure_field_theo} shows the pressure field around the intruder, computed from Eq.~\eqref{eq:papp}, together with the boundary pressure obtained from Eq.~\eqref{eq:p_boundary}. As in the direct simulations reported in the Letter~\cite{letter}, a pressure excess or deficit develops at the corner depending on the flow direction. In this pure Stokes description, however, the two profiles are identical up to a sign, whereas in the simulations~\cite{letter} the excess is typically larger and more extended than the deficit. This asymmetry likely reflects inertial effects neglected here. For instance, convection can induce a strong ejection of material past the corner. In Appendix~\ref{sec:inertial_Stokes}, we compute the fields while taking into account the convective term to first order and show that this asymmetry originates from higher-order mode-coupling terms in the expansion. At first order, it does not affect the torque applied to the intruder computed below, so we do not discuss it further in the main text. At lowest order, we can show that odd viscosity modifies the flow in a similar way to the convective terms and therefore likewise does not contribute to the final torque.

We also note that, in the molecular-dynamics simulations reported in the Letter~\cite{letter}, the intruder is a regular rather than a rounded polygon, and the stress is more strongly localized near its corners than what we find here. This likely reflects the sharp edges of the simulated polygons, which act as stress-concentration points because of their divergent curvature~\cite{Moffatt2019Singularities}, but are difficult to capture theoretically.

\subsection{Torque applied on the intruder}
Now that the boundary fields have been determined, we can compute the torque exerted on the intruder by the bath and by the chiral edge currents. Using Eq.~\eqref{eq:traction_force}, we obtain
\begin{equation}
    F_\varphi=\oint ds [\bm  r_0\times \hat{\bm  n}]_z\left(\hat{\bm  n}\cdot\mathsf\Sigma\big|_{\partial \mathcal B}\cdot\hat{\bm  n}\right),
    \label{eq:Force_boundary}
\end{equation}
where we recall that the full boundary stress has been projected onto its normal component to isolate the traction transmitted to the intruder. In particular, although the bath torque density $\tau_0$ generically produces a finite torque around the intruder, $-2\mathcal A \tau_0$, the intruder does not couple to this contribution directly because it is purely tangential. Instead, the torque is transmitted indirectly through the pressure and velocity gradients generated by the edge currents induced by $\tau_0$. This mechanism is distinct from the chiral driving produced by chiral intruder-bath interactions in the dilute regime, where the leading contribution is purely torque-like, $F^{(\Delta)} \propto 2\mathcal A\Delta$. 

We now quantify this effect from the boundary stress, including the viscous contribution. In our case, the intruder is parametrized by:
\begin{equation}
    \bm r_0(\theta)=R(\theta)\hat{\bm e}_r,\quad R(\theta)=R_0\big[1+\varepsilon \cos (m(\theta - \phi))\big],
\end{equation}
which implies:
\begin{equation}
\begin{gathered}
    ds=\sqrt{R^2+R'^2}d\theta,\qquad\hat{\bm n}=\frac{R\hat{\bm e}_r-R' \hat{\bm e}_\theta}{\sqrt{R^2+R'^2}},\\ [\bm r_0\times \hat{\bm n}]_z=-\frac{RR'}{\sqrt{R^2+R'^2}}
\end{gathered}
\end{equation}
and:
\begin{equation}
    F_\varphi=-\int_0^{2\pi} d\theta  \mathsf \Sigma_{nn}(\theta)R(\theta)R'(\theta), 
    \label{eq:F_torque}
\end{equation}
\begin{equation}
    \begin{split}
        R(\theta)R'(\theta)=&-\varepsilon m R_0^2\sin(m(\theta-\phi))\\
        &-\varepsilon^2 m R_0^2\sin(m(\theta-\phi))\cos(m(\theta-\phi))\\ &+\mathcal O(\varepsilon^3)
    \end{split}
\end{equation}
We have:
\begin{equation}
    \Sigma_{nn}=\Sigma_{rr}+2\varepsilon m\sin (m(\theta-\phi))\Sigma_{r\theta}+\mathcal O(\varepsilon^2),
\end{equation}
leading to:
\begin{equation}
    F_\varphi=\pi\varepsilon^2\tau_0 R_0^2(m^2-1)(m-2), \qquad m\geq 2.
    \label{eq:forceforce}
\end{equation}
It vanishes for a circular intruder $\varepsilon=0$, which therefore does not experience any torque. Although edge currents still develop around it, they cannot generate the anisotropic pressure gradients required to rotate the object. Interestingly, the torque also vanishes for an ellipse-like perturbation ($m=2$ and $\varepsilon\ll 1$). At first sight, this is surprising in view of the asymmetric pressure field in Fig.~\ref{fig:pressure_field_theo}(a). In this particular case, however, the viscous contribution to the torque exactly cancels the pressure contribution.

In the Letter, we measured the torque exerted by the bath on a fixed regular $m$-gon. A sharp-cornered intruder is not directly accessible within our theoretical framework. We therefore approximate the regular $m$-gon by the smooth shape $R(\theta)=R_0\big(1 + \varepsilon\cos(m\theta)\big)$ and choose $\varepsilon$ as large as possible while preserving convexity, \emph{i.e.} up to the threshold at which the curvature
\begin{equation}
    \kappa(\theta)=\frac{R^2+2(R')^2-RR''}{\big(R^2+(R')^2\big)^{3/2}},
\end{equation}
first vanishes. This yields:
\begin{equation}
    \kappa(\theta)=0\Rightarrow \varepsilon=\dfrac{1}{1+m^2}, \qquad m\geq 2,
\end{equation}
and together with Eq.~\eqref{eq:forceforce}
\begin{equation}
    F_\varphi=\pi\tau_0 R_0^2 \frac{(m^2-1)(m-2)}{(1+m^2)^2}.
\end{equation}
The result is nonmonotonic: $F_\varphi$ increases from $m=2$ to a maximum at $m\simeq 5$, then decays as $m^{-1}$ for large $m$ since, with this parameterization, $\epsilon(m\to\infty)$ leads to a circle. In the Letter~\cite{letter}, however, the maximum occurs near $m\simeq 7$ and the large-$m$ decay scales as $m^{-2}$. These discrepancies are unsurprising, since this theory relies on a very crude representation of the sharp edges used in our simulations in the Letter.

\subsection{Rotational damping}

The rotational analogue of the Stokes drag can also be computed for a rotated intruder, which we assume here to remain spatially fixed. Once again, however, the calculation differs from the standard no-slip case because the boundary conditions are slip-like. In particular, as for torque generation, a perfectly smooth circle experiences no rotational damping from the surrounding fluid. Rotational drag, therefore, arises only for a noncircular object and only through the normal traction.

In the body frame, the impermeability condition reads
\begin{equation}
    \left(\bm u(\bm r)-\bm V^{(c)}\right)\cdot \hat{\bm n}=0\Rightarrow \bm u(\bm r)\cdot \hat{\bm n}=\Omega[\bm \rho_0\times \hat{\bm n}]_z,\quad \bm r\in \partial\mathcal B,
\end{equation}
We denote by $\bm u^{(\Omega)}$ and $\mathtt p^{(\Omega)}$ the additional flow and pressure induced by the rotation. By linearity, since inertia is neglected, they satisfy the Stokes equations
\begin{equation}
    -\bm\nabla \mathtt p^{(\Omega)}+\eta \bm\nabla^2 \bm u^{(\Omega)}=0, \qquad \bm \nabla\cdot \bm u^{(\Omega)}=0,\qquad \bm r\in \mathcal{D},
    \label{eq:solved_already}
\end{equation}
with boundary conditions:
\begin{equation}
    \bm u^{(\Omega)}\cdot \hat{\bm n}=\Omega[\bm \rho_0\times \hat{\bm n}]_z, \qquad \hat{\bm t}\cdot\mathsf \Sigma^{(\Omega)}\cdot \hat{\bm n}=0, \qquad \bm r\in \partial\mathcal B,
    \label{eq:BC_last_last}
\end{equation}
where the torque density $\tau_0$ is already included in the previous solution for the nonrotating case. An equation similar to Eq.~\eqref{eq:solved_already} was solved in the previous section and admits Eq.~\eqref{eq:here} as the general solution. Introducing the stream function through
\begin{equation}
    u_r^{(\Omega)}=\frac1r\partial_\theta \psi^{(\Omega)},\qquad u_\theta^{(\Omega)}=-\partial_r \psi^{(\Omega)},
\end{equation}
and enforcing the boundary conditions~\ref{eq:BC_last_last}, we obtain
\begin{equation}
    \begin{gathered}
    \psi^{(\Omega)}(r,\theta)=-\frac{\varepsilon\Omega}{2}\cos (m(\theta-\phi))\times\\
     \Big[(m+1)R_0^m r^{2-m}-(m-1)R_0^{m+2}r^{-m}\Big]+ \mathcal O(\varepsilon^2),
    \end{gathered}
    \label{eq:phi_omega}
\end{equation}
where the boundary conditions were expanded at first order in $\varepsilon$. The torque $F_\varphi^{(\Omega)}$ exerted on the intruder by its rotation is then computed from Eq.~\eqref{eq:F_torque} together with Eq.~\eqref{eq:phi_omega}, which yields
\begin{equation}
    F_\varphi^{(\Omega)}=-2\pi \eta\varepsilon^2m(m^2-1)R_0^2\Omega +\mathcal O(\varepsilon^3).
\end{equation}
We therefore identify the rotational drag coefficient as:
\begin{equation}
\Gamma_{\varphi\varphi}=2\pi \eta\varepsilon^2m(m^2-1)R_0^2,
\end{equation}
with the total torque applied on a weakly non-circular object being:
\begin{equation}
    \begin{split}
    I\dot\Omega=&F_\varphi-\Gamma_{\varphi\varphi}\Omega + \mathcal O(\varepsilon^3)\\
    =&\pi\varepsilon^2\tau_0 R_0^2(m^2-1)(m-2) - 2\pi \eta\varepsilon^2m(m^2-1)R_0^2\Omega,
    \end{split}
\end{equation}
for $m\geq 2$. The stationary angular velocity is then:
\begin{equation}
    \Omega=\dfrac{\tau_0}{2\eta}\dfrac{m-2}{m},\qquad m\geq 2.
\end{equation}

The dependence on $\varepsilon$ drops out of the final result, although the calculation is controlled only for weak shape deformations. As $m\gg1$, the angular velocity approaches a constant because the driving and damping vanish with the same scaling. Note that $m\to \infty$ is a singular limit since neither damping nor torque is applied.

Once again, odd viscosity is subleading as it only adds a $\sin$ term to the streamfunction Eq.~\eqref{eq:phi_omega} at order $\mathcal O(\varepsilon)$, which does not contribute to the torque.

It would be interesting to obtain an even or odd response in the cross-coupling coefficient $\Gamma_{x\varphi}$. This, however, would require a polar shape and the inclusion of the convective term for the odd response. We leave this more involved problem for future work and instead propose an analysis based on the reciprocity relations.

\section{Linear response: Integral formulas for the damping matrix}
\label{sec:linear_formulas}

Before concluding, we derive a general expression for the damping matrix as a formal expression obtained from linear response. In passive fluids, Lorentz reciprocity relations imply that the damping matrix in linear response is symmetric~\cite{kim2013microhydrodynamics}; here, we will see that the interplay between inertia and edge currents naturally leads to an antisymmetric contribution. It is a very general result that sheds some light on the generation of odd responses by the edge currents. Our derivation follows Ref.\onlinecite{choudhary2019inertial} and uses the same technique. For the linear response to be well defined in two dimensions, we need to include a weak linear damping term in the chiral Navier-Stokes equation $-\gamma\rho_b\bm u$. It can be set to 0 for a 3D system.

We consider an intruder of arbitrary shape whose steady rigid-body velocity is $\bm U_0=(0,0,\Omega_0)$, \emph{i.e.}, a purely rotating state with no translation, as for the non-circular intruders discussed above. We work in the co-rotating body frame, in which the reference flow $(\bm u_0(\bm r),\mathtt p_0(\bm r))$ is stationary. We perturb the intruder's generalized velocity slightly and see how the flow reacts:
\begin{equation}
\begin{split}
\bm u(\bm r)&=\bm u_{0}(\bm r)+\delta U_i \bm u^{(i)}(\bm r)+\mathcal O(\delta U^2),\\ \mathtt p(\bm r)&=\mathtt p_{0}(\bm r)+\delta U_i \mathtt p^{(i)}(\bm r)+\mathcal O(\delta U^2),\end{split}
\label{eq:pertubation_linear}
\end{equation}
where we recall $i\in\{x,y,\varphi\}$. Here $\bm u^{(i)}$ is dimensionless, and the notation $\delta U_i \bm u^{(i)}$ (\textit{without} implicit summation) must be understood as the linear correction to the velocity field when the intruder is perturbed by a small velocity (or angular velocity) change $\delta U_i$, with $\delta \bm U=(\delta V_x,\delta V_y,\delta \Omega)$. If $(\bm u,\mathtt p)$ satisfies the steady damped chiral Navier-Stokes equations, then the first-order fields $(\bm u^{(i)},\mathtt p^{(i)})$ obey the Stokes problem driven by a linearized convective term and a Coriolis-like force 
\begin{equation}
\begin{split}
-\bm \nabla \mathtt p^{(i)}+\eta \bm \nabla^2 \bm u^{(i)}&-\gamma\rho_b\bm u^{(i)}=\rho_b\Big[(\bm u_{0}\cdot \bm \nabla)\bm u^{(i)}+\\&(\bm u^{(i)}\cdot \bm \nabla)\bm u_{0} +2\Omega_0 \hat{\bm z}\times \bm u^{(i)}\Big],\\
\bm u^{(i)}\cdot \hat{\bm n}= e_i, \qquad \hat{\bm t}\cdot &\mathsf \Sigma^{(i)}\cdot \hat{\bm n}=0, \qquad \bm r\in \partial \mathcal B,
\label{eq:NS_reciprocity}
\end{split}
\end{equation}
where the last term in the bulk equation is the gyroscopic contribution that arises in the body frame. We recall that
\begin{equation}
e_x=\hat n_x, \qquad e_y=\hat n_y, \qquad e_\varphi=[\bm r_0\times \hat{\bm n}]_z,
\end{equation}
and
\begin{equation}
\mathsf \Sigma^{(i)}=-\mathtt p^{(i)}\bm 1+\eta\Big(\bm \nabla\otimes \bm u^{(i)}+(\bm \nabla\otimes \bm u^{(i)})^T\Big).
\end{equation}

We define the linear-response relations:
\begin{equation}
\delta  F_i=-\Gamma_{ij}\delta U_j, \quad \Gamma_{ij}=\oint_{\partial \mathcal B} ds  \hat{\bm n}\cdot \mathsf \Sigma^{(j)}\cdot \hat{\bm n}e_i.
\label{eq:Kij_boundary_definition}
\end{equation}
The first two components of $\delta F_i$ are the force, while the third is the torque. Eq.~\eqref{eq:Kij_boundary_definition} is now well defined in two dimensions for any finite $\gamma>0$. Indeed, the damping introduces the screening length $\ell_\gamma=\sqrt{\eta/(\gamma\rho_b)}$, so that the perturbation fields decay at infinity. In the limit $\gamma\to0$, the translational sector remains singular because of the Stokes paradox, whereas the angular sector is regular. We also note that $\Gamma_{\varphi\varphi}$ obtained previously from the torque response at finite $\Omega_0$ must coincide, by linearity, with the one obtained here by perturbing around that state.

Multiplying the $j$-th mode of Eq.~\eqref{eq:NS_reciprocity} by $\bm u^{(i)}$ and integrating over space, we isolate by integration by parts a boundary term on $\partial\mathcal B$ proportional to $\Gamma_{ij}$. This yields
\begin{equation}
\begin{split}
\Gamma_{ij}=\,\,&2\eta\int_{\mathcal D} d\bm r  \mathsf E^{(i)}:\mathsf E^{(j)}+\gamma\rho_b\int_{\mathcal D} d\bm r  \bm u^{(i)}\cdot \bm u^{(j)}\\
&+\rho_b\int_{\mathcal D} d\bm r  \bm u^{(i)}\cdot \Big[(\bm u_{0}\cdot \bm \nabla)\bm u^{(j)}+(\bm u^{(j)}\cdot \bm \nabla)\bm u_{0}\\
&+2\Omega_0 \hat{\bm z}\times \bm u^{(j)}\Big],
\end{split}
\label{eq:Kij_bulk_formula}
\end{equation}
with the strain rate of the perturbation:
\begin{equation}
\mathsf E^{(i)}=\frac12\Big(\bm \nabla\otimes \bm u^{(i)}+(\bm \nabla\otimes \bm u^{(i)})^T\Big).
\end{equation}
For $\gamma>0$, the velocity fields vanish at infinity, and the integration by parts is therefore justified. Placing the system in a finite box is not a viable alternative, since it would induce edge currents.

We now decompose the damping matrix into symmetric and antisymmetric parts,
\begin{equation}
\Gamma_{ij}^{\rm S}=\frac{\Gamma_{ij}+\Gamma_{ji}}{2}, \qquad \Gamma_{ij}^{\rm A}=\frac{\Gamma_{ij}-\Gamma_{ji}}{2}.
\end{equation}
Defining
\begin{equation}
\begin{gathered}
\mathsf S^{(0)}=\frac12\Big(\bm \nabla\otimes \bm u_{0}+(\bm \nabla\otimes \bm u_{0})^T\Big), \\\mathsf W^{(0)}=\frac12\Big(\bm \nabla\otimes \bm u_{0}-(\bm \nabla\otimes \bm u_{0})^T\Big),
\end{gathered}
\end{equation}
we obtain:
\begin{equation}
\begin{split}
\Gamma_{ij}^{\rm S}=&2\eta\int_{\mathcal D} d\bm r \mathsf E^{(i)}:\mathsf E^{(j)}+\gamma\rho_b\int_{\mathcal D} d\bm r \bm u^{(i)}\cdot \bm u^{(j)}\\
&+\rho_b\int_{\mathcal D} d\bm r \bm u^{(i)}\cdot \mathsf S^{(0)}\cdot \bm u^{(j)},
\end{split}
\label{eq:Kij_s}
\end{equation}
and
\begin{equation}
\begin{split}
\Gamma_{ij}^{\rm A}=&\frac{\rho_b}{2}\int_{\mathcal D} d\bm r \, \Big[\bm u^{(i)}\cdot (\bm u_{0}\cdot \bm \nabla)\bm u^{(j)}-\bm u^{(j)}\cdot (\bm u_{0}\cdot \bm \nabla)\bm u^{(i)}\Big]\\
&-\rho_b\int_{\mathcal D} d\bm r \bm u^{(i)}\cdot \mathsf W^{(0)}\cdot \bm u^{(j)}\\
&-2\rho_b\Omega_0\int_{\mathcal D} d\bm r  \big[\bm u^{(i)}\times \bm u^{(j)}\big]_z\\
=&\frac{\rho_b}{2}\int_{\mathcal D} d\bm r \, \Big[\bm u^{(i)}\cdot (\bm u_{0}\cdot \bm \nabla)\bm u^{(j)}-\bm u^{(j)}\cdot (\bm u_{0}\cdot \bm \nabla)\bm u^{(i)}\Big]\\&-\frac{\rho_b}{2}\int_{\mathcal D} d\bm r (\omega_0 + 4\Omega_0)\big[\bm u^{(i)}\times \bm u^{(j)}\big]_z,
\end{split}
\label{eq:Kij_a}
\end{equation}
where $\omega_0=[\bm \nabla\times \bm u_{0}]_z$ is the 2D vorticity.

Eqs.~\eqref{eq:Kij_s} and \eqref{eq:Kij_a} are the integral representations of the linear-response matrix around a steady reference state, regularized by a weak linear damping. The symmetric part $\Gamma_{ij}^{\rm S}$ is governed by viscous dissipation $\eta$, by the explicit damping term $\gamma\rho_b$, and by the symmetric part of the reference-flow gradient. The antisymmetric part, in contrast, arises only from a non-zero base field $\bm u_0$ and inertial terms, as can be seen from its linear dependence on $\rho_b$. Thus, although often stated, odd viscosity is not required \emph{per se} to obtain an odd response~\cite{khain2024trading}, provided one goes beyond the Stokes, or low-Reynolds-number, regime. When inertia and rotation are accounted for, an odd response can be understood as a Magnus-like force.

Because of the Stokes paradox, the limit $\gamma\to0$ is singular for the translational response in two dimensions. Thus, for finite $\gamma$, Eqs.~\eqref{eq:Kij_s} and \eqref{eq:Kij_a} are well-defined integral formulas, while their undamped two-dimensional limit should be understood with care. In sectors where the limit is regular, such as the angular response, one may safely take $\gamma\to0$.

\section{Discussion and conclusion}\label{sec:discussion_conclusion}

We derived both the odd response and the torque exerted on an intruder by the bath's chirality. Rather than treating odd viscosity as the primary manifestation of broken parity and time-reversal symmetry, we focused on edge currents generated by the torque density $\tau$ in the bulk, which, in our incompressible description, only generated a tangential forcing at the boundary of the intruder. This immediately highlights a limitation of the Boltzmann-Lorentz approach used at the beginning. In that framework, chirality in the bath is essentially invisible, except possibly through a non-Gaussian velocity distribution, and edge currents, among other possible structures in the bath, are absent from the description. The hydrodynamic description is therefore complementary to the dilute one.

A natural extension of our description is to incorporate stress fluctuations into the fluid hydrodynamics to derive an effective Langevin noise acting on the intruder~\cite{grant1983fluctuating}. A complete account of the hydrodynamic regime should also systematically include odd viscosity and partial-slip boundary conditions.

The dilute theory applies in the large-Knudsen-number limit, whereas the hydrodynamic description pertains to small Knudsen numbers. Each regime, therefore, captures only part of the underlying physics.  For instance, hydrodynamics hides the microscopic intruder-bath collision rule by reducing it partially to a boundary condition, thereby losing essential microscopic information. In particular, a faithful hydrodynamic description should retain the restitution coefficient $\alpha$, which is difficult to implement at the Navier-Stokes level because collisions are not explicitly resolved and $\bm u\cdot\hat{\bm n}=0$ already enforces a normal condition on the flow. A boundary layer is therefore required. In addition, we expect an energy-depleted region around the intruder. Capturing it would require reintroducing the temperature field~\cite{johnson1987frictional, jenkins1986boundary} whose gradients around an intruder might produce a driving. A related issue is the force exerted on confining walls. In nonequilibrium systems, this force is generally not determined by a bulk pressure and therefore is not, in general, a state function. Although a homogeneous bulk \emph{virial} pressure can be expressed in terms of bulk fields, it does not necessarily coincide with the mechanical pressure controlling wall forces. This distinction was first identified in an asymmetric granular piston, whose two faces have different restitution coefficients~\cite{costantini2008noise, talbot2010analysis}, and was later emphasized in active matter~\cite{solon2015pressure,junot2017active}. Thus, a complete theory of the intruder, incorporating both kinetic and hydrodynamic effects, is substantially more involved.

A more promising route is likely a hybrid one: working at the Boltzmann-equation level while relaxing the assumption of vanishing intruder-bath correlations. Approaches of this type can, for example, recover Stokes drag directly from kinetic theory, without first performing a Chapman-Enskog expansion~\cite{scharf1970stokes, cukier1980microscopic, masters1981molecular}. Intermediate descriptions may be built from hybrid schemes, such as adding the Stokes drag phenomenologically to a kinetic equation~\cite{puertas2025tracer}, or supplementing hydrodynamics with Knudsen boundary layers to capture the kinetic region around the intruder~\cite{hadjiconstantinou2006limits}. Such approaches may provide explicit expressions for the odd response of a chiral intruder in an achiral nonequilibrium bath that are predicted on symmetry-based approaches~\cite{Passive2025HargusPRE}, but are too subtle to be captured by either of our present theories. 

More broadly, while the emergence of edge currents is now well understood, their microscopic origin and description are model-dependent. In our case, they arise directly from the torque density itself generated by transverse forces within the bath. Such transverse forces are a relatively recent modeling strategy for chiral systems~\cite{caporusso2024phase,caprini2025bubble, marconi2025spontaneous, huang2025anomalous, guo2025chirality, maire2025hyperuniformity, guo2025diffusion, guo2026tuning, maire2026kinetic, ghimenti2023sampling, ghimenti2024irreversible, ghimenti2024transverse,abdoli2026dynamicaldensityfunctionaltheory, metzger2026equationstateedgeflow}. They should be understood as effective interactions emerging from underlying chiral dynamics---for instance, particle rotation---that may have been coarse-grained out~\cite{Marconi2026hydrodynamics, markovich2024nonreciprocity}. It is still unclear how essential these explicit chiral degrees of freedom are and how faithfully such effective forces reproduce the true dynamics.

This issue is already visible at the level of the steady state. For instance, overdamped particles obeying
\begin{equation}
    \gamma\partial_t\bm r_i=-(\bm \nabla_i U+\bm z\times\bm\nabla_i U)+\sqrt{2\gamma T}\bm \zeta,
\end{equation}
where $U$ is a scalar potential, has the equilibrium stationary distribution~\cite{ghimenti2023sampling}
\begin{equation}
    P[\{\bm r_i\}] \sim \exp[-U/T],
\end{equation}
even though nonequilibrium currents perpendicular to $\bm \nabla U$ span phase space~\cite{abdoli2020nondiffusive}. By contrast, when chirality does not arise solely from such peculiar transverse forces, the steady state is generally not Gibbsian. This raises the question of whether our coarse-graining discards too much microscopic structure. For edge currents themselves, this may not be crucial as they persist in both settings and are not directly tied to the stationary distribution. In any case, for underdamped particles, the stationary state is not Gibbsian~\cite{Marconi2026hydrodynamics}.

This question can also be addressed directly at the hydrodynamic level~\cite{markovich2024nonreciprocity}. In our framework, the torque density generated by transverse forces enters the equations through a term of the form $\bm \varepsilon \cdot \bm \nabla \tau$, where $\tau$ depends on local hydrodynamic fields such as the density and temperature~\cite{maire2026kinetic}. For spinning particles, by contrast, chirality is carried by the particle spin itself, so no torque density appears directly. Instead, edge currents are generated by a term proportional to the rotational viscosity $\eta_R$~\cite{soni2019odd, jia2022incompressible, neville2025breakup}. This term is controlled by the mismatch between the fluid vorticity and the local spin rate $\mathit{\Omega}_S(\bm r, t)$~\cite{tsai2005chiral}:
\begin{equation}
    \partial_t\bm u\propto\eta_R\bm\varepsilon\cdot\bm\nabla \Big(2\mathit\Omega_S(\bm r, t) - [\bm\nabla\times \bm u(\bm r, t)]_z \Big).
\end{equation}
Like the torque density, it arises from interactions, with $\tau,\eta_R \sim n^2$, rather than from the microscopic kinetic stress~\cite{maire2026kinetic,tsuzuki2026retained}. Under adiabatic elimination of $\mathit\Omega_S$, the torque density $\tau$ used in our paper is proportional to the steady-state value of $\mathit\Omega_S$~\cite{markovich2024nonreciprocity}, driven internally or externally by a torque. Although retaining the spin field is unlikely to alter the qualitative physics identified here~\cite{lou2022odd, mecke2025obstacle}, this must still be verified.

It would therefore be interesting to extend our intruder calculation to a theory that explicitly models $\mathit\Omega_S$. Such an extension, however, is substantially more involved than the Stokes-flow problem considered here and would likely require a numerical treatment. However, our coarse-grained description forced us to retain convection to evade Lorentz reciprocity and obtain an odd response. It is plausible that, once the spin field is kept explicitly, convection is no longer required to generate an antisymmetric response.

A third class of models exhibiting edge currents consists of self-propelled particles with a chiral bias in their propulsion~\cite{metzger2026equationstateedgeflow,langford2025phase,Passive2025HargusPRE,hargus2025odd}. Here too, the currents arise from an antisymmetric stress, now generated by activity. In the dilute regime, analytical progress is difficult because particles accumulate and stick at boundaries, and a Boltzmann-Lorentz description cannot capture such correlated collisions, although kinetic approaches based on the Boltzmann equation have recently been developed for self-propelled particles~\cite{soto2024kinetic,pinto2025hydrodynamic,soto2025self}. In the dense regime, hydrodynamic equations should again provide a description of the intruder's dynamics.

Finally, there is a fourth class of models in which edge transport can arise, but in this case, it is tied to the topology of the linear excitation spectrum. In these systems, a \emph{bulk} Coriolis/Lorentz-like body force: $\partial_t\bm u\propto \omega_S(\bm\varepsilon\cdot\bm u)$ opens a gap in the acoustic-wave spectrum, while the odd viscosity makes it possible to define a nontrivial topological invariant~\cite{PhysRevLett.122.128001, shankar2022topological, fujii2025gauge}. Through the bulk-boundary correspondence, the boundary then supports a chiral edge mode determined by the bulk spectrum. Although these mechanisms are often discussed together, this one is distinct from the torque-density and spin-vorticity scenarios described above, which attribute stationary edge flows to \emph{antisymmetric stresses} (although edge currents and topological modes can coexist~\cite{kuroda2026designing,edwards2026robusttopologicallyprotectededge}). In contrast, the topological case only guarantees the existence of an excitable mode with a preferred propagation direction on the boundaries and is therefore weaker than the previously studied mechanisms. To the best of our knowledge, intruder dynamics in such systems remain unexplored.

\section{Acknowledgments}
RM acknowledges useful discussions with Lorenzo Caprini, Umberto Marini Bettolo Marconi, and Alessandro Petrini.

\appendix

\section{Effective collision rule between the intruder and an active spinner}\label{app:spinner}

Let us temporarily endow a bath particle, now called a \emph{spinner}, with a finite diameter $\sigma$ and an angular velocity $\omega$. The velocity of its contact point is then
\begin{equation}
\bm v^{(c)}=\bm v-\frac{\sigma}{2}\omega \hat{\bm t},
\end{equation}
so that the actual relative velocity at contact becomes
\begin{equation}
\begin{gathered}
\bm g^{\rm sp}=\bm v^{(c)}-\bm V^{(c)}=\bm g-\frac{\sigma}{2}\omega \hat{\bm t},\qquad g_n^{\rm sp}=g_n,\\ g_t^{\rm sp}=g_t-\frac{\sigma}{2}\omega.
\end{gathered}
\end{equation}
For a smooth collision ($J_t=0$), Eq.~\eqref{eq:g_to_invert} gives
\begin{equation}
{g_{t}^{\rm sp}}'=g_t^{\rm sp}-\frac{\kappa_n\kappa_t}{I}\frac{1+\alpha}{\lambda_n}g_n, \qquad\text{(smooth)}.
\end{equation}
Therefore, when $\kappa_n\kappa_t\neq 0$, the quantity that is left invariant by a smooth collision is not $g_t^{\rm sp}$ itself, but rather
\begin{equation}
\chi^{\rm sp}= g_t^{\rm sp}-\frac{\kappa_n\kappa_t}{I \lambda_n}g_n.
\label{eq:chi_spinner}
\end{equation}
Indeed, using $g_n'=-\alpha g_n$, we can show that $J_t=0$ implies ${\chi^{\rm sp}}'=\chi^{\rm sp}$.

A natural phenomenological rough extension, consistent with the smooth limit, is then to impose a restitution law on $\chi^{\rm sp}$ rather than on $g_t^{\rm sp}$, namely
\begin{equation}
g_n'=-\alpha g_n,\qquad {\chi^{\rm sp}}'=\beta \chi^{\rm sp},
\qquad -1\le \beta \le 1.
\label{eq:rough_spinner_rule}
\end{equation}
This yields
\begin{equation}
\begin{gathered}
J_t=\frac{1-\beta}{\Lambda_t}\chi^{\rm sp}=\frac{1-\beta}{\Lambda_t}\left(g_t^{\rm sp}-\frac{\kappa_n\kappa_t}{I \lambda_n}g_n\right),\\\Lambda_t = \lambda_t-\frac{\kappa_n^2\kappa_t^2}{I^2\lambda_n}>0,
\end{gathered}
\label{eq:Jt_spinner_consistent}
\end{equation}
together with
\begin{equation}
J_n=\frac{1+\alpha}{\lambda_n}g_n-\frac{\kappa_n\kappa_t}{I \lambda_n}J_t.
\label{eq:Jn_spinner_consistent}
\end{equation}
$\lambda_t$ may now also include the moment of inertia of the bath particles. This does not change the argument made. By construction, $\beta=1$ gives $J_t=0$, \emph{i.e.}, the correct smooth limit. Moreover, when the intruder is circular, $\kappa_n\kappa_t=0$, so that $\chi^{\rm sp}=g_t^{\rm sp}$, $\Lambda_t=\lambda_t$, and Eq.~\eqref{eq:Jt_spinner_consistent} reduces to the standard hard-disk expression $J_t=(1-\beta)g_t^{\rm sp}/\lambda_t$~\cite{brilliantov2010kinetic}. We also note that the energy change at collision is given by:
\begin{equation}
\Delta E=-\frac{1-\alpha^2}{2\lambda_n}g_n^2-\frac{1-\beta^2}{2\Lambda_t}\left(\chi^{\rm sp}\right)^2\le 0,
\label{eq:energy_balance_spinner}
\end{equation}
which is negative, as expected for a dissipative collision. This expression has the same form as that for rough hard disks, except for the appropriate tangential variable. To the best of our knowledge, the formulation of a consistent dissipative collision rule for non-spherical hard particles is new.

Since the bath particles are active spinners whose angular velocities rapidly relax between collisions to a prescribed value $\omega_0$, we may eliminate their rotational degree of freedom by setting $\omega\simeq \omega_0$ at impact. Eq.~\eqref{eq:Jt_spinner_consistent} then becomes
\begin{equation}
J_t=\frac{1-\beta}{\Lambda_t}\left(g_t-\frac{\sigma}{2}\omega_0-\frac{\kappa_n\kappa_t}{I\lambda_n}g_n
\right).
\label{eq:Jt_spinner_eliminated}
\end{equation}
Thus, after eliminating the bath-particle rotation, the tangential impulse splits into a velocity-dependent rough contribution and a constant spin-induced term.

In the regime where the latter dominates:
\begin{equation}
\left|\frac{\sigma\omega_0}{2}\right|\gg |g_t|,\qquad\left|\frac{\kappa_n\kappa_t}{I\lambda_n}g_n\right|\ll\left|\frac{\sigma\omega_0}{2}\right|,
\end{equation}
an effective collision rule for the tangential sector is obtained:
\begin{equation}
J_t\simeq -\frac{1-\beta}{\Lambda_t}\frac{\sigma\omega_0}{2}.
\end{equation}
For weak normal-tangential mixing,
\begin{equation}
\frac{\kappa_n^2\kappa_t^2}{I^2\lambda_n}\ll \lambda_t,
\end{equation}
this further simplifies to
\begin{equation}
J_t\simeq-\frac{1-\beta}{\lambda_t}\frac{\sigma\omega_0}{2}.
\end{equation}
This has the same structure as Eq.~\eqref{eq:J_t}, with the identification
\begin{equation}
\Delta_{\rm eff}\simeq -\frac{1-\beta}{4}\sigma\omega_0.
\end{equation}
Therefore, Eq.~\eqref{eq:J_t} can be interpreted as the minimal coarse-grained remnant of the active rotation of the bath. We note that after eliminating the spin degrees of freedom in the bath, the reduced collision rule \eqref{eq:Jt_spinner_eliminated} becomes effectively driven, so the corresponding energy change at collision is no longer constrained to be negative. This is similar to coarse-grained models for confined granular gases~\cite{brito2026dynamicpropertiescollisionalmodel}.

\section{Non-diagonal kinetic energy, ratchet forces and fluctuation-dissipation theorem}
\label{sec:fluctuationDissipation}
\subsection{Non-diagonal kinetic temperature}

In the linear theory, the velocity covariance $\mathsf S^{\rm lin}$ satisfies:
\begin{equation}
(\Upgamma\cdot\mathsf M^{-1})\cdot \left(\mathsf M\cdot \mathsf S^{\rm lin}\cdot \mathsf M\right)+\left(\mathsf M\cdot\mathsf S^{\rm lin}\cdot\mathsf M\right)\cdot\mathsf (\Upgamma\cdot\mathsf M^{-1})^{T}=\mathsf D.
\label{eq:simple_covarianc}
\end{equation}
When $m\Delta^2\ll T_b$, a simple relation is found
\begin{equation}
\mathsf S^{\rm lin} = T_I\mathsf M^{-1} ,\qquad T_I=\dfrac{\mathcal K_3}{2\mathcal K_1}\dfrac{(1+\alpha)}{2}T_b,
\end{equation}
although the kinetic temperature of the intruder $T_I$ is generally not equal to the kinetic temperature of the bath $T_b=m\int d\bm v \bm v^2f(\bm v)/2$ if $f$ is non-Gaussian or $\alpha<1$. Such equipartition breakdown is typical of granular fluids~\cite{gonzalez2025diffusion,barrat2002lack}. Despite this feature, we can cleanly associate the same kinetic energy to each degree of freedom since the velocity covariance is proportional to the identity. Indeed, $M \langle V_y^2\rangle = M \langle V_x^2\rangle = I\langle \Omega^2\rangle = T_I$.

However, when $\Delta$ is not negligible, the covariance is no longer proportional to $\mathsf M^{-1}$. To see this, we write in the body frame
\begin{equation}
\tilde{\mathsf S}^{\rm lin} = T_I\mathsf M^{-1}+\delta\tilde{\mathsf S},
\label{eq:remaining_contribution}
\end{equation}
and use the decomposition
\begin{equation}
\tilde{\Upgamma}=\tilde{\Upgamma}^{(\alpha)}+\tilde{\Upgamma}^{(\Delta)},\qquad \tilde{\mathsf D}=\tilde{\mathsf D}^{(\alpha)}+\tilde{\mathsf D}^{(\Delta)},
\end{equation}
with
\begin{equation}
\tilde{\mathsf D}^{(\alpha)}=2T_I\tilde{\Upgamma}^{(\alpha)},\qquad\tilde{\mathsf D}^{(\Delta)}=2T_\Delta \tilde{\Upgamma}^{(\Delta),S},
\end{equation}
where we defined:
\begin{equation}
     T_\Delta\equiv (1+\alpha)T_b, \qquad \tilde{\Upgamma}^{(\Delta),S}\equiv\dfrac{\tilde{\Upgamma}^{(\Delta)}+\tilde{\Upgamma}^{(\Delta)T}}{2}
\end{equation}
Substituting Eq.~\eqref{eq:remaining_contribution} into Eq.~\eqref{eq:simple_covarianc}, and using that $T_I\mathsf M^{-1}$ solves the $\Delta=0$ problem, we obtain:
\begin{equation}
\mathsf M^{-1}\cdot\tilde{\Upgamma}^{(\alpha)}\cdot\delta\tilde{\mathsf S} + \delta\tilde{\mathsf S}\cdot\tilde{\Upgamma}^{(\alpha)}\cdot\mathsf M^{-1} = 2(T_\Delta-T_I)\mathsf M^{-1}\cdot\tilde{\Upgamma}^{(\Delta), S}\cdot\mathsf M^{-1}.
\label{eq:sylvester_app}
\end{equation}
We now define:
\begin{equation}
\begin{gathered}
\mathsf X\equiv \sqrt{\mathsf M}\cdot\delta\tilde{\mathsf S}\cdot \sqrt{\mathsf M}, \qquad \mathsf L\equiv  \sqrt{\mathsf M^{-1}}\cdot\tilde{\Upgamma}^{(\alpha)}\cdot\sqrt{\mathsf M^{-1}}, \\\mathsf G\equiv  \sqrt{\mathsf M^{-1}}\cdot\Big(\tilde{\Upgamma}^{(\Delta)}+\tilde{\Upgamma}^{(\Delta)T}\Big)\cdot\sqrt{\mathsf M^{-1}},
\end{gathered}
\end{equation}
to rewrite Eq.~\eqref{eq:sylvester_app} in the simple form:
\begin{equation}
\mathsf L\cdot\mathsf X+\mathsf X\cdot\mathsf L=(T_\Delta-T_I)\mathsf G.
\end{equation}
Since $\mathsf L$ is symmetric, it can be diagonalized with eigenvalues $\lambda_{1, 2, 3}$. By a change of basis, we therefore write:
\begin{equation}
\begin{gathered}
\mathsf L=\mathsf O \cdot \mathrm{diag}(\lambda_1,\lambda_2,\lambda_3)\cdot  \mathsf O^T,\\\widetilde{\mathsf X}=\mathsf O^T\cdot\mathsf X\cdot\mathsf O,\qquad\widetilde{\mathsf G}=\mathsf O^T\cdot\mathsf G\cdot\mathsf O.
\end{gathered}
\end{equation}
Componentwise, we have:
\begin{equation}
(\lambda_i+\lambda_j)\widetilde X_{ij}=(T_\Delta-T_I)\widetilde G_{ij}\Rightarrow \widetilde X_{ij}=(T_\Delta-T_I)\frac{\widetilde G_{ij}}{\lambda_i+\lambda_j}.
\end{equation}
This gives the covariance:
\begin{equation}
\tilde{\mathsf S}^{\rm lin} = T_I\mathsf M^{-1} +
(T_\Delta-T_I) \sqrt{\mathsf M^{-1}}\cdot\mathsf O\cdot\left[\frac{\widetilde G_{ij}}{\lambda_i+\lambda_j}\right]\cdot\mathsf O^T\cdot\sqrt{\mathsf M^{-1}},
\label{eq:covariance_finiteDelta}
\end{equation}
where the content of the bracket has to be understood as a matrix defined componentwise.

The velocity covariance Eq.~\eqref{eq:covariance_finiteDelta} is not diagonal anymore, hence, we cannot attribute a clear kinetic temperature to each sector. Similarly, the laboratory-frame covariance, and therefore the fluctuations, become orientation dependent:
\begin{equation}
\mathsf S^{\rm lin}(\varphi)=\bar{\mathsf R}(\varphi) \cdot\tilde{\mathsf S}^{\rm lin} \cdot\bar{\mathsf R}^T(\varphi), \qquad \bar{\mathsf R}(\varphi)=
\begin{pmatrix}
\mathsf R(\varphi)&0\\
0&1
\end{pmatrix}.
\end{equation}

\subsection{Ratchet forces}

With this new fluctuating term, the ratchet force and torque $F_i(\varphi)\equiv R_i(\varphi)+N_{ijk}(\varphi)S^{\rm lin}_{jk}(\varphi)$ get modified. 
Using Eq.~\eqref{eq:covariance_finiteDelta}, we find:
\begin{equation}
    \begin{split}    
    \tilde N_{ijk}\tilde S^{\rm lin}_{jk}&=-\frac{1+\alpha}{2}  n_bm \epsilon\oint {ds}  e_i^{(n)}\left(e_j^{(n)}\tilde S^{\rm lin}_{jk}e_k^{(n)}\right)\\&~+\mathcal O(\epsilon^{3/2}).\\   
    &=-\frac{1+\alpha}{2}n_bm\epsilon\oint ds\bar e_i^{(n)}(s)\Big[\mu_n(s) \\&~+ \left(\dfrac{T_\Delta}{T_I}-1\right)\sum_{l, m}\frac{h_l(s)\widetilde G_{lm}h_m(s)}{\lambda_l+\lambda_m}\Big] +\mathcal O(\epsilon^{3/2})
    \end{split}
    \label{eq:contribution_dependent}
\end{equation}
with
\begin{equation}
\bm h(s)=\mathsf O^T\cdot\sqrt{\mathsf M^{-1}}\cdot\bm e^{(n)}(s).
\end{equation}

The second contribution in the bracket of Eq.~\eqref{eq:contribution_dependent} is a new fluctuation-driven ratchet force arising from $\Delta$, which was neglected in the main text. Interestingly, it can generate a non-zero force $\tilde F_x\neq 0$ since \emph{a priori} $\tilde N_{xjk}\delta \tilde S_{jk}^{\rm lin}\neq 0$ need not vanish for a polar shape.

\subsection{Fluctuation-Dissipation relation}

We note that the Fluctuation-Dissipation relations we found earlier are not affected by these computations. Indeed,
\begin{equation}
{\mathsf D}^{(\alpha)}=2T_I{\Upgamma}^{(\alpha)},\qquad{\mathsf D}^{(\Delta)}=2T_\Delta {\Upgamma}^{(\Delta),S},
\label{eq:fluctuation_dissipation_appendix}
\end{equation}
still hold. However, previously $T_I$, which enters Eq.~\eqref{eq:fluctuation_dissipation_appendix} was also found to be the kinetic temperature of the intruder:
\begin{equation}
    M\langle\bm V^2\rangle/2 + \mathcal O(m\Delta^2/T_b) =I\langle \Omega^2\rangle+ \mathcal O(m\Delta^2/T_b)=T_I.
\end{equation}
When $\Delta$ is not negligible with respect to $\sqrt{T_b/m}$, the temperature from the Fluctuation-Dissipation theorem remains $T_I$, but it is no longer the kinetic temperature of the intruder, as we showed.

\section{Geometric dependency of the coefficients in the Langevin equation}\label{app:geometry}

\subsection{Geometric identities}
For $a,b\in\{x,y\}$ and $i, j\in\{x, y, \varphi\}$, define
\begin{equation}
Q_{ab}\equiv \oint {ds} \hat n_a\hat n_b,\quad Q_{ab}^{\rm dev}\equiv Q_{ab}-\frac{\mathcal P}{2}\delta_{ab}, \quad \mathrm{tr}(\mathsf Q)=\mathcal P,
\end{equation}
where $\mathcal P=\oint {ds}$ is the perimeter. We also introduce
\begin{equation}
W_a\equiv \oint {ds}  \kappa_n\hat n_a,\quad S_a\equiv \oint {ds}  \kappa_t\hat n_a, \quad T_a\equiv \oint {ds}  \kappa_n\hat t_a,
\end{equation}
and the scalar moments
\begin{equation}
\begin{gathered}
P_a\equiv \oint {ds}  \kappa_n^2\hat n_a,\qquad H_n\equiv \oint {ds}  \kappa_n^2,\\ H_t\equiv \oint {ds}  \kappa_n\kappa_t,\qquad H_{\varphi}\equiv \oint {ds}  \kappa_n^3 .
\end{gathered}
\label{eq:vectors_to_ref}
\end{equation}
Since $\hat t_a=-\varepsilon_{ab}\hat n_b$ (with $\bm \varepsilon$ the Levi-Civita pseudotensor), we obtain $T_a=-\varepsilon_{ab}W_b$, so $\bm T$ is not independent from $\bm W$.

For a closed boundary, with $\bm r\equiv \bm r_0(s)$ and $\hat{\bm t}=d\bm r/{ds}$, we have
\begin{equation}
\oint {ds}  \hat t_a= \oint dr_a =0,\qquad\oint {ds}  \hat n_a= -\varepsilon_{ab}\oint {ds}  \hat t_b =0.
\label{eq:closedcurve_id1}
\end{equation}
Moreover,
\begin{equation}
\kappa_n=[\bm r\times \hat{\bm n}]_z=\bm r\cdot \hat{\bm t},\qquad\kappa_t=[\bm r\times \hat{\bm t}]_z,
\end{equation}
hence
\begin{equation}
\oint {ds} \kappa_n=\oint {ds}  \bm r\cdot \hat{\bm t}=\oint \bm r\cdot d\bm r=\frac12\oint d(r^2)=0,
\end{equation}
while Green's theorem gives
\begin{equation}
\oint {ds} \kappa_t=\oint (x dy-y dx)=2\mathcal A,
\label{eq:closedcurve_id3}
\end{equation}
with $\mathcal A$ the \emph{signed} area enclosed by the intruder.

\subsection{Force and torque: \texorpdfstring{$F_i$}{Fi}}
Using Eqs.~\eqref{eq:closedcurve_id1}-\eqref{eq:closedcurve_id3}, the translational and angular components (the torque) of the force array follow directly. 

For the chiral contribution, within our approximation $m\Delta^2\ll T_b$,
\begin{equation}
\begin{gathered}
F_a^{(\Delta)}\propto \oint {ds}  \hat t_a =0+\mathcal O(m\Delta^2/T_b),\\
\quad F_{\varphi}^{(\Delta)}\propto \oint {ds}  \kappa_t =2\mathcal A+\mathcal O(m\Delta^2/T_b).
\end{gathered}
\label{eq:result_disprovedinAppendix}
\end{equation}
As shown in Appendix~\ref{sec:fluctuationDissipation}, relaxing the condition $m\Delta^2 \ll T_b$ gives rise to a ratchet-like force, so that $F_a^{(\Delta)} \neq 0$ for a polar shape and the torque is no longer set solely by the intruder area. These corrections are, however, subleading and have cumbersome expressions. We omit them from now on.

For the normal/dissipative contribution,
\begin{equation}
F_a^{(\alpha)}\propto\oint {ds}  \mu_n \hat n_a=\frac1I\oint {ds}  \kappa_n^2\hat n_a=\frac1I P_a,
\end{equation}
and
\begin{equation}
F_{\varphi}^{(\alpha)}\propto\oint {ds}  \mu_n \kappa_n=\frac1I\oint {ds}  \kappa_n^3=\frac1IH_{\varphi}.
\end{equation}
Therefore,
\begin{equation}
F_a^{(\alpha)}\propto (T_b-T_I) P_a,
\qquad
F_{\varphi}^{(\alpha)}\propto (T_b-T_I) H_{\varphi}.
\end{equation}
A nonzero $F_a^{(\alpha)}$ requires both nonequilibrium dynamics, either directly in the bath or in the interaction between the bath and the intruder $T_b\neq T_I$, and a polar shape since $P=\oint ds \kappa_n^2\hat{\bm n}$ is zero otherwise. Similarly, $H_{\varphi}=\oint ds [\bm r_0\times\hat{\bm n}]_z^3$ is a pseudoscalar, and a nonzero torque $F_{\varphi}^{(\alpha)}$ requires both nonequilibrium dynamics and a chiral shape. We note that this form of the ratchet force, driven by an effective temperature difference between the intruder and the bath, was already identified in Refs.~\onlinecite{costantini2007granular, cleuren2007granular} and closely resembles the rectification of thermal fluctuations in Brownian motors coupled to two baths at different temperatures~\cite{Meurs_Van2004}.

\subsection{Translational damping: \texorpdfstring{$\Gamma_{ab}$}{Gab}.}
The $\alpha$-sector is simple:
\begin{equation}
\Gamma_{ab}^{(\alpha)}\propto \oint {ds}  \hat n_a\hat n_b = Q_{ab},
\end{equation}
which is symmetric and therefore does \emph{not} generate odd transport, even for a chiral intruder. This differs from Ref.~\onlinecite{Passive2025HargusPRE}, where odd transport arises in an achiral nonequilibrium bath with a chiral intruder. As we explain below, the vanishing found here follows from the assumption of uncorrelated collisions.

The $\Delta$-sector reads
\begin{equation}
\Gamma_{ab}^{(\Delta)}\propto\oint {ds}  \hat t_a\hat n_b=(\bm\varepsilon\cdot \mathsf Q)_{ab}.
\end{equation}
Since $\mathsf Q$ is symmetric, it is natural to split this into symmetric and antisymmetric parts:
\begin{equation}
\Gamma_{ab}^{(\Delta),{\rm S}}\equiv\frac{\Gamma_{ab}^{(\Delta)}+\Gamma_{ba}^{(\Delta)}}{2},\qquad\Gamma_{ab}^{(\Delta),{\rm A}}\equiv\frac{\Gamma_{ab}^{(\Delta)}-\Gamma_{ba}^{(\Delta)}}{2}.
\end{equation}
Using $\mathsf Q=\frac{\mathcal P}{2}\bm 1+\mathsf Q^{\rm dev}$ yields
\begin{equation}
\Gamma_{ab}^{(\Delta)}\propto\frac{\mathcal P}{2}\varepsilon_{ab}+(\bm \varepsilon\cdot \mathsf Q^{\rm dev})_{ab},
\end{equation}
with
\begin{equation}
\Gamma_{ab}^{(\Delta),{\rm A}}\propto\frac{\mathcal P}{2}\varepsilon_{ab},\qquad\Gamma_{ab}^{(\Delta),{\rm S}}\propto(\bm\varepsilon\cdot \mathsf Q^{\rm dev})_{ab}.
\end{equation}
Hence the antisymmetric part is universal and survives even for an isotropic body, whereas the symmetric part requires translational anisotropy. 

This decomposition is physically important. Only the symmetric part contributes to the deterministic power dissipated by friction (neglecting, without loss of generality, the orientational sector):
\begin{equation}
V_a\Gamma_{ab}V_b=V_a\Gamma_{ab}^{\rm S}V_b,
\end{equation}
because $V_a\Gamma_{ab}^{\rm A}V_b=0$. Thus, the universal perimeter term $\propto \mathcal P \varepsilon_{ab}$ is reactive rather than dissipative since it rotates the translational velocity but does not damp its magnitude. We note that $\Gamma_{ab}^{(\alpha)}$ is positive semidefinite,
\begin{equation}
V_a\Gamma_{ab}^{(\alpha)}V_b \propto\oint {ds}  \left(V_a e_a^{(n)}\right)^2 \ge 0,
\end{equation}
and thus dissipative. The same is not true of $\Gamma_{ab}^{(\Delta)}$: because it changes sign with $\Delta$, it can be positive-definite for a given $\Delta$ and negative-definite for $-\Delta$. This is nevertheless compatible with stability, since in our $\epsilon$ expansion the $\Delta$ sector is subleading to the $\alpha$ sector.

\subsection{Translational-rotation damping: \texorpdfstring{$\Gamma_{a\varphi}$}{GaO}.}
The mixed blocks follow directly from the definitions:
\begin{align}
\Gamma_{\varphi a}^{(\alpha)}&\propto\oint {ds}  \kappa_n \hat n_a=  W_a,\quad\Gamma_{a\varphi}^{(\alpha)}\propto\oint {ds}  \hat n_a\kappa_n=  W_a,
\end{align}
so the $\alpha$-sector is symmetric in the translation-rotation couplings,
\begin{equation}
\Gamma_{a\varphi}^{(\alpha)}=\Gamma_{\varphi a}^{(\alpha)}.
\end{equation}

For the $\Delta$-sector, however,
\begin{align}
\Gamma_{\varphi a}^{(\Delta)}\propto\oint {ds}  \kappa_t\hat n_a= S_a,\\\Gamma_{a\varphi}^{(\Delta)}\propto\oint {ds}  \kappa_n\hat t_a= T_a=\varepsilon_{ab}W_b.
\end{align}
These two vectors are generically distinct, so the chiral sector makes $\Gamma$ non-symmetric.
Its symmetric and antisymmetric mixed parts are
\begin{align}
\Gamma_{a\varphi}^{(\Delta),{\rm S}}=\Gamma_{\varphi a}^{(\Delta),{\rm S}}\propto \frac{T_a+S_a}{2},\\\Gamma_{a\varphi}^{(\Delta),{\rm A}}=-\Gamma_{\varphi a}^{(\Delta),{\rm A}}\propto \frac{T_a-S_a}{2}.
\end{align}

Since $\bm W$, $\bm S$, and $\bm T$ transform as vectors under rotations, any nontrivial discrete rotational symmetry $C_{n\geq 2}$ imposes:
\begin{equation}
\bm W=\mathsf R(2\pi/n)\cdot \bm W,
\end{equation}
and therefore:
\begin{equation}
\bm W=\bm S=\bm T=0.
\end{equation}
Thus, nonzero translation-rotation couplings require a polar object. Conversely, the absence of a $C_{n\geq 2}$ symmetry is necessary, but not sufficient, and accidental cancellation may happen.

Mirror symmetry about the polar axis further constrains the allowed directions of these vectors. For example, if the polar axis is along $y$ such that the object is invariant against $(x, y)\to (-x, y)$, we find $\bm W=(W_x, 0)$, $S=(0, S_y)$ and $\bm T=(0, T_y)$.

Our results contrast with the conclusion of Refs.~\onlinecite{hargus2025odd, Passive2025HargusPRE}, where such cross couplings were argued to vanish in an achiral bath for a polar object, such as a wedge. Their symmetry argument applies to bodies with nontrivial rotational symmetry, but not to generic polar intruders, as our explicit calculation shows.

Indeed, we find that the cross-couplings between the translational velocity of the intruder $\bm V = (V_x, V_y)$ and its angular velocity $\Omega$: $\Gamma_{a\varphi}$ (with $a\in\{x, y\}$) are proportional to the vector:
\begin{equation}
W_a\equiv \oint ds \kappa_n \hat{n}_a ,
\end{equation}
and therefore vanishes whenever symmetry forbids a nonzero vector of that type. This is immediate for any $C_{n\ge2}$ object. Since $\bm W$ transforms as an ordinary vector under rigid rotations of the body, invariance under a rotation $\mathsf R(2\pi/n)$ implies
\begin{equation}
\bm W=\mathsf R(2\pi/n)\cdot \bm W.
\end{equation}
For $n\ge2$, the only planar vector left invariant by a nontrivial rotation is the zero vector. Hence
\begin{equation}
\bm W=0 \qquad \Rightarrow \qquad
\Gamma_{a\varphi}^{(\alpha)}=\Gamma_{\varphi a}^{(\alpha)}=0 .
\end{equation}
This covers, in particular, disks and all apolar shapes.

By contrast, a single mirror symmetry does \emph{not} force $\bm W$ to vanish. Let the mirror axis be the body frame $y$ axis, so that reflection acts as
\begin{equation}
(x,y,\varphi)\mapsto (-x,y,-\varphi).
\end{equation}
Under this reflection, $\hat n_x$ is odd, $\hat n_y$ is even, and $\kappa_n=[\bm r_0\times \hat{\bm n}]_z$ is odd. Therefore $\kappa_n \hat n_x$ is even while  $\kappa_n \hat n_y$ is odd. After integration along the boundary, the odd contribution must vanish, but the even one is generically allowed:
\begin{equation}
W_y=0,\qquad W_x\neq 0.
\end{equation}
Equivalently, in the body frame, the reflection operator is
\begin{equation}
\mathsf P_\mid\equiv\rm{diag}(-1,1,-1),
\end{equation}
and invariance of the linear law $\bm F=-\Upgamma\cdot \bm U$ requires
\begin{equation}
\mathsf P_\mid\cdot\tilde{\Upgamma}\cdot \mathsf P_\mid^{-1}=\tilde{\Upgamma},
\end{equation}
which yields the general structure
\begin{equation}
\tilde{\Upgamma}=
\begin{pmatrix}
\Gamma_{xx} & 0 & \Gamma_{x\varphi}\\
0 & \Gamma_{yy} & 0\\
\Gamma_{\varphi x} & 0 & \Gamma_{\varphi\varphi}
\end{pmatrix}.
\end{equation}
Hence a polar body, such as a wedge or an isosceles triangle considered in Ref.~\onlinecite{Passive2025HargusPRE}, may have nonzero translation-rotation couplings~\cite{voss2018hydrodynamic}, but only in the direction perpendicular to the symmetry axis: $\Gamma_{x\varphi},\Gamma_{\varphi x}\neq 0$ but $\Gamma_{y\varphi}=\Gamma_{\varphi y}=0$.

This also clarifies why the argument of Refs.~\onlinecite{hargus2025odd, Passive2025HargusPRE} fails. They claim that, for an achiral object in an achiral nonequilibrium bath, reflection ‘‘preserves the fluctuation in the force but changes the sign of the torque'', so that the force-torque correlator vanishes. However, the force is a \emph{vector}, not a scalar:
\begin{equation}
(\delta F_x,\delta F_y, \delta F_\varphi)\mapsto (-\delta F_x,\delta F_y, -\delta F_\varphi),
\end{equation}
with $\delta \bm F=\bm F - \langle \bm F\rangle$. Therefore
\begin{equation}
\delta F_x \delta F_\varphi \mapsto + \delta F_x \delta F_\varphi, \qquad \delta F_y \delta F_\varphi \mapsto - \delta F_y \delta F_\varphi.
\end{equation}
Only the second correlator, which generates $\Gamma_{y\varphi}$, is forced to vanish by mirror symmetry. The first one is generically nonzero. 

\subsection{Rotation damping: \texorpdfstring{$\Gamma_{\varphi\varphi}$}{Goo}.}
For the rotation damping in the normal sector, we find:
\begin{equation}
\Gamma_{\varphi\varphi}^{(\alpha)}
\propto
\oint {ds}  \kappa_n^2
=H_n>0,
\end{equation}
where positivity is immediate. The chiral contribution is
\begin{equation}
\Gamma_{\varphi\varphi}^{(\Delta)}\propto-\Delta\oint {ds}  \kappa_t\kappa_n=- H_t\Delta.
\end{equation}
The scalar $H_t$ changes sign under reflection and therefore vanishes for any achiral body. A nonzero $\Gamma_{\varphi\varphi}^{(\Delta)}$ thus requires a chiral shape. Once again, $\Gamma^{\Delta}_{\varphi\varphi}$ may become negative and the system may \emph{formally} become unstable since $\Gamma$ might become non-positive definite within our approximations. This apparent instability is again unphysical as it lies outside the regime of validity of our linear theory, which assumes $m\Delta^2\ll T_b$. Higher-order terms can either reverse the sign of this contribution or keep it negative while producing additional nonlinear stabilizing terms of the form $|\Omega|\Omega^m$ with $m>0$. In the latter case, these terms saturate the growth and thus maintain a finite steady-state angular velocity. A simple example is
\begin{equation}
    \dot\Omega = |\Gamma_{\varphi\varphi}|\Omega - |\Gamma_{\varphi\varphi}^{(3)}|\Omega^3 + \sqrt{D_{\varphi\varphi}}\zeta,
\end{equation}
which is reminiscent of the early active-matter models known as the Rayleigh-Helmholtz model~\cite{ganguly2013stochastic}.

\subsection{Diffusion tensor: \texorpdfstring{$D_{ij}$}{Dij}}
The diffusion tensor is constructed from similar geometric objects. In the dissipative sector, we find,
\begin{equation}
D_{ij}^{(\alpha)}=2T_I \Gamma_{ij}^{(\alpha)},
\label{eq:FDT_alpha_final}
\end{equation}
which is a fluctuation-dissipation relation for the $\alpha$-sector. Note, however, that the temperature $T_I$ is not the same as that of the bath $T_b$ for a nonequilibrium dynamics. Moreover, $T_I$ corresponds to the kinetic temperature of the intruder only when $m\Delta^2\ll T_b$, otherwise we show in Appendix~\ref{sec:fluctuationDissipation} that for large $\Delta$, this is not the case anymore. In Appendix~\ref{sec:FDT_breaking_alpha}, we also show how Eq.~\eqref{eq:FDT_alpha_final} is modified when higher-order terms in the van Kampen expansion are retained, and how these terms break this fluctuation-dissipation relation through non-Gaussian effects.

In the chiral sector, Eq.~\eqref{eq:DDelta_from_GammaDelta} shows that only the symmetric part of
$\Gamma^{(\Delta)}$ contributes to the noise covariance~\cite{krommes1993general}:
\begin{equation}
D_{ij}^{(\Delta)}=2(1+\alpha)T_b \Gamma_{ij}^{(\Delta),{\rm S}}.
\end{equation}
This is a fluctuation-dissipation-like relation with temperature $(1+\alpha)T_b$, independent of $\Delta$. Care is required here, since we have already seen that
$\Gamma^{(\Delta)}$ can be negative definite. Therefore, this decomposition does not guarantee that $D_{ij}^{(\Delta)}$ is positive definite, and thus $D_{ij}^{(\Delta)}$ cannot by itself be regarded as a
\emph{bona fide} covariance. By contrast, the total covariance, $\mathsf D^{(\Delta)} + \mathsf D^{(\alpha)}$, must be positive semidefinite.

Interestingly, the translational block is
\begin{equation}
D_{ab}^{(\Delta)}\propto\oint {ds} (\hat n_a\hat t_b+\hat t_a\hat n_b)=2(\varepsilon\cdot \mathsf Q^{\rm dev})_{ab},
\end{equation}
so the universal antisymmetric perimeter piece cancels identically from the diffusion tensor.
Hence an isotropic body may have $\Gamma_{ab}^{(\Delta)}\neq0$ but still $D_{ab}^{(\Delta)}=0$.

\section{Breakdown of the equilibrium fluctuation-dissipation and higher-order equilibrium relations}
\label{sec:FDT_breaking_alpha}

\subsection{Generality}
In this Appendix, we clarify when the fluctuation-dissipation-like relations found in the main text hold. Since detailed balance is broken at the level of the Boltzmann-Lorentz equation, we cannot expect an equilibrium-like relation to persist at all orders in the expansion $\epsilon$. For simplicity, we focus on the case $\Delta=0$. The analysis can be extended to the full dynamics, but this adds little physical insight. Accordingly, we omit the explicit superscript $(\alpha)$ from now on, since the $(\Delta)$ sector is set to zero.

We will show that the relation $D_{ij}=2T_I\Gamma_{ij}$ is only the first (and approximate) element of a broader hierarchy satisfied at equilibrium by the full Kramers-Moyal expansion. At higher orders, it picks up a non-Gaussian contribution and the equilibrium-like relation fails. Similarly, higher-order equilibrium-like ‘‘fluctuation'' relations generally fail in our system already at order $\mathcal O(\varepsilon)$, since the distribution becomes non-Gaussian.

\subsection{Expected hierarchy at equilibrium}

Consider the Kramers-Moyal equation truncated at third order, Eq.~\eqref{eq:KM_order_eps},
\begin{equation}
\frac{\partial P}{\partial t}=-\partial_i\left[A_i P\right] + \frac{1}{2}\partial_i\partial_j\left[B_{ij}P\right]-\frac{1}{6}\partial_i\partial_j\partial_k\left[C_{ijk}P\right],
\end{equation}
where we recall that $\partial_i\equiv\partial_{\Pi_i}$. The truncation should be understood as the neglect of higher-order terms in a van Kampen expansion. We also discarded the reversible advective current, which will not play any role here.

At equilibrium, the stationary momentum distribution remains Gaussian:
\begin{equation}
P_{\rm eq}(\bm \Pi)\sim \exp\left[-\frac{\bm\Pi\cdot \mathsf M^{-1}\cdot\bm\Pi}{2T}\right],
\label{eq:Peq_appendix}
\end{equation}
\emph{independently} of the intruder size~\cite{PLYUKHIN2005198} and the non-Gaussian noise describing its dynamics. Here, $T$ is the equilibrium temperature. The equilibrium condition is not merely the Einstein relation, but detailed balance, and hence, the vanishing of the irreversible probability current in the full Kramers-Moyal operator:
\begin{equation}
A_i P_{\rm eq}-\frac{1}{2}\partial_j\left(B_{ij}P_{\rm eq}\right) +\frac{1}{6}\partial_j\partial_k\left(C_{ijk}P_{\rm eq}\right)=0.
\end{equation}
We obtain the exact third-order equilibrium identity
\begin{equation}
A_i=\frac{1}{2P_{\rm eq}}\partial_j\left(B_{ij}P_{\rm eq}\right)-\frac{1}{6P_{\rm eq}}\partial_j\partial_k\left(C_{ijk}P_{\rm eq}\right).
\label{eq:general_hierarchy_appendix}
\end{equation}

Now, if we assume the same low-velocity structure as in the main text, justified by the expansion in $\mathcal O(\epsilon)$,
\begin{equation}
\begin{gathered}
A_i(\bm U)=R_i-\Gamma_{ij}U_j+N_{ijk}U_jU_k + \mathcal O(\epsilon^{3/2}), \\
B_{ij}(\bm U)=D_{ij}+E_{ijk}U_k+\mathcal O(\epsilon^{3/2}), \\
C_{ijk}(\bm U)=F_{ijk}+\mathcal O(\epsilon^{3/2}),
\end{gathered}
\end{equation}
Expanding Eq.~\eqref{eq:general_hierarchy_appendix} in powers of $\bm U$ yields, order by order, a hierarchy of equations that must be satisfied at equilibrium:
\begin{align}
D_{ij}&=2T\Gamma_{ij}+ \mathcal O(\epsilon^{3/2}),
\label{eq:eq_hierarchy_linear_appendix}\\
R_i&=\frac{1}{2} E_{ijk} M^{-1}_{jk}+\frac{1}{6T}F_{ijk} M^{-1}_{jk} + \mathcal O(\epsilon^{3/2}),
\label{eq:eq_hierarchy_constant_appendix}\\
N_{ijk}&=-\frac{1}{2T}E_{ijk}-\frac{1}{6T^2}F_{ijk}+ \mathcal O(\epsilon^{3/2}).
\label{eq:eq_hierarchy_quadratic_appendix}
\end{align}
The first one is the well-known Einstein relation. As we will see later, it is, in fact, not exact at higher orders in $\epsilon$, due to strong non-Gaussian noise. The second relation is a non-trivial relation between the ‘‘bare'' force and the non-Gaussian and multiplicative fluctuations. We will show how these are related to ratchet effects. The last relation is a generalized fluctuation-dissipation relation for the multiplicative part of the noise and the nonlinear damping ($2TN_{ijk}=E_{ijk}$ is the Einstein relation for the multiplicative part of the noise), corrected by the third Kramers-Moyal moment $F_{ijk}$ equivalent to a non-Gaussian noise. All these relations are non-trivial and are imposed by the requirement of detailed balance of the process and Gaussianity of the stationary distribution (but not of the stochastic process). 

We now want to see if Eqs.~\eqref{eq:eq_hierarchy_linear_appendix}-\eqref{eq:eq_hierarchy_quadratic_appendix} hold for our intruder.

\subsection{Violation of detailed balance and the higher-order fluctuation relations in a nonequilibrium bath}

From the coefficients derived in the main text, we have at $\Delta=0$:
\begin{align}
R_i&=\frac{1+\alpha}{2}n_bmT_b\epsilon \oint ds \mu_n e_i^{(n)},
\label{eq:R_delta0_appendix}\\
\Gamma_{ij}&=2(1+\alpha)\mathcal K_1 n_b\sqrt{mT_b}\,\epsilon^{1/2}\oint ds e_i^{(n)}e_j^{(n)},
\label{eq:Gamma_delta0_appendix}\\
N_{ijk}&=-\frac{1+\alpha}{2}n_bm\epsilon \oint ds e_i^{(n)}e_j^{(n)}e_k^{(n)},
\label{eq:N_delta0_appendix}\\
D_{ij}&=(1+\alpha)^2\mathcal K_3 n_bT_b\sqrt{mT_b}\epsilon^{1/2}\oint ds e_i^{(n)}e_j^{(n)},
\label{eq:D_delta0_appendix}\\
E_{ijk}&=\frac{3}{2}(1+\alpha)^2 n_bmT_b\epsilon\oint ds e_i^{(n)}e_j^{(n)}e_k^{(n)},
\label{eq:E_delta0_appendix}\\
F_{ijk}&=-(1+\alpha)^3\mathcal K_4n_bmT_b^2\epsilon\oint dse_i^{(n)}e_j^{(n)}e_k^{(n)}.
\label{eq:C_delta0_appendix}
\end{align}
We recall that
\begin{equation}
T_I=\frac{\mathcal K_3}{2\mathcal K_1}\frac{1+\alpha}{2}T_b
\label{eq:TI_appendix_here}
\end{equation}
is the kinetic temperature of the intruder in the linear regime when $\Delta=0$.

For notational convenience, we define
\begin{equation}
\begin{gathered}
\mathcal J_{ijk}\equiv n_bm\epsilon\oint dse_i^{(n)}e_j^{(n)}e_k^{(n)},\\
\mathcal J_i\equiv \mathcal J_{ijk}\mathsf M^{-1}_{jk}=n_bm\epsilon\oint ds\mu_n e_i^{(n)}.
\end{gathered}
\end{equation}
Then Eqs.~\eqref{eq:R_delta0_appendix}-\eqref{eq:C_delta0_appendix} become
\begin{align}
R_i&=\frac{1+\alpha}{2}T_b\mathcal J_i,
\label{eq:R_compact_appendix}\\
N_{ijk}&=-\frac{1+\alpha}{2}\mathcal J_{ijk},
\label{eq:N_compact_appendix}\\
E_{ijk}&=\frac{3}{2}(1+\alpha)^2T_b\mathcal J_{ijk},
\label{eq:E_compact_appendix}\\
F_{ijk}&=-(1+\alpha)^3\mathcal K_4 T_b^2\mathcal J_{ijk}.
\label{eq:C_compact_appendix}
\end{align}

A natural question is whether the full equilibrium hierarchy also survives after replacing the equilibrium temperature $T$ in Eqs.~\eqref{eq:eq_hierarchy_constant_appendix} and \eqref{eq:eq_hierarchy_quadratic_appendix} by the effective temperature $T_I$. The answer is negative in general. This is due to the non-Gaussianity of the intruder's probability distribution. We first analyze the non-trivial relations previously found for an equilibrium system.

\subsubsection{Higher-order fluctuation relation n°1}

Using Eqs.~\eqref{eq:R_compact_appendix}, \eqref{eq:E_compact_appendix}, and \eqref{eq:C_compact_appendix}, we find
\begin{equation}
\begin{split}
\delta R_i&\equiv R_i-\frac12E_{ijk}\mathsf M^{-1}_{jk}-\frac{1}{6T_I}F_{ijk}\mathsf M^{-1}_{jk}\\
&=\frac{1+\alpha}{12}T_b\left[6-9(1+\alpha)+8(1+\alpha)\frac{\mathcal K_1\mathcal K_4}{\mathcal K_3}\right]\mathcal J_i.
\end{split}
\label{eq:deltaR_appendix}
\end{equation}
This quantity vanishes only under special conditions. In particular, for a Gaussian bath, we have $\mathcal K_3=2\mathcal K_1$ and  $\mathcal K=3/2$ such that Eq.~\eqref{eq:deltaR_appendix} reduces to
\begin{equation}
\delta R_i=\frac{1-\alpha}{2}T_In_bm\oint ds\mu_n e_i^{(n)}.
\label{eq:deltaR_gaussian_appendix}
\end{equation}
Hence, the higher-order fluctuation-dissipation identity required at equilibrium is recovered only for $\alpha=1$. Interestingly, $\delta R_i$ is exactly the ratchet force/torque that we found earlier (for a Gaussian bath):
\begin{equation}
     F_i^{(\alpha)}=\frac{1+\alpha}{2}n_bm(T_b-T_I)\oint {ds}  \mu_n e_i^{(n)}
\end{equation}
This follows from Eqs.~\eqref{eq:eq_hierarchy_linear_appendix} and Eq.~\eqref{eq:eq_hierarchy_quadratic_appendix}. 

Alternatively, we can ask which temperature must be inserted into Eq.~\eqref{eq:eq_hierarchy_linear_appendix} for the higher-order fluctuation relation to be satisfied. At this order, we obtain
\begin{equation}
    T_{(1)}=\frac{2(1+\alpha)^2\mathcal K_4}{3(1+3\alpha)}T_b,
\end{equation}
with $T_{(1)}=T_b$ at equilibrium and, generically, $T_{(1)}\neq T_I$. The relation can therefore still be enforced at this order, provided we use a temperature different from that entering the Einstein relation. We will see later that this ceases to be possible at higher orders in $\epsilon$, where no scalar temperature can restore the equilibrium-like form.

\subsubsection{Higher-order fluctuation relation n°2}
Similarly,
\begin{equation}
\begin{split}
\delta N_{ijk} &\equiv N_{ijk}+\frac{1}{2T_I}E_{ijk}+\frac{1}{6T_I^2}F_{ijk}\\
&= (1+\alpha) \left[-\frac12+\frac{3\mathcal K_1}{\mathcal K_3} -\frac{8\mathcal K_1^2\mathcal K_4}{3\mathcal K_3^2}\right]\mathcal J_{ijk}.
\end{split}
\label{eq:deltaN_appendix}
\end{equation}
For a Gaussian bath, $\mathcal K_3=2\mathcal K_1$ and $\mathcal K_4=3/2$, so
\begin{equation}
\delta N_{ijk}=0,
\end{equation}
for any $\alpha$. For a generic non-Gaussian bath, Eq.~\eqref{eq:deltaN_appendix} does not vanish. However, for a Gaussian bath, the quadratic-drift identity remains valid \emph{at this order} even in the presence of inelastic collisions. It would be interesting to investigate whether this term can be interpreted as a higher-order ‘‘ratchet effect'', since $\delta R_i$ may be related to an effective ratchet force.

As for the previous fluctuation relation, we may again ask which temperature makes the relation hold. In this case, we find
\begin{equation}
    T_{(2)}^{\pm}=\frac{1+\alpha}{12}\Bigl(9\pm\sqrt{81-48\mathcal K_4}\Bigr)T_b.
\end{equation}
There are now two positive candidates. The equilibrium branch is the minus sign, for which $T_{(2)}^-(\alpha=1,\mathcal K_4=3/2)=T_b$. Thus, the fluctuation relation can again be enforced at this order by a suitable choice of temperature. However, this condition is more restrictive: for $\mathcal K_4>81/48$, no real solution exists, and no scalar temperature can restore the relation.

\subsubsection{Einstein relation}
Using Eqs.~\eqref{eq:TI_appendix_here}, \eqref{eq:Gamma_delta0_appendix} and \eqref{eq:D_delta0_appendix}, we recover:
\begin{equation}
D_{ij}=2T_I\Gamma_{ij}+\mathcal O(\epsilon^{3/2}),
\end{equation}
which matches the equilibrium result Eq.~\eqref{eq:eq_hierarchy_linear_appendix}. Therefore, at order $\mathcal O(\epsilon)$, we found 3 fluctuation relations that can all be satisfied if we choose a different temperature for each. 

This correspondence only holds at order $\mathcal O(\epsilon)$. We show it here using Einstein's relation. At order $\mathcal O(\epsilon^{3/2})$, the equilibrium fluctuation-dissipation relation departs from the simple Einstein's relation since at the next order, the equilibrium current involves additional terms. More precisely, at order $\mathcal O(\epsilon^{3/2})$:
\begin{equation}
    \begin{gathered}
        B_{ij}(\bm U)=D_{ij}+E_{ijk}U_k+H_{ijk\ell}U_kU_\ell+\mathcal O(\epsilon^2),\\
        C_{ijk}(\bm U)=F_{ijk}+L_{ijk\ell}U_\ell+\mathcal O(\epsilon^2),\\
        G_{ijk\ell}(\bm U)=K_{ijk\ell}+\mathcal O(\epsilon^2),
    \end{gathered}
\end{equation}
where $G_{ijkl}$ is the fourth Kramers-Moyal coefficient. Performing once again the same analysis as before, but at this order, leads to a generalization of Einstein's relation for non-Gaussian noise:
\begin{equation}
\Gamma_{ij}=\frac{D_{ij}}{2T}-H_{i\ell kj}\mathsf M^{-1}_{\ell k}-\frac{L_{i\ell kj}\mathsf M^{-1}_{\ell k}}{2T}-\frac{K_{i\ell kj}\mathsf M^{-1}_{\ell k}}{8T^2}+\mathcal O(\epsilon^2),
\label{eq:advanced_FD}
\end{equation}
We insist that Eq.~\eqref{eq:advanced_FD} is an equilibrium relation solely based on detailed balance at the level of the van-Kampen expansion and the Gaussianity of the probability distribution. 

If we replace $T$ by $T_I$ and compute the difference between the left-hand side and right-hand side of Eq.~\eqref{eq:advanced_FD}, which we call $\delta \Gamma_{ij}$, we obtain:
\begin{equation}
    \delta \Gamma_{ij}=(1+\alpha)\mathcal K_1\left[3+5\alpha-2(1+\alpha)\frac{\mathcal K_1\mathcal K_5}{\mathcal K_3^2}\right]\mathcal V_{ij},
\end{equation}
with $\mathcal V_{ij}\propto \oint ds \mu_n e_i^{(n)}e_j^{(n)}.$
which vanishes once $\alpha= 1$ and the $\mathcal K_n$ are taken from a Gaussian.

In contrast to the previous $\mathcal O(\epsilon)$ examples, Eq.~\eqref{eq:advanced_FD} cannot be restored for our nonequilibrium dynamics by any choice of $T$. A scalar temperature, therefore, ceases to be an adequate description at order $\mathcal O(\epsilon^{3/2})$.

We note that this was already apparent at order $\mathcal O(\epsilon)$ if we kept both the $\Delta$ and $\alpha$ sector and tried to find a full fluctuation-dissipation relation:
\begin{equation}
2T\left(\Gamma_{ij}^{(\alpha)}+\Gamma_{ij}^{(\Delta)}\right)\neq D_{ij}^{(\alpha)}+D_{ij}^{(\Delta)}, \quad\forall T,
\end{equation}
instead of a sector-wise one.

\subsection{Discussion}

To summarize, the fluctuation-dissipation theorem derived in the main text for each sector ceases to hold beyond leading order in the van Kampen expansion. At lowest order, all fluctuation relations can still be recovered by introducing an effective temperature for each sector~\cite{cugliandolo2011effective}. At higher orders, however, this description breaks down as the probability distribution departs progressively from a Gaussian form when increasingly fine details of the system are retained~\cite{sarracino2010granular}.

Perturbatively, this non-Gaussianity remains straightforward to work with whenever the nonequilibrium stationary distribution $P_{\rm ss}\neq P_{\rm eq}$ can still be expanded in powers of $\bm U$, since we may still start from Eq.~\eqref{eq:general_hierarchy_appendix}. The resulting fluctuation relations, however, involve derivatives of $P_{\rm ss}$ which cannot be parametrized by a single scalar quantity such as the temperature. A better approach would be to work non-perturbatively with the probability distribution~\cite{agarwal1972fluctuation, PhysRevLett.103.090601, marconi2008fluctuation}.

Beyond Gaussianity, we assumed detailed balance on the Kramers-Moyal expansion. This assumption cannot remain valid at arbitrary order, since the Master equation itself does not satisfy detailed balance. This is particularly evident in the $\Delta$ sector, where the odd response generates odd currents. The correct starting point is therefore not the condition $J_i=0$, but the weaker global balance condition $\partial_i J_i = 0$. This framework can still be developed perturbatively, although additional tensors must then enter the fluctuation relations in order to enforce global balance.

\section{Explicit coefficients for the Langevin equation of non-trivial shapes}\label{app:finite}
In this appendix, we derive the coefficients entering the Langevin equation of intruders of various shapes in a dilute bath.

\subsection{Body frame and laboratory frame}

We denote the body frame quantities with a tilde. For example, an intruder with a triangle shape will have on average $\langle F_{y}\rangle\equiv\langle F_{y}^{\rm laboratory}\rangle=0$ because of isotropy, but if we always rotate the intruder to a fixed angle, then $\tilde F_y$, the force in a given direction in the body frame, might not be 0 on average, for example because the intruder is propelled toward its apex.

By isotropy of space (or direct computations from the previous sections), we can obtain all relevant quantities in the laboratory frame by applying a rotation in the $\mathbb R^2$ part of $\mathbb R^2\oplus \mathbb R$. We define the generalized rotation matrix:
\begin{equation}
    \bar{\mathsf R}(\varphi)=\mathrm{diag}(\mathsf R(\varphi),1).
    \label{eq:generalized_rotation_matrix}
\end{equation}
The transformation rule is:
\begin{equation}
\begin{gathered}
\Upgamma(\varphi)=\bar{\mathsf R}(\varphi)\cdot\tilde{\Upgamma}\cdot\bar{\mathsf R}(\varphi)^T, \qquad \mathsf D(\varphi)=\bar{\mathsf R}(\varphi)\cdot\tilde{\mathsf D}\cdot\bar{\mathsf R}(\varphi)^T,\\ \bm F(\varphi)=\bar{\mathsf R}(\varphi)\cdot \tilde{\bm F},
\end{gathered}
\end{equation}
where a tilde denotes a quantity evaluated in the body frame. We note that the quantities in the body frame are independent of the actual angle of the intruder $\varphi$. To be more precise, we could parametrize the intruder by any angle $\varphi_0$, in which case, we would have: $\bm F(\varphi)=\bar{\mathsf R}(\varphi - \varphi_0)\cdot \tilde{\bm F}(\varphi_0)$ for example.

Finally, since the intruder rotates during a free flight, the velocity precesses in the body frame
\begin{equation}
\begin{pmatrix}
\dot{\tilde V}_x\\ \dot{\tilde V}_y\\ \dot \Omega
\end{pmatrix}_{\rm free}=
\begin{pmatrix}
\Omega\tilde V_y\\
-\Omega\tilde V_x\\
0
\end{pmatrix},
\end{equation}
as obtained from differentiating
\begin{equation}
    \tilde{\bm V}(t)=\mathsf R\left[-\varphi(t)\right]\cdot \bm V
\end{equation}
Therefore, the body frame dynamics is not a plain Ornstein-Uhlenbeck process but instead:
\begin{equation}
\mathsf M\cdot\dot{\tilde{\bm U}}=\tilde{\bm F}-\tilde{\Upgamma} \cdot\tilde{\bm U}+
\begin{pmatrix}
M\Omega\tilde V_y\\
-M\Omega\tilde V_x\\
0
\end{pmatrix}+\sqrt{\tilde{\mathsf D}}\cdot\bm \zeta(t).
\label{eq:additional_force}
\end{equation}

\subsection{Finite bath-particle size and Minkowski sum}

The Boltzmann-Lorentz description \eqref{eq:Lorentz} was written for point-like bath particles. If the bath particles are hard disks of diameter $\sigma$, their centers cannot approach the intruder closer than a distance $\sigma/2$. The relevant excluded region for the bath-particle centers is therefore not the intruder shape $\mathcal B$, but its Minkowski sum with the hard-particles:
\begin{equation}
\mathcal B_{\sigma}\equiv \mathcal B\oplus D_{\sigma}=\left\{\bm r+\bm u:\bm r\in\mathcal B, |\bm u|\le \sigma/2\right\},
\end{equation}
where $D_{\sigma}$ is the disk of diameter $\sigma$.

Hence, when the bath particles have a finite size, the collision \emph{rate} is controlled by the boundary of $\mathcal B_{\sigma}$, because this is the surface swept by the centers of the colliding bath particles. The transition kernel is therefore naturally written as
\begin{equation}
\begin{split}
W_{\sigma}(\bm U'|\bm U,\bm Y)&=n_b\oint_{\partial\mathcal B_{\sigma}} {ds}_\sigma\int d\bm v f(\bm v)\Theta(-g_n)(-g_n)\times\\\delta&\left(\bm V'-\bm V-\frac{\bm J}{M}\right)\delta\left(\Omega'-\Omega-\frac{[\bm r\times \bm J]_z}{I}\right).
\end{split}
\label{eq:W_sigma}
\end{equation}
However, although the integration domain is $\partial\mathcal B_{\sigma}$, the lever arm entering the collision rule, in quantities such as $\bm J$ and $g_n$, is still that associated with the actual point of contact on the intruder:
\begin{equation}
    \bm r_{\rm lever~ arm}=\bar{\bm r}-\frac{\sigma}{2}\hat{\bm n},\quad\text{with}\quad \bar{\bm r}\in\partial\mathcal B_\sigma.
\end{equation}
Thus, with finite-size bath particles, the coefficients are controlled neither by the bare shape alone nor by the Minkowski sum alone, but by a combination of both. The excluded boundary defines the integration domain, whereas the lever arm's position is determined only by the intruder. In the Letter~\cite{letter}, we focus on intruders large enough that the bath particles can be treated as effectively point-like.

\subsection{Regular polygon with \texorpdfstring{$n$}{n} vertices}\label{sec:circular}

A simple intruder is a polygon with $n$ vertices $\bm v_j$ placed at a distance $R$ from the origin:
\begin{equation}
\bm v_j=R\left(\cos({2\pi j}/{n}),\sin({2\pi j}/{n})\right),\qquad j=0,\dots,n-1.
\end{equation}
The case $n=3$ yields an equilateral triangle, $n=4$ a square, and in the limit $n\to\infty$ a circle is recovered.

Its side length $\ell$, perimeter $\mathcal P$, and area $\mathcal A$ are
\begin{equation}
\begin{gathered}
\ell=2R\sin\left(\frac{\pi}{n}\right),\qquad \mathcal P=2nR\sin\left(\frac{\pi}{n}\right),\\\mathcal A=\frac{n}{2}R^2\sin\left(\frac{2\pi}{n}\right).
\end{gathered}
\end{equation}

The regular polygon has a non-trivial discrete rotational symmetry ($C_{n\geq 2}$) and is achiral. Therefore,
\begin{equation}
P_a=W_a=S_a=T_a=0,\qquad H_t=H_{\varphi}=0,
\end{equation}
while the translational tensor is isotropic:
\begin{equation}
Q_{ab}=\oint ds\hat n_a \hat n_b=\frac{\mathcal P}{2}\delta_{ab},\qquad\oint ds\hat t_a\hat n_b=\frac{\mathcal P}{2}\varepsilon_{ab}.
\end{equation}
We also find:
\begin{equation}
H_n=\oint ds\kappa_n^2=n\frac{\ell^3}{12}=\frac{2}{3}nR^3\sin^3\left(\frac{\pi}{n}\right).
\end{equation}

The body frame force is thus purely rotational:
\begin{equation}
\begin{gathered}
\tilde F_x=\tilde F_y=0,\\\tilde F_{\varphi}=\tilde F_\varphi^{(\Delta)}=2\mathcal K_1n_b\Delta\sqrt{mT_b}nR^2\sin\left(\frac{2\pi}{n}\right).
\end{gathered}
\end{equation}
The dissipative contribution proportional to $T_b-T_I$ vanishes identically because the intruder is neither chiral nor polar.

It is useful to define
\begin{equation}
\Gamma_0^{(\alpha)}\equiv(1+\alpha)\mathcal K_1 n_b\sqrt{mT_b}.
\end{equation}
With this notation, the drift matrix reads
\begin{equation}
\Upgamma=\Upgamma^{(\alpha)}+\Upgamma^{(\Delta)},
\end{equation}
with
\begin{equation}
{\Upgamma}^{(\alpha)}=
\Gamma_0^{(\alpha)}
\begin{pmatrix}
\mathcal P & 0 & 0\\
0 & \mathcal P & 0\\
0 & 0 & 2H_n
\end{pmatrix},
\end{equation}
and
\begin{equation}
{\Upgamma}^{(\Delta)}=\dfrac{
n_b m\Delta}{2}
\begin{pmatrix}
0 & \mathcal P & 0\\
-\mathcal P & 0 & 0\\
0 & 0 & 0
\end{pmatrix},
\end{equation}
in the body frame. We note that $H_{n}(n\to\infty)=0$, that is, for a circle, there is no orientational damping from the chiral sector. This is expected since for a circle, the only effect of $\Delta$ is to increase $\Omega$ at collision.

The diffusion matrix also splits as
\begin{equation}
\mathsf D=\mathsf D^{(\alpha)}+\mathsf D^{(\Delta)}.
\end{equation}
The $\alpha$-sector obeys the fluctuation-dissipation relation
\begin{equation}
{\mathsf D}^{(\alpha)}=2T_I {\Upgamma}^{(\alpha)}=2T_I \Gamma_0^{(\alpha)}
\begin{pmatrix}
\mathcal P & 0 & 0\\
0 & \mathcal P & 0\\
0 & 0 &  2H_n
\end{pmatrix},
\end{equation}
while:
\begin{equation}
{\mathsf D}^{(\Delta)}=0
\end{equation}
Indeed, for a regular polygon, the translational deviatoric part vanishes, and since $H_t=0$, there is no chiral correction to the rotational diffusion either.

Since the translation and orientational degrees of freedom decouple, the long-time translational diffusivity tensor is easily written from a generalized Green-Kubo as:
\begin{equation}
\mathcal D_{ab}=\lim_{t'\to\infty}\int_0^\infty dt \langle V_a(t') V_b(t'+t)\rangle=\mathcal D_{\rm even} \delta_{ab}+\mathcal D_{\rm odd} \varepsilon_{ab},
\end{equation}
with
\begin{equation}
\mathcal D_{\rm even}=\frac{D_V}{2(\gamma^2+\omega^2)},\qquad\mathcal D_{\rm odd}=-\frac{D_V \omega}{2\gamma(\gamma^2+\omega^2)},
\end{equation}
the regular and odd diffusivity, respectively.  $D_V$ is the variance of the translational noise:
\begin{equation}
\mathsf D_{\rm tr}=D_V
\begin{pmatrix}
1&0\\
0&1
\end{pmatrix}, \qquad D_V=2T_I\Gamma_0^{(\alpha)}\mathcal P
\end{equation}
and $\gamma$ and $\omega$ the diagonal and off-diagonal contributions to the translational damping matrix:
\begin{equation}
\Upgamma_{\rm tr}=
\begin{pmatrix}
\gamma & \omega\\
-\omega & \gamma
\end{pmatrix},\qquad \gamma = \Gamma_0^{(\alpha)}\mathcal P,\qquad \omega =n_bm \mathcal P\Delta/2
\end{equation}

We note that the odd diffusivity diverges as $1/n_b$. If instead, as in the Letter~\cite{letter}, $\omega$ depends on $n_b^2$ rather than $n_b$ due to collective effects, then $\mathcal D_{\rm odd}\sim \mathcal O(1)$ as $n_b\to0$. 

\begin{widetext}
\subsection{Chiral and achiral wheel}

\begin{figure*}
    \centering
    \includegraphics[width=0.99\linewidth]{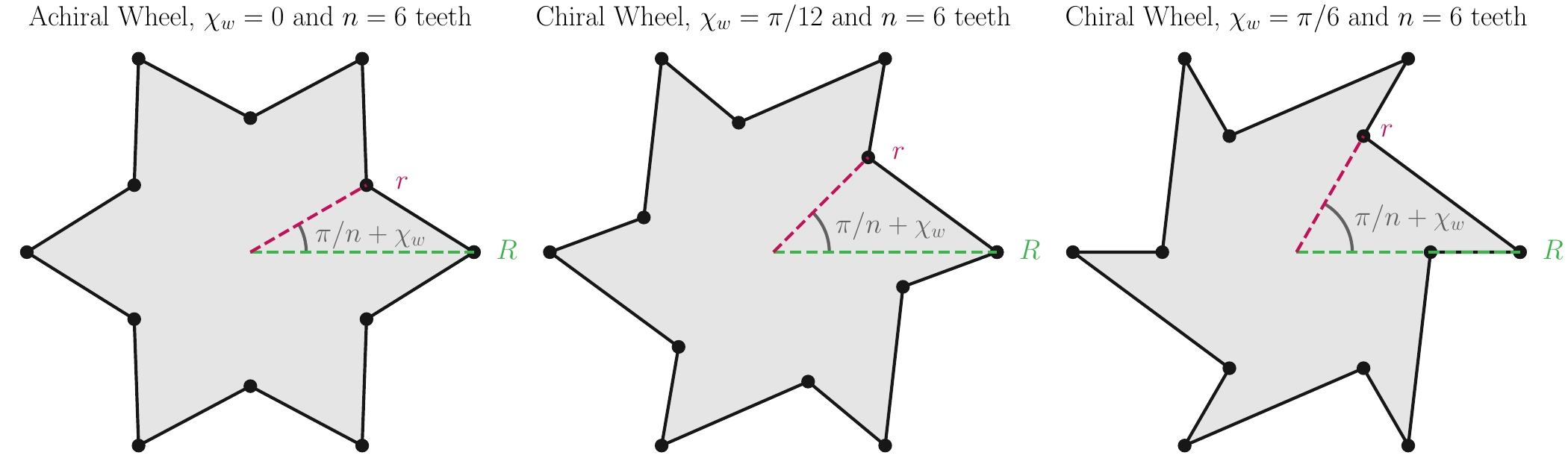}
    \caption{Examples of wheels as given by Eqs.~\eqref{eq:wheel_def} and \eqref{eq:chi_w_def}, for various $\chi_w$ at fixed number of teeth $n$.}
    \label{fig:wheel_fig}
\end{figure*}

We now consider a possibly chiral intruder, modeled as a wheel with $n$ teeth, \emph{i.e.}, a $2n$-vertex polygon, with alternating outer and inner radii $R$ and $r$. The vertices are
\begin{equation}
\bm v_{2j}=R(\cos 2\pi j/n,\sin 2\pi j/n),\qquad\bm v_{2j+1}=r(\cos(2\pi j/n+\delta),\sin(2\pi j/n+\delta)),\qquad j=0,\dots,n-1.
\label{eq:wheel_def}
\end{equation}
The wheel is achiral when the inner vertices lie halfway between successive outer ones, that is, when $\delta=\pi/n$. It is therefore convenient to introduce the shape-chirality parameter
\begin{equation}
\chi_{\rm w}\equiv \delta-\frac{\pi}{n},\qquad\theta_\pm\equiv \frac{\pi}{n}\pm \chi_{\rm w},
\label{eq:chi_w_def}
\end{equation}
so that mirror reflection corresponds to $\chi_{\rm w}\to-\chi_{\rm w}$. Various wheels are shown in Fig.~\ref{fig:wheel_fig}. 

The two edge lengths are
\begin{equation}
\ell_\pm=\sqrt{R^2+r^2-2Rr\cos(\theta_\pm)},
\end{equation}
hence the perimeter is
\begin{equation}
\mathcal P(\chi_{\rm w})=n(\ell_++\ell_-).
\end{equation}
The enclosed area is
\begin{equation}
\mathcal A(\chi_{\rm w})=nRr\sin \left(\frac{\pi}{n}\right)\cos\chi_{\rm w}.
\end{equation}

We recall:
\begin{equation}
H_n\equiv \oint {ds} \kappa_n^2,\qquad H_{\varphi}\equiv \oint {ds} \kappa_n^3, \qquad H_t\equiv \oint {ds} \kappa_t\kappa_n.
\end{equation}
On a straight edge of length $\ell$, if $s'$ is the arclength from one endpoint, then $\kappa_n(s')=\kappa_0-s'$, so
\begin{equation}
\mathcal H_{q}=\int_{\rm edge}\kappa_n^q {ds}=\frac{\kappa_0^{q+1}-(\kappa_0-\ell)^{q+1}}{q+1}.
\end{equation}

For the wheel, define
\begin{equation}
u_+\equiv R(R-r\cos(\theta_+)),\qquad u_-\equiv r(r-R\cos(\theta_-)),
\end{equation}
so that the initial values of $\kappa_n$ on the two edge types are
\begin{equation}
\kappa_{0+}=\frac{u_+}{\ell_+},\qquad \kappa_{0-}=\frac{u_-}{\ell_-}.
\end{equation}
Then
\begin{equation}
\mathcal H_q(\chi_{\rm w})=n\left[\frac{\kappa_{0+}^{ q+1}-(\kappa_{0+}-\ell_+)^{q+1}}{q+1}+\frac{\kappa_{0-}^{ q+1}-(\kappa_{0-}-\ell_-)^{q+1}}{q+1}\right].
\end{equation}
In particular,
\begin{equation}
\mathcal H_2=H_n(\chi_{\rm w})=n\left[\left(\frac{u_+^2}{\ell_+}-u_+\ell_+ +\frac{\ell_+^3}{3}\right)+\left(\frac{u_-^2}{\ell_-}-u_-\ell_- +\frac{\ell_-^3}{3}\right)\right],
\end{equation}
and
\begin{equation}
\mathcal H_3=H_{\varphi}(\chi_{\rm w})=n\left[\left(\frac{u_+^3}{\ell_+^2}-\frac{3}{2}u_+^2+u_+\ell_+^2-\frac{\ell_+^4}{4}\right)+\left(\frac{u_-^3}{\ell_-^2}-\frac{3}{2}u_-^2+u_-\ell_-^2-\frac{\ell_-^4}{4}\right)\right].
\end{equation}

For the mixed moment, note that on an edge joining radii $a$ and $b$ with included angle $\theta$, we have $\kappa_t=ab\sin(\theta)/\ell$, which is constant along that edge. This gives
\begin{equation}
H_t(\chi_{\rm w})=n\left[\frac{Rr\sin(\theta_+)}{\ell_+}\left(u_+-\frac{\ell_+^2}{2}\right)+\frac{Rr\sin(\theta_-)}{\ell_-}\left(u_--\frac{\ell_-^2}{2}\right)\right].
\end{equation}

By symmetry,
\begin{equation}
\mathcal P(-\chi_{\rm w})=\mathcal P(\chi_{\rm w}),\qquad\mathcal A(-\chi_{\rm w})=\mathcal A(\chi_{\rm w}),\qquad H_n(-\chi_{\rm w})=H_n(\chi_{\rm w}),
\end{equation}
whereas
\begin{equation}
H_{\varphi}(-\chi_{\rm w})=-H_{\varphi}(\chi_{\rm w}),\qquad H_t(-\chi_{\rm w})=-H_t(\chi_{\rm w}).
\end{equation}
In particular,
\begin{equation}
H_{\varphi}(0)=H_t(0)=0.
\end{equation}

For $n\ge 3$, the wheel has $C_n$ symmetry, so there is no translational force in the body frame:
\begin{equation}
\tilde F_x=\tilde F_y=0.
\end{equation}
The force is therefore purely rotational,
\begin{equation}
\tilde{\bm F}=
\begin{pmatrix}
0\\
0\\
\tilde F_{\varphi}
\end{pmatrix},
\end{equation}
with
\begin{equation}
\tilde F_{\varphi}=4\mathcal K_1 n_b\Delta\sqrt{mT_b}\mathcal A(\chi_{\rm w})+\frac{1+\alpha}{2}\left(T_b-T_I\right)\frac{n_b m}{I} H_{\varphi}(\chi_{\rm w}).
\end{equation}
The first term is present even for an achiral wheel as long as the collision is chiral $\Delta\neq 0$ and is controlled only by the area of the intruder. The second one is the geometric chirality contribution, which is odd in $\chi_{\rm w}$.

As previously done, we define
\begin{equation}
\Gamma_0^{(\alpha)}\equiv (1+\alpha)\mathcal K_1 n_b\sqrt{mT_b}.
\end{equation}
and find:
\begin{equation}
\Upgamma=\Upgamma^{(\alpha)}+\Upgamma^{(\Delta)},\quad {\Upgamma}^{(\alpha)}=
\Gamma_0^{(\alpha)}
\begin{pmatrix}
\mathcal P(\chi_{\rm w}) & 0 & 0\\
0 & \mathcal P(\chi_{\rm w}) & 0\\
0 & 0 & 2H_n(\chi_{\rm w})
\end{pmatrix},\quad {\Upgamma}^{(\Delta)}=\dfrac{
n_b m\Delta}{2}
\begin{pmatrix}
0 & \mathcal P(\chi_{\rm w}) & 0\\
-\mathcal P(\chi_{\rm w}) & 0 & 0\\
0 & 0 & -2H_t(\chi_{\rm w})\end{pmatrix}
\end{equation}
in the body frame.

The diffusion matrix likewise splits with the effective fluctuation-dissipation relaxation for the $\alpha$ sector:
\begin{equation}
\mathsf D=\mathsf D^{(\alpha)}+\mathsf D^{(\Delta)},\qquad
{\mathsf D}^{(\alpha)}=2T_I {\Upgamma}^{(\alpha)}=2T_I \Gamma_0^{(\alpha)}
\begin{pmatrix}
\mathcal P(\chi_{\rm w}) & 0 & 0\\
0 & \mathcal P(\chi_{\rm w}) & 0\\
0 & 0 &  2H_n(\chi_{\rm w})
\end{pmatrix}.
\end{equation}
The leading chiral correction only affects the rotational noise intensity,
\begin{equation}
{\mathsf D}^{(\Delta)}=2(1+\alpha)n_bmT_b\Delta
\begin{pmatrix}
0 & 0 & 0\\
0 & 0 & 0\\
0 & 0 & -H_t(\chi_{\rm w})
\end{pmatrix}.
\end{equation}
In particular, the translational diffusion tensor remains isotropic.

We obtain the mean angular velocity:
\begin{equation}
\langle \Omega\rangle=\frac{F_{\varphi}}{\Gamma_{\varphi\varphi}}=\frac{\displaystyle 4 \mathcal K_1 n_b\Delta\sqrt{mT_b} \mathcal A(\chi_{\rm w})+\frac{1+\alpha}{2}\left(T_b-T_I\right)\frac{n_b m}{I} H_{\varphi}(\chi_{\rm w})}{2\Gamma_0^{(\alpha)} H_n(\chi_{\rm w})-n_bm\Delta H_t(\chi_{\rm w})},
\end{equation}
which has a contribution from the chiral driving $\Delta$ and a contribution from the chiral shape, which vanishes for an achiral wheel: $H_\varphi(\chi_{\rm w} = 0) = 0$.

\subsection{Polar triangle}

Finally, we consider an isosceles triangle whose symmetry axis is the body frame $y$ axis (pointing from the base toward the apex, or the opposite). With the center of mass at the origin,
a convenient choice of vertices is
\begin{equation}
A=\left(-\frac a2,-\frac h3\right),\qquad
B=\left(\frac a2,-\frac h3\right),\qquad
C=\left(0,\frac{2h}{3}\right),
\end{equation}
and the perimeter is
\begin{equation}
\mathcal P=a+2\ell,
\qquad
\ell\equiv \sqrt{h^2+\frac{a^2}{4}}.
\end{equation}

In the body frame, the force $\tilde{\bm F}$ is
\begin{equation}
\tilde{\bm F}=
\begin{pmatrix}
0\\
\displaystyle
\frac{1+\alpha}{2}\left(T_b-T_I\right)
n_b m\frac{1}{I}
\frac{a h^2 (4h^2-3a^2)}{9(a^2+4h^2)}
\\
\displaystyle
2\mathcal K_1n_b\Delta\sqrt{mT_b}ah
\end{pmatrix}.
\end{equation}
We recall that in the laboratory frame, the torque $F_{\varphi}$ is unchanged, while the translational force $\bm F_{\rm{tr}}=(F_x,F_y)$ is obtained from a rotation since it is a vector: $\bm F_{\text{tr}}(\varphi)=\mathsf R(\varphi)\cdot\tilde{\bm F}_{\text{tr}}$. We note that when the shape is polar ($ h/a\neq \sqrt{3}/2$), there is a nonzero force along the apex, but recall that while $\tilde{F}_y\neq 0$, the average force must be 0 by isotropy: $\langle F_a\rangle=0$ with $a=\{x, y\}$.

The drift matrix decomposes as $\Upgamma=\Upgamma^{(\alpha)}+\Upgamma^{(\Delta)}$, with
\begin{equation}
\tilde{\Upgamma}^{(\alpha)}=
\Gamma_0
\begin{pmatrix}
\dfrac{4h^2}{\ell} & 0 &
\dfrac{h(3a^2-4h^2)}{6\ell}\\
0 & \dfrac{a(a+2\ell)}{\ell} & 0\\
\dfrac{h(3a^2-4h^2)}{6\ell} & 0 &
\dfrac{3a^4+6a^3\ell+16h^4}{36\ell}
\end{pmatrix}
\end{equation}
and
\begin{equation}
\tilde{\Upgamma}^{(\Delta)}=
n_b m\Delta
\begin{pmatrix}
0 & \dfrac{a(a+2\ell)}{2\ell} & 0\\
-\dfrac{2h^2}{\ell} & 0 & -\dfrac{h(3a^2-4h^2)}{12\ell}\\
0 & \dfrac{ah(\ell-a)}{3\ell} & 0
\end{pmatrix},
\end{equation}
\emph{in the body frame}, where
\begin{equation}
\Gamma_0\equiv (1+\alpha)\mathcal K_1n_b\sqrt{mT_b}.
\end{equation}

Therefore, in the body frame and neglecting the pseudo-forces of order $\mathcal O(U^2)$ in Eq.~\eqref{eq:additional_force}, we find
\begin{equation}
\langle \tilde{V}_y\rangle=\frac{\sqrt{mT_b}}{18\mathcal K_1 I}\dfrac{T_b-T_I}{T_b}\frac{\ell h^2(4h^2-3a^2)}{(a+2\ell)(a^2+4h^2)},
\end{equation}
\begin{equation}
\langle \tilde{V}_x\rangle=-\frac{4 a\ell(3a^2-4h^2)}{(1+\alpha)\big(a^4+8a^3\ell+8a^2h^2+16h^4\big)}\Delta,
\end{equation}
and
\begin{equation}
\langle\Omega\rangle=\frac{96 a\ell h}{(1+\alpha)\big(a^4+8a^3\ell+8a^2h^2+16h^4\big)}\Delta.
\end{equation}
We see that for a polar shape, $\tilde V_x\neq0$ although $\tilde F_x=0$ because $\Omega\neq 0$ and $\Gamma_{x\varphi}\neq 0$.

\section{Oseen problem}\label{app:oseen}

\subsection{Passive Oseen solution}

We solve Eq.~\eqref{eq:oseen_translational} perturbatively in the chiral Reynolds number. We first introduce
\begin{equation}
    R=\frac{r}{R_0},\quad X=\frac{x}{R_0},\quad  \bm u^{(T)}=U_\infty\bm Q,\quad \bar{\mathtt p}^{(T)}=\rho_bU_\infty^2\mathit\Pi, \quad\mathrm{Re}_\infty=\frac{\rho_b U_\infty R_0}{\eta}, \quad \delta = \mathrm{Re}_\infty/2,\quad \mathrm{Re}_{\chi}\equiv \frac{\rho_b c}{\eta}.
\end{equation}
Eq.~\eqref{eq:oseen_translational} becomes
\begin{equation}
    \begin{split}
    \partial_X\bm Q=&-\bm\nabla\mathit\Pi+\frac{1}{2\delta}\left[\bm\nabla^2\bm Q -\frac{\mathrm{Re}_\chi}{R}(\bm\nabla\times\bm Q)_z\hat{\bm e}_r\right],\qquad \bm\nabla\cdot\bm Q=0.
    \end{split}
    \label{eq:stokes_adim}
\end{equation}
We write $\bm Q=\bm Q^{(0)}+\bm Q^{(1)}+\mathcal O(\mathrm{Re}_\chi^2)$, where $\bm Q^{(1)}=\mathcal O(\mathrm{Re}_\chi)$. The passive field $\bm Q^{(0)}$ obeys the standard Oseen equation. We decompose it as
\begin{equation}
    \bm Q^{(0)}=\bm\nabla\Phi+\bm Q'',\qquad \Phi(R,\theta)=A_0\log R+A_1\frac{\cos\theta}{R},
    \label{eq:1}
\end{equation}
and use the Oseen representation~\cite{batchelor2000introduction}
\begin{equation}
    \bm Q''=s\hat{\bm e}_x-\frac{1}{2\delta}\bm\nabla s,
    \qquad
    s=C_0e^{\delta R\cos\theta}K_0(\delta R).
    \label{eq:2}
\end{equation}
This gives
\begin{align}
    Q_r^{(0)}&=\frac{A_0}{R}-A_1\frac{\cos\theta}{R^2}
    +\frac{C_0}{2}e^{\delta R\cos\theta}
    \left[K_1(\delta R)+\cos\theta K_0(\delta R)\right],
    \label{eq:Qr_general}\\
    Q_\theta^{(0)}&=-A_1\frac{\sin\theta}{R^2}
    -\frac{C_0}{2}e^{\delta R\cos\theta}K_0(\delta R)\sin\theta.
    \label{eq:Qtheta_general}
\end{align}
For $\delta\ll1$, the boundary conditions yield
\begin{equation}
    A_1=0,\qquad C_0=-\frac{2}{\Lambda},\qquad A_0=\frac{1}{\delta\Lambda},\qquad \Lambda=1+\log\left(\frac{2}{\delta}\right)-\gamma_E.
\end{equation}
The corresponding passive pressure is
\begin{equation}
    \bar{\mathtt p}^{(0)}(r,\theta) = -\frac{2\eta U_\infty}{R_0\Lambda}\frac{\cos\theta}{R},
\end{equation}
and the passive normal stress on the intruder is
\begin{equation}
    \Sigma_{rr}^{(0)}(R_0,\theta)=-p_\infty +\frac{4\eta U_\infty}{R_0\Lambda}\cos\theta+\mathcal O\left(\frac{\eta U_\infty}{R_0\Lambda^2},\frac{\eta U_\infty}{R_0}\delta\log\delta\right).
\end{equation}
It therefore produces the usual two-dimensional Oseen drag with perfect-slip boundary conditions,
\begin{equation}
    \bm F_{\partial\mathcal B}^{(0)}=\frac{4\pi\eta U_\infty}{\Lambda}\hat{\bm e}_x+\mathcal O\left(\frac{\eta U_\infty}{\Lambda^2}, \eta U_\infty\delta\log\delta\right).
\end{equation}

\subsection{Chiral correction}

Taking the curl of Eq.~\eqref{eq:stokes_adim}, we obtain
\begin{equation}
    \left(\bm\nabla^2-2\delta\partial_X\right)\Omega+\frac{\mathrm{Re}_\chi}{R^2}\partial_\theta\Omega=0,\qquad \Omega=(\bm\nabla\times\bm Q)_z.
    \label{eq:oseen_vorticity_chiral}
\end{equation}
The second term describes the advection of the translational vorticity by the circular edge current. In the inner Oseen region, $1\lesssim R\ll\delta^{-1}$, the total passive translational flow $\bm U_\infty+\bm u^{(T,0)}$ has streamfunction
\begin{equation}
    \psi^{(0)}(R,\theta)=\frac{U_\infty R_0}{\Lambda}R\log R\sin\theta+\mathcal O\left(\frac{U_\infty R_0}{\Lambda^2},U_\infty R_0\delta R\log(\delta R)\right),
\end{equation}
and vorticity
\begin{equation}
    \omega^{(0)}(R,\theta)=-\frac{2U_\infty}{R_0\Lambda}\frac{\sin\theta}{R}+\mathcal O\left(\frac{U_\infty}{R_0\Lambda^2}, \frac{U_\infty}{R_0}\delta\log(\delta R)\right).
    \label{eq:omega_inner_oseen}
\end{equation}
At first order in $\mathrm{Re}_\chi$, the Oseen advection of the correction is subleading in the inner region. We introduce
\begin{equation}
    \psi^{(1)}(R,\theta)=U_\infty R_0\Phi(R)\cos\theta,
\end{equation}
so that
\begin{equation}
    u_r^{(1)}=-U_\infty\frac{\Phi(R)}{R}\sin\theta, \qquad u_\theta^{(1)}=-U_\infty\Phi'(R)\cos\theta.
\end{equation}
Using Eq.~\eqref{eq:omega_inner_oseen}, the curl of the first-order momentum equation becomes
\begin{equation}
    \mathcal L_1^2\Phi=-\frac{2\mathrm{Re}_\chi}{\Lambda R^3},\qquad \mathcal L_1=\frac{d^2}{dR^2}+\frac{1}{R}\frac{d}{dR}-\frac{1}{R^2}.
\end{equation}
The homogeneous boundary conditions are
\begin{equation}
    \Phi(1)=0,\qquad\left[\Phi''-\frac{\Phi'}{R}+\frac{\Phi}{R^2}\right]_{R=1}=0.
\end{equation}
Since
\begin{equation}
    \mathcal L_1^2\left[R(\log R)^2\right]=-\frac{8}{R^3},
\end{equation}
the inner solution is
\begin{equation}
    \Phi(R)=\frac{q}{2}\left(R-\frac{1}{R}\right)+CR\log R+qR(\log R)^2,\qquad q=\frac{\mathrm{Re}_\chi}{4\Lambda}.
    \label{eq:inner_chiral_streamfunction}
\end{equation}
The constant $C$ is fixed by matching to the outer Oseen region. At $R\sim\delta^{-1}$, $\log R=\Lambda+\mathcal O(1)$, and the two logarithmically growing terms in Eq.~\eqref{eq:inner_chiral_streamfunction} must combine into a bounded outer solution. Hence
\begin{equation}
    C=-q\Lambda+\mathcal O(q)=-\frac{\mathrm{Re}_\chi}{4}+\mathcal O\left(\frac{\mathrm{Re}_\chi}{\Lambda}\right).
    \label{eq:C_chiral_matching}
\end{equation}
This matching also shows that the perturbative expansion requires $\mathrm{Re}_\chi\Lambda\ll1$.

The modified pressure associated with Eq.~\eqref{eq:inner_chiral_streamfunction} is
\begin{equation}
    \bar{\mathtt p}^{(1)}(R,\theta)=\frac{2\eta U_\infty}{R_0}\frac{C-q+2q\log R}{R}\sin\theta.
\end{equation}
The physical pressure differs from the modified pressure by the gradient contribution absorbed in Eq.~\eqref{eq:vorticity}. At first order in $c$,
\begin{equation}
    \mathtt p^{(1)}=\bar{\mathtt p}^{(1)}+\rho_b\frac{c}{r}v_\theta^{(0)},
\end{equation}
where $\bm v^{(0)}=\bm U_\infty+\bm u^{(T,0)}$ and
\begin{equation}
    v_\theta^{(0)}(R_0,\theta) =-\frac{U_\infty}{\Lambda}\sin\theta +\mathcal O\left(\frac{U_\infty}{\Lambda^2},U_\infty\delta\log\delta\right).
\end{equation}
The first-order normal stress is therefore
\begin{equation}
    \Sigma_{rr}^{(1)}(R_0,\theta)=\frac{4\eta U_\infty}{R_0}(q-C)\sin\theta+\mathcal O\left(\frac{\rho_bcU_\infty}{R_0\Lambda}\right).
    \label{eq:sigmarr_boundary_circle}
\end{equation}
Using
\begin{equation}
    \bm F_{\partial\mathcal B}=R_0\int_0^{2\pi}d\theta\,\Sigma_{rr}(R_0,\theta)\hat{\bm e}_r,
    \label{eq:F_boundary_def}
\end{equation}
and Eq.~\eqref{eq:C_chiral_matching}, we obtain
\begin{equation}
    \bm F_{\partial\mathcal B}^{(1)}=\pi\rho_bcU_\infty\hat{\bm e}_y+\mathcal O\left(\frac{\rho_bcU_\infty}{\Lambda},\eta U_\infty\mathrm{Re}_\chi^2\right).
\end{equation}
Combining the passive and chiral contributions gives
\begin{equation}
    \bm F_{\partial\mathcal B}=\frac{4\pi\eta U_\infty}{\Lambda}\hat{\bm e}_x+\pi\rho_bcU_\infty\hat{\bm e}_y +\mathcal O\left(\frac{\eta U_\infty}{\Lambda^2},\frac{\rho_bcU_\infty}{\Lambda},\eta U_\infty\mathrm{Re}_\chi^2\right).
    \label{eq:appendix_important}
\end{equation}

\subsection{Velocity and pressure fields}

Putting everything together, we obtain the translational fields in the inner Oseen region, $1\lesssim R\ll\delta^{-1}$, perturbatively in the chiral Reynolds number $\mathrm{Re}_\chi$. It is convenient to introduce the first-order streamfunction correction
\begin{equation}
    \Phi_\chi(R)=\frac{q}{2}\left(R-\frac{1}{R}\right)
    +CR\log R+qR(\log R)^2,\qquad q=\frac{\mathrm{Re}_\chi}{4\Lambda},\qquad C=-\frac{\mathrm{Re}_\chi}{4}+\mathcal O\left(\frac{\mathrm{Re}_\chi}{\Lambda}\right),\qquad R=\frac{r}{R_0},
\end{equation}
whose radial derivative is
\begin{equation}
    \Phi_\chi'(R)=\frac{q}{2}\left(1+\frac{1}{R^2}\right) +C(1+\log R) +q\left[(\log R)^2+2\log R\right].
\end{equation}
The translational contribution to the velocity and pressure fields is therefore
\begin{align}
\bm U_\infty\cdot \hat{\bm e}_r+u_r^{(T)}(r,\theta)&=U_\infty\Bigg[
\cos(\theta)+\frac{1}{\delta\Lambda}\frac{R_0}{r}-\frac{1}{\Lambda}e^{\delta R\cos(\theta)}\Big(K_1(\delta R)+\cos(\theta)K_0(\delta R)\Big)-\frac{\Phi_\chi(R)}{R}\sin(\theta)\Bigg],\\
\bm U_\infty\cdot \hat{\bm e}_\theta+u_\theta^{(T)}(r,\theta)&=U_\infty\Bigg[-1+\frac{1}{\Lambda}e^{\delta R\cos(\theta)}K_0(\delta R)
\Bigg]\sin(\theta)-U_\infty\Phi_\chi'(R)\cos(\theta),\\
\mathtt p^{(T)}(r,\theta)&=-\frac{2\eta U_\infty}{R_0\Lambda}\frac{\cos(\theta)}{R}+\frac{\rho_b cU_\infty}{\Lambda r}e^{\delta R\cos(\theta)}K_0(\delta R)\sin(\theta)+\frac{2\eta U_\infty}{R_0}\frac{C-q+2q\log R}{R}\sin(\theta).
\end{align}
These expressions are valid up to terms of order $\mathcal O(\mathrm{Re}_\chi^2)$ and corrections of order $\mathcal O(\mathrm{Re}_\chi\delta R)$ in the inner region.

Including the circular edge current and its pressure contribution, the full fields are
\begin{align}
u_r(r,\theta)&=U_\infty\Bigg[\cos(\theta)+\frac{1}{\delta\Lambda}\frac{R_0}{r}-\frac{1}{\Lambda}e^{\delta R\cos(\theta)}\Big(K_1(\delta R)+\cos(\theta)K_0(\delta R)\Big)-\frac{\Phi_\chi(R)}{R}\sin(\theta)\Bigg],\label{eq:a}\\
u_\theta(r,\theta)&=-\frac{c}{r}+U_\infty\Bigg[-1+\frac{1}{\Lambda}e^{\delta R\cos(\theta)}K_0(\delta R)\Bigg]\sin(\theta)-U_\infty\Phi_\chi'(R)\cos(\theta),\label{eq:b}\\
\mathtt p(r,\theta)&=p_\infty-\frac{2\eta U_\infty}{R_0\Lambda}\frac{\cos(\theta)}{R}-\rho_bU_\infty c\left[1-\frac{1}{\Lambda}e^{\delta R\cos(\theta)}K_0(\delta R)\right]\frac{\sin(\theta)}{r}+\frac{2\eta U_\infty}{R_0}\frac{C-q+2q\log R}{R}\sin(\theta)-\frac{\rho_b c^2}{2r^2}.\label{eq:c}
\end{align}

\section{Inertial contribution to the fields around a noncircular intruder}\label{sec:inertial_Stokes}
In this Appendix, we derive perturbatively the inertial corrections to the fields induced by a noncircular intruder.

We keep the first convective correction to the Stokes equation and work at first order in both the shape parameter $\varepsilon$ and the ‘‘chiral'' Reynolds number:
\begin{equation}
\mathrm{Re}_{\chi} = \frac{\rho_b c}{\eta},\qquad c =\frac{\tau_0 R_0^2}{2\eta},\qquad x= m(\theta-\phi),
\end{equation}
where we introduce $x$ for notational convenience. The steady incompressible chiral Navier-Stokes equation is
\begin{equation}
-\bm\nabla \mathtt p + \eta \bm\nabla^2 \bm u = \rho_b (\bm u\cdot\bm\nabla)\bm u,\qquad\bm\nabla\cdot\bm u = 0,
\label{eq:ns_chiral}
\end{equation}
with the same impermeability and tangential boundary conditions as in Eq.~\eqref{eq:boundary_condition_stokes}.

We decompose the solution as:
\begin{equation}
\bm u = \bm u^{(S)} + \bm u^{(I)} + \mathcal O(\varepsilon^2)+\mathcal O(\mathrm{Re}_\chi^2), \qquad \mathtt p = \mathtt p^{(S)} + \mathtt p^{(I)} + \mathcal O(\varepsilon^2)+\mathcal O(\mathrm{Re}_\chi^2),
\label{eq:low_reynolds_pertubation}
\end{equation}
where $\bm u^{(S)}$ and $\mathtt p^{(S)}$ are the Stokes fields Eqs.~\eqref{eq:uthetaapp} and \eqref{eq:papp} while $\bm u^{(I)}$ and $\mathtt p^{(I)}$ are the first order inertial corrections, in a low chiral Reynolds expansion. By replacing Eq.~\eqref{eq:low_reynolds_pertubation} into Eq.~\eqref{eq:ns_chiral}, we find:
\begin{equation}
    -\bm \nabla \mathtt p^{(I)}+\eta \bm\nabla^2 \bm u^{(I)}=\rho_b(\bm u^{(S)}\cdot\bm\nabla)\bm u^{(S)} + \mathcal O(\varepsilon^2)+\mathcal O(\mathrm{Re}_\chi^2).
    \label{eq:oseen}
\end{equation}
In the following, we drop the $\mathcal O(\varepsilon^2)+\mathcal O(\mathrm{Re}_\chi^2)$ contribution.

It is useful to further expand the fields:
\begin{equation}
    \bm u=\underbrace{\vphantom{\frac12}\bm u^{(0,S)}+\varepsilon\bm u^{(1,S)}}_{\bm u^{(S)}}+\underbrace{\mathrm{Re}_\chi\left(\bm u^{(0,I)}+\varepsilon\bm u^{(1,I)}\right)}_{\bm u^{(I)}},\qquad \mathtt p=\underbrace{\vphantom{\frac12}\mathtt p^{(0,S)}+\varepsilon \mathtt p^{(1,S)}}_{\mathtt p^{(S)}}+\underbrace{\mathrm{Re}_\chi\left(\mathtt p^{(0,I)}+\varepsilon \mathtt p^{(1,I)}\right)}_{\mathtt p^{(I)}},
\end{equation}
where $\bm u^{(0, I)}$ and $\mathtt p^{(0, I)}$ are the inertial contributions for a circular intruder. They can be computed (even non-perturbatively) and lead to:
\begin{equation}
 \mathtt p^{(0, I)}(r) = -\frac{\rho_b c^2}{2r^2}, \qquad \bm u^{(0, I)}(r)=0,
\end{equation}
which corresponds to a centrifugal pressure shift that can also be found via Bernoulli's equation.

To obtain the inertial corrections for non-circular shapes, we write the full inertial contribution in terms of a streamfunction
\begin{equation}
 u_r^{(I)} = \frac{1}{r} \partial_\theta \psi^{(I)},\qquad u_\theta^{(I)} = -\partial_r \psi^{(I)},
 \label{eq:inertia_stream}
\end{equation}
and taking the curl of Eq.~\eqref{eq:oseen}, we obtain
\begin{equation}
 \eta \bm\nabla^4\psi^{(I)} = -\rho_b\left[\bm\nabla\times\left((\bm u^{(S)}\cdot\bm\nabla)\bm u^{(S)}\right)\right]_z= 2\varepsilon \rho_b c^2 (m-1)^2(m+2)R_0^{m-2}r^{-m-2}\sin(x).
 \label{eq:chiral_stream_inertia}
\end{equation}
Based on the form of Eq.~\eqref{eq:chiral_stream_inertia}, we seek a $m$-fold symmetric separable solution:
\begin{equation}
 \psi^{(I)}(r,\theta)=\varepsilon P(r)\sin(x),
\end{equation}
with $P(r)$ decaying to 0 as $r\to\infty$. The boundary conditions
\begin{equation}
    u_r(R(\theta),\theta)=0,\qquad\Sigma_{r\theta}(R(\theta),\theta)=\tau_0,
\end{equation}
must still be imposed. The Stokes solution already incorporates $\tau_0$, and we therefore find:
\begin{equation}
 u_r^{(I)}(R_0,\theta)=0,\qquad\Sigma^{(I)}_{r\theta}(R_0,\theta)=0.
\end{equation}
These boundary conditions readily lead to the solution:
\begin{equation}
 \psi^{(I)}(r,\theta)= \varepsilon \frac{\rho_b c^2 R_0^{m-2}(m-1)(m+2)}{4\eta m}\left[r^{2-m}\left(\log\left(\frac{r}{R_0}\right)-\frac{m-1}{2m}\right)+\frac{m-1}{2m}R_0^2r^{-m} \right]\sin(x).
\end{equation}
The fields follow from Eq.~\eqref{eq:inertia_stream}:
\begin{align}
 u_r^{(I)}(r,\theta) &= \varepsilon \frac{\rho_b c^2 R_0^{m-2}(m-1)(m+2)}{4\eta}r^{1-m}\left[\log\left(\frac{r}{R_0}\right)+\frac{m-1}{2m}\left(\frac{R_0^2}{r^2}-1\right)\right]\cos(x), \label{eq:ur_inertial_weak_shape}\\
 u_\theta^{(I)}(r,\theta)&= \varepsilon \frac{\rho_b c^2 R_0^{m-2}(m-1)(m+2)}{4\eta m}r^{1-m}\left[(m-2)\log\left(\frac{r}{R_0}\right) -\frac{m^2-m+2}{2m}+\frac{m-1}{2}\frac{R_0^2}{r^2}\right]\sin(x),\label{eq:uth_inertial_weak_shape}\\
 \mathtt p^{(I)}(r,\theta)&= -\frac{\rho_b c^2}{2r^2}+\varepsilon \rho_b c^2R_0^{m-2}\Bigg[\frac{(m-1)^2(m+2)}{m}\frac{\log(r/R_0)}{r^m}-\frac{(m-1)^2(m+1)(m+2)}{2m^2}\frac{1}{r^m} \\
 &\hspace{4.cm}+\frac{(m-2)(m+1)}{2}\frac{R_0^2}{r^{m+2}}\Bigg]\cos(x).\label{eq:p_inertial_weak_shape}
\end{align}
The full fields on the boundary have simple expressions:
\begin{align}
 u_r\big|_{\partial \mathcal B}&= \varepsilon\frac{\tau_0R_0}{2\eta}m\sin(x),\\
 u_\theta\big|_{\partial \mathcal B}&=-\frac{\tau_0R_0}{2\eta}+\varepsilon \frac{\tau_0R_0}{\eta}\frac{m-1}{m}\cos(x)-\varepsilon \frac{\rho_b \tau_0^2 R_0^3}{16\eta^3 }\frac{(m-1)(m+2)}{m^2}\sin(x), \\
 \mathtt  p\big|_{\partial \mathcal B}&= p_\infty-\frac{\rho_b\tau_0^2R_0^2}{8\eta^2}+\varepsilon \tau_0\frac{(m-1)^2(m+2)}{m}\sin(x)-\varepsilon \frac{\rho_b\tau_0^2R_0^2}{8\eta^2}\frac{(m-1)(2m^2-m-2)}{m^2}\cos(x).
\end{align}
The key difference with the purely Stokes result is that inertia generates a new $\cos(x)$ harmonic in the pressure and a new $\sin(x)$ harmonic in the tangential boundary velocity. These contributions are quadratic in the chiral driving and therefore remain unchanged under reversal of the circulation. They also produce a slight shift in the location of the pressure maxima, but they are insufficient to explain the asymmetric distribution observed in the simulations. Those effects lie beyond our present approximation and will arise at order $\mathcal O(\varepsilon\mathrm{Re}_\chi^2)$ or $\mathcal O(\varepsilon^2\mathrm{Re}_\chi)$, where mode-coupling will generate both $\cos(2x)$ and $\sin(2x)$, which spoils the symmetry: $p(x)\leftrightarrow-p(x+\psi)$ with $|\psi|<2\pi/m$. Note that inertia is really required, as Stokes terms in $\mathrm {Re}_\chi^0\varepsilon^n$ are not sufficient to generate an asymmetric pressure because they would produce only sinusoidal harmonics, which always keep the $p(x)\leftrightarrow -p(x+\psi)$ symmetry. 

Finally, since they are in quadrature with the previous contributions (a $\sin$ for the tangential velocity field and a $\cos$ for the pressure field), they will not contribute to the torque applied onto the intruder.

\end{widetext}
\bibliography{bib}

\clearpage
\onecolumngrid

\section*{Supplemental Material: List of symbols}

\subsection*{General}

\begin{symbollist}
\sym{$\bm Y(t)=(\bm X(t), \varphi(t))$}{Intruder ‘‘configuration array'' with center-of-mass position and orientation.}
\sym{$\bm U(t)=(\bm V(t), \Omega(t))$}{Intruder ‘‘velocity array'' with translational and angular velocities.}
\sym{$M$}{Intruder mass.}
\sym{$I$}{Intruder moment of inertia.}
\sym{$\mathcal P$}{Intruder perimeter.}
\sym{$\mathcal A$}{Intruder signed area.}
\sym{$\mathsf R(\varphi)$}{$2\times 2$ rotation matrix of angle $\varphi$.}
\sym{$\bar{\mathsf R}(\varphi)=\mathrm{diag}(\mathsf R(\varphi),1)$}{Generalized rotation matrix.}
\sym{$s$}{Boundary parameter with $\oint ds \equiv \int_0^1 ds  |\partial_s \bm \rho_0|$.}
\sym{$\bm \rho_0(s)$}{Boundary parametrization of the intruder in the body frame.}
\sym{$\bm r_0(s,t)=\mathsf{R}(\varphi(t))\cdot\bm{\rho}_0(s)$}{Boundary point position relative to the center of mass.}
\sym{$\bm r^{(c)}(s,t)=\bm{X}(t)+\bm r_0(s, t)$}{Position of a boundary point in the laboratory frame.}
\sym{$\bm V^{(c)}=\bm V+\Omega\hat{\bm z}\times \bm r_0$}{Intruder velocity at the contact point or at the boundary point considered.}
\sym{$\hat{\bm n}, \hat{\bm t}$}{Outward unit normal $\bm n$ and unit tangent, $\hat{\bm t}=\hat{\bm z}\times \hat{\bm n}$ to the intruder boundary.}
\sym{$a,b$}{Translational Cartesian indices, $a,b\in\{x,y\}$.}
\sym{$i,j,k$}{Generalized indices in $i, j, k\in\{x,y,\varphi\}$.}
\sym{$\bm \varepsilon$, $\varepsilon_{ab}$}{$2\times2$ Levi-Civita pseudotensor}
\sym{$\tilde{(\cdot)}$}{Quantity expressed in the body frame.}
\sym{$\bm F = (F_x, F_y, F_\varphi)$}{ Intruder ‘‘force array'' with $F_x$ and $F_y$ the force on the intruder and $F_\varphi$ the torque.}
\sym{$\Upgamma$}{Generalized damping matrix.}
\sym{$\mathsf D$}{Noise covariance.}
\end{symbollist}

\subsection*{Dilute regime}

\begin{symbollist}
\sym{$m$}{Bath-particle mass.}
\sym{$\bm v,\bm v'$}{Pre- and post-collisional bath-particle velocities.}
\sym{$n_b$}{Bath number density.}
\sym{$f(\bm v)$}{Bath velocity distribution.}
\sym{$T_b=m\int d\bm v\bm v^2 f(\bm v)/2$}{Bath kinetic temperature.}
\sym{$\alpha$}{Bath-intruder normal restitution coefficient.}
\sym{$\Delta$}{Chiral tangential kick in the bath-intruder collision rule.}
\sym{$\bm J$}{Impulse transferred during a collision, $J_n=\bm J\cdot \hat{\bm n}$ and $J_t=\bm J\cdot \hat{\bm t}$.}
\sym{$\bm g=\bm v-\bm V^{(c)}$}{Relative velocity at contact, $g_n=\bm g\cdot \hat{\bm n}$ and $g_t=\bm g\cdot \hat{\bm t}$}
\sym{$a=m^{-1}+M^{-1}$}{Inverse sum of mass (not to be confused with the triangle base length later).}
\sym{$\kappa_n=\lbrack \bm r_0\times \hat{\bm n}\rbrack_z$}{Geometric lever-arm factor.}
\sym{$\kappa_t=\lbrack \bm r_0\times \hat{\bm t}\rbrack_z$}{Tangent geometric lever-arm factor.}
\sym{$\lambda_n=a+\kappa_n^2/I$}{Effective normal inverse reduced mass.}
\sym{$\lambda_t=a+\kappa_t^2/I$}{Effective tangential inverse reduced mass.}
\sym{$\mu_n={M}^{-1}+{\kappa_n^2}/{I}$}{Geometric mass factor.}
\sym{$P(\bm U,\bm Y,t)$}{Intruder probability density in velocity and configuration space.}
\sym{$W(\bm U'|\bm U,\bm Y)$}{Boltzmann-Lorentz transition rate.}
\sym{$\bm \Pi=(MV_x,MV_y,I\Omega)$}{Generalized momentum array.}
\sym{$\mathcal M_{\ell_x\ell_y\ell_\varphi}$}{Kramers-Moyal jump moments.}
\sym{$\Delta \Pi$}{Change in generalized momentum at collision.}
\sym{$\bm G,\bm H$}{Geometric vectors entering the linear decomposition $\Delta \bm \Pi=\bm G  g_n+\bm H$.}
\sym{$\bm e^{(n)}=(\hat n_x,\hat n_y,\kappa_n)$}{Generalized normal array.}
\sym{$\bm e^{(t)}=(\hat t_x,\hat t_y,\kappa_t)$}{Generalized tangential array.}
\sym{$\phi(v_n)$}{Marginal of the bath velocity distribution along the normal direction.}
\sym{$\psi(c)$}{Rescaled dimensionless normal-velocity distribution.}
\sym{$\mathcal K_r=\int_0^\infty dc  c^r\psi(c)$}{Positive half moments of $\psi(c)$.}
\sym{$v_{\rm th}=\sqrt{T_b/m}$}{Thermal velocity scale.}
\sym{$\epsilon$}{Bookkeeping parameter for the van Kampen small-mass expansion, $m\to \epsilon m$.}
\sym{$\mathcal I_r(U_n)$}{Normal-velocity integrals appearing in the jump moments.}
\sym{$U_n=\bm U\cdot \bm e^{(n)}=\bm V\cdot \hat{\bm n}+\Omega \kappa_n$}{Normal contact velocity of the intruder.}
\sym{$A_i$}{First Kramers-Moyal coefficient (generalized drift).}
\sym{$B_{ij}$}{Second Kramers-Moyal coefficient.}
\sym{$C_{ijk}$}{Third Kramers-Moyal coefficient.}
\sym{$R_i$}{Velocity-independent part of the drift.}
\sym{$\Gamma_{ij}$}{Linear damping/drift matrix in the dilute Langevin equation.}
\sym{$N_{ijk}$}{Quadratic drift tensor.}
\sym{$D_{ij}$}{Additive diffusion/noise covariance tensor in the reduced Fokker-Planck equation.}
\sym{$E_{ijk}$}{Multiplicative-noise correction tensor.}
\sym{$\mathsf M=\mathrm{diag}(M,M,I)$}{Generalized mass array.}
\sym{$\mathsf S$}{Covariance matrix of the velocity array.}
\sym{$\mathsf S^{\rm lin}$}{Covariance matrix in the linearized theory.}
\sym{$T_I$}{Intruder kinetic temperature in the weak-chiral-driving limit. It also enters the Fluctuation-Dissipation relation of the $(\alpha)$ sector.}
\sym{$\bm \zeta$}{Gaussian white noise entering the dilute Langevin equation.}
\sym{$\bm F^{(\alpha)}$}{Contribution to $\bm F$ from the normal/dissipative sector.}
\sym{$\bm F^{(\Delta)}$}{Contribution to $\bm F$ from the chiral collision sector.}
\sym{$\Upgamma^{(\alpha)}$}{Damping from the normal/dissipative sector.}
\sym{$\Upgamma^{(\Delta)}$}{Damping from the chiral collision sector.}
\sym{$\mathsf D^{(\alpha)}$}{Diffusion tensor from the normal/dissipative sector.}
\sym{$\mathsf D^{(\Delta)}$}{Diffusion tensor from the chiral collision sector.}
\sym{$\Gamma_{ij}^{(\Delta),{\rm S}}$}{Symmetric part of $\Gamma_{ij}^{(\Delta)}$.}
\sym{$\Gamma_{ij}^{(\Delta),{\rm A}}$}{Antisymmetric part of $\Gamma_{ij}^{(\Delta)}$.}
\sym{$Q_{ab}=\oint ds\hat n_a\hat n_b$}{Translational geometric tensor.}
\sym{$Q_{ab}^{\rm dev}=Q_{ab}-\dfrac{\rm{Tr}(\mathsf Q)}{2}\delta_{ab}$}{Deviatoric part of $Q_{ab}$, $Q_{ab}^{\rm dev}=Q_{ab}-\dfrac{\mathcal P}{2}\delta_{ab}$.}
\sym{$W_a=\oint ds\kappa_n \hat n_a$}{Geometric vector.}
\sym{$S_a=\oint ds\kappa_t \hat n_a$}{Geometric vector.}
\sym{$T_a=\oint ds  \kappa_n \hat t_a$}{Geometric vector.}
\sym{$P_a=\oint ds \kappa_n^2 \hat n_a$}{Geometric vector.}
\sym{$H_n=\oint ds  \kappa_n^2$}{Scalar geometric moment.}
\sym{$H_t=\oint ds  \kappa_n\kappa_t$}{Scalar geometric moment.}
\sym{$H_\varphi=\oint ds  \kappa_n^3$}{Pseudoscalar geometric moment.}
\sym{$\mathcal D_{ab}=\mathcal D_{\rm even}\delta_{ab}+\mathcal D_{\rm odd}\varepsilon_{ab}$}{Long-time translational diffusivity tensor and its even and odd contribution.}
\sym{$D_V$}{Translational noise variance for the polygon example.}
\sym{$\gamma$}{Diagonal translational damping coefficient in the polygon example.}
\sym{$\omega$}{Off-diagonal translational damping coefficient in the polygon example.}
\sym{$\Gamma_0^{(\alpha)}=(1+\alpha)\mathcal K_1n_b\sqrt{mT_b}$}{Shape-independent prefactor in several dilute examples.}
\sym{$R$}{Circumradius for a regular polygon and outer radius for the wheel.}
\sym{$r$}{Inner radius of the wheel example.}
\sym{$\ell$}{Side length of a regular polygon; for the triangle, $\ell=\sqrt{h^2+a^2/4}$.}
\sym{$n$}{Number of vertices (regular polygon) or number of teeth (wheel).}
\sym{$\delta$}{Angular offset of the inner vertices in the wheel construction.}
\sym{$\chi_{\rm w}=\delta-\pi/n$}{Shape-chirality parameter of the wheel.}
\sym{$\theta_\pm=\pi/n\pm \chi_{\rm w}$}{Angle between two vertices for the wheel.}
\sym{$\ell_\pm$}{Two alternating edge lengths of the wheel.}
\sym{$u_\pm$}{Auxiliary geometric quantities in the wheel formulas.}
\sym{$\kappa_{0\pm}$}{Initial values of $\kappa_n$ on the two edge types of the wheel.}
\sym{$\mathcal H_q$}{Edge integral used to compute $H_n$ and $H_\varphi$ for the wheel.}
\sym{$a$}{Triangle base length in the triangle example.}
\sym{$h$}{Triangle height in the triangle example.}
\end{symbollist}

\subsection*{Dense regime}

\begin{symbollist}
\sym{$\alpha_b$}{Bath-bath restitution coefficient in the chiral bath collision rule.}
\sym{$\Delta_b$}{Bath-bath transverse kick in the chiral bath collision rule.}
\sym{$\bm v_{\alpha\beta}$}{Relative velocity of bath particles $\alpha$ and $\beta$.}
\sym{$\hat{\bm \sigma}_{\alpha\beta}$}{Unit vector joining the centers of bath particles $\alpha$ and $\beta$ at contact.}
\sym{$n(\bm r,t)$}{Hydrodynamic number-density field of the bath.}
\sym{$\bm u(\bm r,t)$}{Hydrodynamic velocity field of the bath.}
\sym{$\rho_b$}{Bath mass density, $\rho_b=mn=mn_b$ in the homogeneous incompressible state.}
\sym{$\mathsf \Sigma^{\rm h}$}{Homogeneous stress tensor.}
\sym{$\mathsf \Sigma^\eta$}{Viscous stress tensor.}
\sym{$p$}{Homogeneous thermodynamic pressure entering $\mathsf \Sigma^{\rm h}$.}
\sym{$\tau$}{Torque density in the bath stress.}
\sym{$\mathsf \Sigma^{\rm virial}$}{Virial stress tensor.}
\sym{$\bm F_{\alpha\beta}$}{Microscopic impulsive force between bath particles $\alpha$ and $\beta$.}
\sym{$\eta$}{Shear viscosity.}
\sym{$\eta_o$}{Odd viscosity.}
\sym{$\eta_R$}{Rotational viscosity coefficient.}
\sym{$\mathtt p$}{Mechanical pressure enforcing incompressibility.}
\sym{$\tau_0=\tau(n_b)$}{Constant bulk torque density in the homogeneous and incompressible fluid outside the intruder.}
\sym{$\mathcal B$}{Interior region occupied by the intruder.}
\sym{$\partial \mathcal B$}{Intruder boundary.}
\sym{$\mathcal D$}{Exterior fluid domain.}
\sym{$\delta_{\partial\mathcal B}$}{Dirac delta distribution supported on the boundary.}
\sym{$\lambda$}{Lagrange multiplier enforcing impermeability in the traction decomposition.}
\sym{$\xi$}{Tangential friction coefficient at the boundary; $\xi=0$ for a smooth intruder.}
\sym{$\bm f_{\partial\mathcal B}$}{Actual traction transmitted by the fluid to the intruder.}
\sym{$\ell_\gamma=\sqrt{\eta/\gamma}$}{Damping length.}
\sym{$\gamma$}{External/substrate damping coefficient used to regularize edge currents near flat walls.}
\sym{$\delta y$}{Distance to a flat wall in the large-separation edge-current solution.}
\sym{$R_0$}{Radius of the circular intruder, or mean radius in the weakly deformed-shape calculation.}
\sym{$\bm U_\infty$}{Uniform flow imposed at infinity in the Oseen problem.}
\sym{$U_\infty$}{Magnitude of the far-field uniform flow.}
\sym{${\rm Re}_\infty=\rho_b U_\infty R_0/\eta$}{Reynolds number based on the imposed far-field flow.}
\sym{$\delta={\rm Re}_\infty/2$}{Oseen parameter.}
\sym{$c=\tau_0R_0^2/(2\eta)$}{Chiral flow amplitude for a circular intruder.}
\sym{$\bm u^{(T)}$}{Translational disturbance field in the Oseen decomposition.}
\sym{$\bm u^{(\chi)}$}{Chiral azimuthal flow induced by the torque density around a circular intruder.}
\sym{$\mathtt p^{(T)}$}{Pressure contribution associated with the translational disturbance.}
\sym{$\mathtt p^{(\chi)}$}{Pressure contribution associated with the chiral flow.}
\sym{$\bar{\mathtt p}^{(T)}$}{Shifted translational pressure after absorbing gradient terms.}
\sym{$R=r/R_0$}{Dimensionless radial coordinate in the Oseen calculation.}
\sym{$\bm X=\bm x/R_0$}{Dimensionless Cartesian coordinate in the Oseen calculation.}
\sym{$\bm Q=\bm u^{(T)}/U_\infty$}{Dimensionless translational disturbance field.}
\sym{$\mathit\Pi=\bar{\mathtt p}^{(T)}/\rho_b U_\infty^2$}{Dimensionless translational pressure.}
\sym{$s$}{Auxiliary scalar field in the Oseen representation.}
\sym{$\bar s$}{Reduced scalar field solving the modified Helmholtz equation in the Oseen representation.}
\sym{$\Lambda=1+\log(2/\delta)-\gamma_E$}{Oseen logarithmic factor.}
\sym{$\gamma_E$}{Euler's constant.}
\sym{$K_0,K_1$}{Modified Bessel functions of the second kind.}
\sym{$\Upgamma_{\rm trans}$}{Effective translational drag matrix in the hydrodynamic/Oseen calculation.}
\sym{$\hat{\bm e}_r,\hat{\bm e}_\theta$}{Polar basis vectors.}
\sym{$u_r,u_\theta$}{Radial and azimuthal components of the fluid velocity.}
\sym{$\varepsilon$}{Small shape-deformation amplitude in the weakly noncircular dense calculation (distinct from the dilute bookkeeping parameter $\epsilon$).}
\sym{$m$}{Mode number of the boundary deformation $R(\theta)=R_0[1+\varepsilon \cos(m(\theta-\phi))]$ (distinct from the dilute bath-particle mass).}
\sym{$\phi$}{Orientation of the weak boundary deformation.}
\sym{$R(\theta)$}{Radius of the weakly noncircular intruder as a function of polar angle.}
\sym{$h(\theta)=R_0\cos(m(\theta-\phi))$}{Boundary-shape perturbation.}
\sym{${\rm Re}_\chi=\rho_b c/\eta$}{Chiral Reynolds number.}
\sym{$\bm u^{(S)},\mathtt p^{(S)}$}{Stokes-order velocity and pressure fields.}
\sym{$\bm u^{(I)},\mathtt p^{(I)}$}{First inertial corrections to the velocity and pressure fields.}
\sym{$\mathit\Omega_S(\bm r,t)$}{Local spin field of the chiral fluid when spin is retained explicitly.}
\end{symbollist}

\end{document}